\documentclass[11pt]{article}
\pdfoutput=1
\usepackage{amsmath,amssymb,epsfig,amsfonts}
\usepackage{graphicx}
\usepackage{subfig}
\usepackage[usenames, dvipsnames]{color}
\usepackage{hyperref}
\usepackage{cite}
\usepackage{multirow}
\usepackage{slashed}
\usepackage{verbatim} 
\usepackage{enumitem}
\usepackage{rotating}
\usepackage{xcolor}
\usepackage{multirow}
\usepackage{bm}
\usepackage[normalem]{ulem}
\usepackage{cleveref}
\usepackage{booktabs} 
\usepackage{adjustbox}

\usepackage{tikz}
\usepackage{diagbox}
\usetikzlibrary{positioning}
\usetikzlibrary{calc}
\usetikzlibrary{decorations.pathreplacing,calligraphy}

\usepackage{xstring}
\usetikzlibrary{decorations.pathmorphing} 
\usetikzlibrary{decorations.markings} 
\usetikzlibrary{arrows} 
\usetikzlibrary{shapes} 
\usetikzlibrary{matrix} 
\usepackage[autostyle]{csquotes}

\tikzstyle{brane}=[draw]
\tikzset{roundnode/.style={circle, draw=black, thick, fill=white, minimum size=5mm}}
\tikzset{D7/.style={circle, draw=black, inner sep=0pt, fill=white, minimum size=2.5mm}}
\tikzset{hasse/.style={circle, fill,inner sep=2pt}}
\tikzset{flavor/.style={regular polygon,regular polygon sides=4,inner sep=2.5pt, draw}}
\tikzset{gauge/.style={circle, draw,inner sep=2.5pt}}
\tikzset{gaugeb/.style={circle, draw,fill=black,inner sep=2.5pt}}
\tikzset{gauger/.style={circle, draw,fill=red,inner sep=2.5pt}}
\tikzset{gaugeg/.style={circle, draw,fill=green,inner sep=2.5pt}}
\tikzset{bd/.style={circle, draw=black, inner sep=0pt, fill=black, minimum size=2mm}}
\tikzset{wd/.style={circle, draw=black, inner sep=0pt, fill=white, minimum size=2mm}}
\tikzset{Dynkin/.style={circle, draw=black, inner sep=0pt, fill=white, minimum size=2mm}}
\tikzstyle{ligne}=[draw, thick] 
\tikzset{doublearrow/.style={ draw=black!75, color=black!75, thick, double distance=3pt, }} 
\newcommand{\midarrow}{\tikz 
\draw[-triangle 90] (0,0) -- +(.1,0);} 
\newcommand{\midarrowrev}{\tikz 
\draw[-triangle 90] +(.1,0) -- (0,0);}

\makeatletter

\newcommand{\example}{\noindent{\bf Example. }}

\def\mbf{\mathbf}

\newcommand{\be}{\begin{equation}}
\newcommand{\ee}{\end{equation}}
\newcommand{\ba}{\begin{aligned}}
\newcommand{\ea}{\end{aligned}}

\newcommand{\su}{\mathfrak{su}}
\newcommand{\so}{\mathfrak{so}}
\newcommand{\e}{\mathfrak{e}}
\newcommand{\uu}{\mathfrak{u}}
\newcommand{\spp}{\mathfrak{sp}}

\newcommand{\bea}{\begin{eqnarray}}
\newcommand{\eea}{\end{eqnarray}}

\newcommand{\Z}{{\mathbb Z}}
\def\Tr{\mathop{\mathrm{Tr}}\nolimits}

\def\ii{{\text{i}}}
\def\half{{\frac{1}{2}}}

\def\p{\partial}

\def\unit{{1\kern-.65ex {\rm l}}}
\def\1{{1\kern-.65ex {\rm l}}}

\def\CN{{\cal N}}

\makeatother

\graphicspath{ {figs/} { figs/}}

\usepackage{subfiles}

\begin{document}

\baselineskip=18pt  
\numberwithin{equation}{section}  
\allowdisplaybreaks  
\thispagestyle{empty}

\vspace*{0.8cm} 
\begin{center}
{
{\huge (5d RG-flow) Trees in the Tropical Rain Forest}
}

 \vspace*{1.5cm}
Marieke van Beest$^1$, Antoine Bourget$^2$, Julius Eckhard$^1$,  Sakura Sch\"afer-Nameki$^1$\\

 \vspace*{1.0cm} 
{\it ${}^1$ Mathematical Institute, University of Oxford, \\
Andrew-Wiles Building,  Woodstock Road, Oxford, OX2 6GG, UK}\\
\smallskip
{\it ${}^2$ Theoretical Physics Group, The Blackett Laboratory, Imperial College London, \\
Prince Consort Road London, SW7 2AZ, UK}\\

\vspace*{0.8cm}
\end{center}
\vspace*{.5cm}

\noindent
5d superconformal field theories (SCFTs) can be obtained from 6d SCFTs by circle compactification and mass deformation. 
Successive decoupling of hypermultiplet matter and RG-flow generates a decoupling tree of descendant 5d SCFTs. 
In this paper we determine the magnetic quivers and Hasse diagrams, that encode the Higgs branches of 5d SCFTs, for entire decoupling trees. Central to this undertaking is the approach in \cite{vanBeest:2020kou}, which, starting from the generalized toric polygons (GTPs) dual to 5-brane webs/tropical curves, provides a systematic and succinct derivation of magnetic quivers and their Hasse diagrams.  The decoupling in the GTP description is straightforward, and generalizes the standard flop transitions of curves in toric polygons. 
We apply this approach to a large class of 5d KK-theories, and compute the Higgs branches for their descendants. In particular we determine the decoupling tree for all rank 2 5d SCFTs. 
For each tree, we also identify the flavor symmetry algebras from the magnetic quivers, including non-simply-laced flavor symmetries.

\newpage
\tableofcontents

\section{Introduction}
\label{sec:intro}

Our understanding of 5d superconformal field theories (SCFTs) derives in large part from our ability to construct and study these theories in string theory. 5d $\mathcal{N}=1$ SCFTs can arise in M-theory by compactification on a (non-compact) canonical Calabi-Yau three-fold singularity, or in type IIB string theory as the world-volume theory of a 5-brane-web. The moduli space of the 5d SCFT comprises the Higgs branch (HB) and (extended) Coulomb branch (CB), mapped out in M-theory respectively by the deformations and resolutions, i.e. K\"ahler deformations, of the canonical singularity. 
Much progress has been made in the exploration of the CB, using the resolutions of singular Calabi-Yau three-folds, allowing the classification of a vast number of theories. This programme was initiated in \cite{Morrison:1996xf} and more recent works include \cite{Intriligator:1997pq,Diaconescu:1998cn,Hayashi:2013lra,Hayashi:2014kca,DelZotto:2017pti,Xie:2017pfl,Jefferson:2017ahm,Jefferson:2018irk,Bhardwaj:2018yhy,Bhardwaj:2018vuu,Apruzzi:2018nre,Closset:2018bjz,Apruzzi:2019vpe,Apruzzi:2019opn,Apruzzi:2019enx,Bhardwaj:2019jtr,Bhardwaj:2019fzv,Bhardwaj:2019ngx,Saxena:2019wuy,Apruzzi:2019kgb,Closset:2019juk,Bhardwaj:2019xeg,Bhardwaj:2020gyu,Eckhard:2020jyr,Bhardwaj:2020kim,Hubner:2020uvb,Bhardwaj:2020ruf,Bhardwaj:2020avz}.

Starting from a 5d SCFT, by turning on mass parameters to decouple hypermultiplets, one can flow to a different SCFT, giving rise to a tree of descendant SCFTs. 
In this paper we will be interested in mapping out the Higgs branch of a large set of 5d SCFTs, extending this analysis to the fixed points they flow to under RG. Compared to the status of the extended CB, it is less clear how one can in general determine the structure of the HB from the geometry. A precise correspondence of the deformation space of the Calabi-Yau singularity and the HB of the associated 5d SCFT has only been achieved for singularities that can be realized as strictly convex toric polygons \cite{Altmann2,Altmann} and, recently, for isolated hyper-surface singularities \cite{Closset:2020scj}.

Where a 5-brane-web description of the SCFT is available \cite{Seiberg:1996bd,Aharony:1997ju,Aharony:1997bh,DeWolfe:1999hj,Benini:2009gi,Bergman:2013aca,Zafrir:2014ywa,Hayashi:2015zka,Hayashi:2015fsa,Bergman:2015dpa,Hayashi:2018lyv,Hayashi:2018bkd,Hayashi:2019yxj,Hayashi:2019jvx,Bergman:2020myx}, the Higgs branch can be characterized in terms of the associated magnetic quiver (MQ) and Hasse diagram \cite{Cremonesi:2015lsa,Ferlito:2016grh,Ferlito:2017xdq,Cabrera:2018ann,Cabrera:2018jxt,Cabrera:2019izd,Bourget:2019aer,Bourget:2019rtl,Cabrera:2019dob,Eckhard:2020jyr,Grimminger:2020dmg,Bourget:2020gzi,Bourget:2020asf,Akhond:2020vhc,Bourget:2020mez}. The magnetic quiver represents a 3d $\mathcal{N}=4$ quiver gauge theory, whose Coulomb branch is conjecturally identified with the Higgs branch of the 5d theory. The 3d CB is well-understood, and 
 can be studied in detail from the representation theoretic point of view
by computing its Hilbert series and highest weight generating function \cite{Cremonesi:2013lqa,Hanany:2014dia,Bullimore:2015lsa}. The Hasse diagram, introduced in \cite{Bourget:2019aer}, is a representation of the foliation structure of the Higgs branch by symplectic leaves, where each leaf is associated to a physical phase of the SCFT.

For any 5d SCFT with a brane-web realization without orientifold-planes there exists a dual representation of the theory in terms of a generalized toric polygon (GTP), or dot diagram \cite{Aharony:1997bh,Benini:2009gi,Bao:2011rc,Taki:2014pba}. In particular, when the dual polygon is actually \textit{toric}, corresponding to a brane-web with a single 5-brane ending on each 7-brane, this polygon precisely represents the canonical Calabi-Yau singularity on which M-theory can be compactified to give rise to the same 5d SCFT. In general, however, it is not known how to associate a GTP with a Calabi-Yau singularity. Indeed this is an interesting question, to which a first step towards an answer was taken in \cite{vanBeest:2020kou}, where a map from a general GTP to the magnetic quiver and Hasse diagram of the 5d theory was formulated, thereby approaching a description of the deformation space associated to the GTP.

In this paper we apply the framework developed in \cite{vanBeest:2020kou} to characterize the Higgs branch of 5d SCFTs that can flow to single gauge node theories with anti-symmetric ($\mathbf{AS}$) and fundamental ($\mathbf{F}$) matter \cite{Jefferson:2017ahm}, as well as $SU(2)^m$ quiver gauge theories, in the IR. We give a detailed description of the descendant trees obtained by decoupling matter multiplets in these theories, providing the associated magnetic quivers.

In particular, starting from a non-convex polygon representing a 5d gauge theory, a so-called pre-GTP (pGTP), we demonstrate how to apply a series of monodromy-preserving edge-moves (dual to Hanany-Witten moves \cite{Hanany:1996ie} in the brane-web) to obtain a GTP, associated to an SCFT. Moreover, we show how the geometric implementation of decoupling hyper-multiplets by performing flop transitions \cite{Apruzzi:2019opn} is realized as a flop of the corresponding matter curve in the pGTP, directly generalizing the application for toric models in \cite{Eckhard:2020jyr}.
Together, these operations clearly allow us to reach any pGTP (i.e. 5d gauge theory) in a given descendant tree and take the strong-coupling limit to obtain the corresponding GTP (SCFT). What is left to do is then simply to apply the method presented in \cite{vanBeest:2020kou} to each GTP to produce the associated magnetic quiver and Hasse diagram.

From the magnetic quivers/Hasse diagrams we furthermore can read off the UV flavor symmetry algebra. In the magnetic quivers, the non-abelian part of the flavor algebra is given by the set of balanced nodes.  There are various new aspects of the flavor symmetries that we discuss beyond the simple balanced nodes. In particular we will give rules how to identify certain non-simply laced flavor symmetries as well, like $\mathfrak{g}_2$.

\subsection{Overview of Results}
\label{sec:results}

We now summarize the theories and our main results, namely the magnetic quivers for all decoupling trees. 
5d SCFTs that have a low-energy description as a gauge theory with a simple gauge group were classified in \cite{Jefferson:2017ahm, Bhardwaj:2020gyu}. They distinguish two types of theories, called \emph{standard} and \emph{exceptional}. In table \ref{table:Standard5dSCFTs} we list the standard gauge theories, whose UV magnetic quivers are shown as well. The exceptional theories are limited to rank of the gauge group $\leq 8$. 

Using the strategy outlined above, we derive the magnetic quivers and flavor symmetry for the sub-marginal theories (obtained by a mass deformation from the 5d marginal theories), i.e. ``tree tops''. We consider 5d single gauge node theories with anti-symmetric ($N_{AS}=0,1,2$) and fundamental matter shown in table \ref{table:Standard5dSCFTs}.

\begin{enumerate}

\item Magnetic Quivers for all Rank 2 Theories:\\ 
In appendix \ref{sec:rank2}, in particular in tables \ref{tab:Rank2SU}, \ref{tab:Rank2Sp2}, \ref{tab:Rank2SU2SU2}, \ref{tab:Rank2G2}, we determine the complete set of magnetic quivers for all rank 2 5d SCFTs. 
The set of all rank 2 theories were determined from the geometries and flavor symmetries were discussed in \cite{Jefferson:2018irk, Apruzzi:2019opn,Apruzzi:2019enx} and the brane-webs in \cite{Hayashi:2018lyv}\footnote{There are additional rank 2 theories obtained by twisted compactifications, which do not have a known web/GTP realization \cite{Bhardwaj:2019jtr, Bhardwaj:2019fzv}. It would be interesting to  extend the results to these theories.}
Here we provide the full rank 2 tree in terms of the Higgs branch data, i.e. magnetic quivers, and thereby resulting flavor algebras, which are in agreement  with those found in \cite{Apruzzi:2019opn}, in particular we are able to extend the realm of magnetic quivers to non-simply laced symmetry algebras.

\item $(D_{N+2},D_{N+2})$ Conformal Matter Trees:\\
The $(D_{N+2},D_{N+2})$-conformal matter theory gives rise to 5d SCFTs that are UV completions of an SQCD theory. We determine a representation of the sub-marginal theory in terms of a GTP, and identify the magnetic quiver and flavor symmetry, finding agreement with \cite{Cabrera:2018jxt}. The tree of descendants can be determined for all, with the most interesting features appearing near the top, so we tabulate the magnetic quivers for a substantial part of the tree tops in table \ref{tab:SQCD}. 
\item Rank $(N-1)$ E-string Trees:\\ 
We then turn to the higher rank E-string, which has both an $SU(N)_{\frac{N+1}{2}} + 1 \mathbf{AS} + 7 \mathbf{F}$ and $Sp(N-1)+ 1 \mathbf{AS} + 7 \mathbf{F}$ gauge theory realization. The 5d descendant tree of the higher rank E-string joins that of the $(D_{N+2},D_{N+2})$-conformal matter theory by decoupling of the anti-symmetric matter. We identify the magnetic quiver for the sub-marginal rank $(N-1)$ E$_8$-theory, as well as those for all its 5d descendants with $N_F \geq 4$ in tables \ref{tab:UVMQsEStringEven} and \ref{tab:UVMQsEStringOdd}.
\item $SU(N)_{\half} + 1\mathbf{AS} + (N+5) \mathbf{F}$ Trees:\\ 
We subsequently study the 5d SCFT with an $SU(N)_{\half} + 1\mathbf{AS} + (N+5) \mathbf{F}$ gauge theory realization, determining the GTP and magnetic quiver for the tree top. The decoupling tree of this class of theories can be determined for all ranks. We provide the magnetic quivers for a significant portion of the tree tops in tables \ref{tab:UVMQs1ASEven} and \ref{tab:UVMQs1ASOdd}.
\item $2 \mathbf{AS}$ Trees: \\
The last single gauge node theories that we discuss are the SCFTs whose tree tops flow to $SU(N)_{\half} + 2\mathbf{AS} + 7 \mathbf{F}$ and $SU(N)_{2} + 2\mathbf{AS} + 6 \mathbf{F}$ in the weak coupling regime. We give the magnetic quivers and manifest flavor symmetries for the complete descendant trees of these theories in tables \ref{tab:UVMQs2ASSU2n} and \ref{tab:UVMQs2ASSU2n+1}. For $N=4,5$ the flavor symmetry exhibits a non-simply laced enhancement. We discuss this phenomenon for the magnetic quivers and tabulate them, along with the enhanced flavor symmetries, in tables \ref{tab:UVMQs2ASSU4} and \ref{tab:UVMQs2ASSU5}.

\item Quiver Trees:\\
Finally, we also consider the class of 5d SCFTs that have IR descriptions as $[ n_1 ]-SU(2)^m-[ n_2 ]$ 5d quiver gauge theories in the IR. These theories connect to the decoupling trees of $(D_{m+3},D_{m+3})$-conformal matter as well as that of the rank 2 E-string. The majority of the quiver theories are UV dual to an SQCD theory (i.e. they share the same strong coupling fixed point). However, for quiver gauge theories with no fundamental matter associated to (at least) one of the $SU(2)$ factors and theta angle $\theta=0$ this is not the case. We supply the GTPs and magnetic quivers for these theories in table \ref{tab:UVMQsSU2m}.

\end{enumerate}

\begin{table}
    \centering
    \begin{tabular}{c|c|c} \toprule 
        Gauge & Matter & Magnetic Quiver for sub-marginal theory \\ \midrule 
        $SU(N)_{0}$ & $2\bm{AS} + 8\bm{F}$ &  
        \begin{tabular}{cc}
               & $N$ even: \cr 
               & \raisebox{-.5\height}{\begin{tikzpicture}[x=1cm,y=.8cm]
\node (g1) at (1,0) [gauge,label=below:{\scalebox{.76}{$1$}}] {};
\node (g2) at (2,0) [gaugeb,label=below:{\scalebox{.76}{$2$}}] {};
\node (g3) at (3,0) [gauge,label=below:{\scalebox{.76}{$\frac{N+2}{2}$}}] {};
\node (g4) at (4,0) [gaugeb,label=below:{\scalebox{.76}{$N$}}] {};
\node (g5) at (5,0) [gauge,label=below:{\scalebox{.76}{$\frac{3N}{2}$}}] {};
\node (g6) at (6,0) [gauge,label=below:{\scalebox{.76}{$2N$}}] {};
\node (g7) at (7,0) [gauge,label=below:{\scalebox{.76}{$\frac{5N}{2}$}}] {};
\node (g8) at (8,0) [gauge,label=below:{\scalebox{.76}{$3N$}}] {};
\node (g9) at (9,0) [gauge,label=below:{\scalebox{.76}{$2N$}}] {};
\node (g10) at (10,0) [gauge,label=below:{\scalebox{.76}{$N$}}] {};
\node (g11) at (8,1) [gauge,label=right:{\scalebox{.76}{$\frac{3N}{2}$}}] {};
\draw (g1)--(g2)--(g3)--(g4)--(g5)--(g6)--(g7)--(g8)--(g9)--(g10);
\draw (g8)--(g11);
\end{tikzpicture}} \\ 
            & $N$ odd: \\
             & \raisebox{-.5\height}{\begin{tikzpicture}[x=1cm,y=.8cm]
\node (g1) at (1,0) [gauge,label=below:{\scalebox{.76}{$1$}}] {};
\node (g2) at (2,0) [gaugeb,label=below:{\scalebox{.76}{$2$}}] {};
\node (g3) at (3,0) [gauge,label=below:{\scalebox{.76}{$\frac{N+3}{2}$}}] {};
\node (g4) at (4,0) [gauge,label=below:{\scalebox{.76}{$N$+1}}] {};
\node (g5) at (5,0) [gauge,label=below:{\scalebox{.76}{$\frac{3N+1}{2}$}}] {};
\node (g6) at (6,0) [gauge,label=below:{\scalebox{.76}{$2N$}}] {};
\node (g7) at (7,0) [gauge,label=below:{\scalebox{.76}{$ \frac{5N-1}{2}$}}] {};
\node (g8) at (8,0) [gauge,label=below:{\scalebox{.76}{$3N$-1}}] {};
\node (g9) at (9,0) [gauge,label=below:{\scalebox{.76}{$2N$-1}}] {};
\node (g10) at (10,0) [gaugeb,label=below:{\scalebox{.76}{$N$-1}}] {};
\node (g11) at (8,1) [gauge,label=right:{\scalebox{.76}{$\frac{3N-1}{2}$}}] {};
\draw (g1)--(g2)--(g3)--(g4)--(g5)--(g6)--(g7)--(g8)--(g9)--(g10);
\draw (g8)--(g11);
\end{tikzpicture}} 
        \end{tabular} \\  \midrule 
        $SU(N)_{3/2}$ & $2\bm{AS} + 7\bm{F}$ &   \begin{tabular}{cc}
            $N$ even:  & \raisebox{-.5\height}{\begin{tikzpicture}[x=1cm,y=.8cm]
\node (g2) at (2,0) [gaugeb,label=below:{\scalebox{.76}{$2$}}] {};
\node (g3) at (3,0) [gauge,label=below:{\scalebox{.76}{$\frac{N+4}{2}$}}] {};
\node (g4) at (4,0) [gauge,label=below:{\scalebox{.76}{$N$+2}}] {};
\node (g5) at (5,0) [gauge,label=below:{\scalebox{.76}{$\frac{3N+4}{2}$}}] {};
\node (g6) at (6,0) [gauge,label=below:{\scalebox{.76}{$2N$+2}}] {};
\node (g7) at (7,0) [gauge,label=below:{\scalebox{.76}{$\frac{5N+4}{2}$}}] {};
\node (g8) at (8,0) [gauge,label=below:{\scalebox{.76}{$3N$+2}}] {};
\node (g9) at (9,0) [gauge,label=below:{\scalebox{.76}{$2N$+1}}] {};
\node (g10) at (10,0) [gaugeb,label=below:{\scalebox{.76}{$N$}}] {};
\node (g11) at (8,1) [gauge,label=right:{\scalebox{.76}{$\frac{3N+2}{2}$}}] {};
\draw (g2)--(g3)--(g4)--(g5)--(g6)--(g7)--(g8)--(g9)--(g10);
\draw (g8)--(g11);
\end{tikzpicture}} \\ 
             $N$ odd:  & \raisebox{-.5\height}{\begin{tikzpicture}[x=1cm,y=.8cm]
\node (g2) at (2,0) [gaugeb,label=below:{\scalebox{.76}{$2$}}] {};
\node (g3) at (3,0) [gauge,label=below:{\scalebox{.76}{$\frac{N+3}{2}$}}] {};
\node (g4) at (4,0) [gaugeb,label=below:{\scalebox{.76}{$N$+1}}] {};
\node (g5) at (5,0) [gauge,label=below:{\scalebox{.76}{$\frac{3N+3}{2}$}}] {};
\node (g6) at (6,0) [gauge,label=below:{\scalebox{.76}{$2N$+2}}] {};
\node (g7) at (7,0) [gauge,label=below:{\scalebox{.76}{$\frac{5N+5}{2}$}}] {};
\node (g8) at (8,0) [gauge,label=below:{\scalebox{.76}{$3N$+3}}] {};
\node (g9) at (9,0) [gauge,label=below:{\scalebox{.76}{$2N$+2}}] {};
\node (g10) at (10,0) [gauge,label=below:{\scalebox{.76}{$N$+1}}] {};
\node (g11) at (8,1) [gauge,label=right:{\scalebox{.76}{$\frac{3N+3}{2}$}}] {};
\draw (g2)--(g3)--(g4)--(g5)--(g6)--(g7)--(g8)--(g9)--(g10);
\draw (g8)--(g11);
\end{tikzpicture}}
        \end{tabular} \\  \midrule 
        $SU(N)_{0}$ & $1\bm{AS} + (N+6)\bm{F}$ &  \begin{tabular}{cc}
            $N$ even:  & \raisebox{-.5\height}{\begin{tikzpicture}[x=1cm,y=.8cm]
\node (g1) at (0,0) [gauge,label=below:{\scalebox{.76}{$1$}}] {};
\node (g2) at (1,0) {$\cdots$};
\node (g3) at (2,0) [gauge,label=below:{\scalebox{.76}{$N$+5}}] {};
\node (g4) at (3,0) [gauge,label=below:{\scalebox{.76}{$\frac{N+8}{2}$}}] {};
\node (g5) at (4,0) [gaugeb,label=below:{\scalebox{.76}{$3$}}] {};
\node (g6) at (2,1) [gaugeb,label=right:{\scalebox{.76}{$\frac{N+4}{2}$}}] {};
\draw (g1)--(g2)--(g3)--(g4)--(g5);
\draw (g3)--(g6);
\end{tikzpicture}} \\ 
             $N$ odd:  & \raisebox{-.5\height}{\begin{tikzpicture}[x=1cm,y=.8cm]
\node (g1) at (0,0) [gauge,label=below:{\scalebox{.76}{$1$}}] {};
\node (g2) at (1,0) {$\cdots$};
\node (g3) at (2,0) [gauge,label=below:{\scalebox{.76}{$N$+5}}] {};
\node (g4) at (3,0) [gaugeb,label=below:{\scalebox{.76}{$\frac{N+7}{2}$}}] {};
\node (g5) at (4,0) [gaugeb,label=below:{\scalebox{.76}{$3$}}] {};
\node (g6) at (2,1) [gauge,label=right:{\scalebox{.76}{$\frac{N+5}{2}$}}] {};
\draw (g1)--(g2)--(g3)--(g4)--(g5);
\draw (g3)--(g6);
\end{tikzpicture}}
        \end{tabular}\\  \midrule 
        $SU(N)_{N/2}$ & $1\bm{AS} + 8\bm{F}$& 
\raisebox{-.5\height}{\begin{tikzpicture}[x=1cm,y=.8cm]
\node (g1) at (1,0) [gaugeb,label=below:{\scalebox{.76}{$1$}}] {};
\node (g2) at (2,0) [gaugeb,label=below:{\scalebox{.76}{$N$-1}}] {};
\node (g3) at (3,0) [gauge,label=below:{\scalebox{.76}{$2N$-2}}] {};
\node (g4) at (4,0) [gauge,label=below:{\scalebox{.76}{$3N$-3}}] {};
\node (g5) at (5,0) [gauge,label=below:{\scalebox{.76}{$4N$-4}}] {};
\node (g6) at (6,0) [gauge,label=below:{\scalebox{.76}{$5N$-5}}] {};
\node (g7) at (7,0) [gauge,label=below:{\scalebox{.76}{$6N$-6}}] {};
\node (g8) at (8,0) [gauge,label=below:{\scalebox{.76}{$4N$-4}}] {};
\node (g9) at (9,0) [gauge,label=below:{\scalebox{.76}{$2N$-2}}] {};
\node (g10) at (7,1) [gauge,label=right:{\scalebox{.76}{$3N$-3}}] {};
\draw (g1)--(g2)--(g3)--(g4)--(g5)--(g6)--(g7)--(g8)--(g9);
\draw (g7)--(g10);
\end{tikzpicture} }
        
         \\  \midrule 
        $SU(N)_{0}$ & $(2N+4)\bm{F}$ & 
\raisebox{-.5\height}{\begin{tikzpicture}[x=1cm,y=.8cm]
\node (g1) at (0,0) [gauge,label=below:{\scalebox{.76}{$1$}}] {};
\node (g2) at (1,0) {$\cdots$};
\node (g3) at (2,0) [gauge,label=below:{\scalebox{.76}{$2N$+2}}] {};
\node (g4) at (3,0) [gauge,label=below:{\scalebox{.76}{$N$+2}}] {};
\node (g5) at (4,0) [gaugeb,label=below:{\scalebox{.76}{$2$}}] {};
\node (g6) at (2,1) [gauge,label=right:{\scalebox{.76}{$N$+1}}] {};
\draw (g1)--(g2)--(g3)--(g4)--(g5);
\draw (g3)--(g6);
\end{tikzpicture} }\\  
         \bottomrule 
    \end{tabular}
    \caption{Marginal or KK-theories with single gauge group, which we will discuss. The magnetic quiver is shown for the theory obtained after decoupling one of the fundamental matter fields. These theories will be referred to as sub-marginal. The white nodes in the quivers are the balanced nodes which encode the flavor symmetry algebra, see (\ref{balancies}). 
    \label{table:Standard5dSCFTs}}
\end{table}

The paper is structured as follows:  In section \ref{sec:review} we give a detailed review of 5d SCFTs, their IR descriptions, as well as their description in terms of brane-webs and generalized toric polygons, and the associated magnetic quivers and Hasse diagram. We then turn to the construction of a GTP from a non-convex polygon, a pre-GTP, in section \ref{sec:SUNGTPs}. Section \ref{sec:Decoupling} exhibits the decoupling of hyper-multiplets in the pre-GTP, as well as decoupling of fundamental matter directly in the magnetic quiver in the IR. In section \ref{sec:SUNasf} we derive the GTPs and magnetic quivers for the tree tops, i.e. the sub-marginal theories, of the single gauge node theories with anti-symmetric and fundamental matter, as well as for $SU(2)^m$ quiver gauge theories. Finally, we discuss the enhancement to non-simply laced flavor symmetries and the corresponding magnetic quivers and Hasse diagrams in section \ref{sec:NSLQuivers}. Notably, we give the complete decoupling tree of all rank 2 theories in appendix \ref{sec:rank2}, and appendix \ref{sec:MQTables} contains the decoupling trees of the SCFTs discussed in section \ref{sec:SUNasf}. Appendix \ref{app:webs} shows the brane-webs from which the pre-GTPs in section \ref{sec:SUNasf} are obtained. Finally, we give details of the derivation of the magnetic quiver for the higher rank E-string in appendix \ref{app:HigherRankEString}.


\section{5d SCFTs, Higgs Branches and Magnetic Quivers}
\label{sec:review}

\subsection{5d SCFTs and Gauge Theories}
\label{sec:5dSCFTs}

We consider 5d $\mathcal{N}=1$ SCFTs which admit a weakly coupled gauge theory description. Field theoretically, the best way to understand this is by starting with an SCFT in the UV and turning on a Super Yang-Mills term for a gauge group. This is a relevant operator as the gauge coupling $g$ has negative mass dimension in 5d and thus triggers an RG-flow. We say that two weakly coupled gauge theory descriptions are ``dual'' if they have the same UV-completion, i.e. they are two deformations of the same SCFT. From the gauge theory perspective we can reach the UV limit by taking the gauge coupling to infinity $g\to \infty$.

Let us consider a specific 5d $\mathcal{N}=1$ gauge theory with gauge group $G$ of rank $r$, associated to a Lie algebra $\mathfrak{g}$ with generators $T_i$, and matter hypermultiplets transforming in some representation $\bm{R}$ of $G$. We take $G$ to be simple for now, but will relax this condition to semi-simple later. The vector multiplet contains the gauge field $A$ and a real scalar whose vevs $\phi^i$, $i=1,\dots,r$ parametrise the Coulomb branch. The Coulomb branch is described by the effective Lagrangian
\be
\mathcal{L}_{\text{CB}}=\tau_{ij} \left( \mathrm{d} \phi^i \wedge \star \mathrm{d} \phi^j + \mathrm{d} A^i \wedge \star  \mathrm{d} A^j\right) + \frac{c_{ijk}}{24\pi^2} \mathrm{d} A^i \wedge \mathrm{d} A^j \wedge A^k + \dots\,,
\ee
where $\tau_{ij}$ and $c_{ijk}$ are the effective gauge coupling and Chern-Simons level, respectively, and the $A^i$ are the gauge fields for the Cartan subgroup $U(1)^r \subset G$. They can be conveniently obtained from a single function, the prepotential $\mathcal{F}(\phi^i)$
\be
\tau_{ij} = \frac{\p^2 \mathcal{F}}{\p \phi^i \p \phi^j}\,, \qquad c_{ijk} = \frac{\p^3 \mathcal{F}}{\p \phi^i \p \phi^j \p \phi^k}\,.
\ee
The IMS prepotential \cite{Intriligator:1997pq} of a pure 5d $\mathcal{N}=1$ gauge theory with gauge group $G$ is given by 
\be \label{1-loop}
\mathcal{F}_{\text{IMS}}=\frac{1}{2g_0^2}h_{ij} \phi^i \phi^j +\frac{k}{6}d_{ijk} \phi^i \phi^j \phi^k + \frac{1}{6} \sum_{\alpha \in \Phi_+} \left|\alpha \cdot \phi\right|^3\,,
\ee
where $h_{ij}=\Tr\left(T_iT_j\right)$ and the $\alpha \in \Phi_+$ are the positive roots of $\mathfrak{g}$. For $G=SU(N)$, $N\geq3$ the symmetric tensor $d_{ijk}=\half\Tr\left(T_i \left\{T_j,T_k\right\}\right)$ is non-zero and $k$ is the bare Chern-Simons level.
We choose a Weyl wedge such that $\alpha \cdot \phi>0$ for all $\alpha$.
To couple the theory to matter we add the term
\be \label{FMatter}
\mathcal{F}_{\text{matter}}=\sum_{\bm{R}} \sum_{I=1}^{N_R} \mathcal{F}_{\bm{R}}\,, \qquad \mathcal{F}_{\bm{R}}=-\frac{1}{12}\sum_{a=1}^{\dim\bm{R}} \left|\lambda^a \cdot \phi-m_I\right|^3\,,
\ee
where $\bm{R}$ runs over the irreducible representations of $G$ and the $\lambda^a$ are the weights of $\bm{R}$. The matter content of such a theory consists of $N_R$ hypermultiplets in the representation $\bm{R}$ with masses $m_I$. In this paper we will primarily be concerned with theories with $N_{AS}$ anti-symmetric matter multiplets and $N_F$ fundamentals. These theories will be denoted 
\be \label{Gauge+Matter}
G_k + N_{AS} \bm{AS} + N_F \bm{F}\,.
\ee

Another class of theories we encounter are quiver gauge theories, consisting of multiple simple gauge groups, coupled by bifundamental matter. The prepotential of a quiver $G_1 - G_2$ with two constituents is given by
\be
\mathcal{F}= \mathcal{F}_{\text{IMS}}^{(1)} + \mathcal{F}_{\text{IMS}}^{(2)} - \frac{1}{12} \sum_{a=1}^{\dim \bm{F}_1}  \sum_{b=1}^{\dim \bm{F}_2} \left|\lambda^a \cdot \phi^{(1)} + \lambda^b \cdot \phi^{(2)} - m_B\right|^3\,,
\ee
where $m_B$ is the mass of the bifundamental matter field. Of course, this generalises to arbitrary quivers with additional matter.

An important feature of the prepotential is its ability to describe the gauge theory in various phases. We see that the matter contribution \eqref{FMatter} depends on the relative values of the Coulomb branch parameters $\phi^i$ and the flavor masses $m_I$. Together with the bare gauge coupling\footnote{Or multiple gauge couplings in case of a quiver gauge theory.} 
\be
\label{defm0}
m_0 = \frac{1}{g_0^2}
\ee
they form the extended Coulomb branch $\mathcal{K}$.
Thus, the prepotential, and consequently the Lagrangian, depends on the position along $\mathcal{K}$. 

The flavor symmetry $\mathfrak{g}_{\text{F,cl}}$ of a weakly coupled gauge theory is easy to determine. Given a theory with a single gauge group $G$, $N_{R}$ hypermultiplets in representation $\bm{R}$ contribute as
\be \label{GFclReps}
\bm{R}\text{ complex } \Rightarrow \mathfrak{u}(N_R)\,, \qquad \bm{R}\text{ real } \Rightarrow \mathfrak{sp}(N_R)\,, \qquad \bm{R}\text{ quaternionic } \Rightarrow \mathfrak{so}(2N_R)\,.
\ee
For the theories in \eqref{Gauge+Matter} the flavor symmetry algebras are given in table \ref{tab:classflavorsym}.
\begin{table}
	\center
\begin{tabular}{c||c|c|c|c|c}
	$G$ & $SU(2)$ & $SU(3)$ & $SU(4)$ & $SU(N\geq5)$ & $Sp(N\geq2)$ \\
	\hline
	$\mathfrak{g}_{\text{F,cl}}$ & $\mathfrak{so}(2N_F)$ & $\mathfrak{u}(N_F)$ & $\mathfrak{u}(N_F) \oplus \mathfrak{sp}(N_{AS})$ & $\mathfrak{u}(N_F) \oplus \mathfrak{u}(N_{AS})$ & $\mathfrak{so}(2N_F)\oplus \mathfrak{sp}(N_{AS})$
\end{tabular}
	\caption{Classical flavor symmetry algebras for gauge theories of the form \eqref{Gauge+Matter}.
		\label{tab:classflavorsym}}
\end{table}
In a quiver gauge theory each bifundamental contributes an additional $U(1)$ to the classical flavor symmetry.\footnote{For $SU(2)-SU(2)$ quivers this is instead $SU(2)$.}
Furthermore, each simple gauge group constituent has a topological flavor symmetry $U(1)_T$ with current
\be 
J_T=\frac{1}{8\pi^2} \star \Tr\left( \mathrm{d} A\wedge \mathrm{d} A\right)\,.
\ee
In the strongly coupled limit the flavor symmetry algebra is enhanced
\be
\mathfrak{g}_{\text{F,cl}} \oplus \bigoplus_{j=1}^n \mathfrak{u}(1)_T^{(j)} \to \mathfrak{g}_{\text{F}}\,,
\ee
where $n$ is the number of simple gauge group constituents.

\begin{table}
\center
\begin{tabular}{c||c|c|c||c|c|c}
$G$ & $I_2(\bm{F})$ & $I_2(\bm{AS})$ & $I_2(\bm{Sym})$ & $A(\bm{F})$ & $A(\bm{AS})$ & $A(\bm{Sym})$\\
\hline
$SU(N)$ & $\half$ & $\frac{N-2}{2}$ & $\frac{N+2}{2}$ & 1 & $N-4$ & $N+4$\\
\hline
$Sp(N)$ & $\half$ & $N-1$ & $N+1$ & - & - & -
\end{tabular}
\caption{Second index and triangle anomaly coefficient for various representations of $G$.
\label{tab:IndexAnomaly}}
\end{table}

In this paper we will be interested in explicitly integrating out matter. Consider a theory with a single gauge group and matter in representation $\bm{R}$ of mass $m_I$. Taking $m_I\to\pm\infty$\footnote{In fact it is sufficient to choose $|m_I|$ large enough such that all the terms in \eqref{FMatter} have the same sign.} corresponds to decoupling the hypermultiplet. In this limit $\mathcal{F}_{\bm{R}}$ simplifies to
\be
\ba
\lim_{m_I\to\pm\infty}\mathcal{F}_{\bm{R}}=\pm\frac{1}{12}&\left(- m_I^3\ \dim\bm{R} + 3 m_I^2 \underbrace{\left(\sum_{a} \lambda^a_i\right)}_{0} \phi^i \right.\\
&\left.\quad- 3 m_I \underbrace{\left(\sum_a  \lambda^a_i \lambda^a_j\right)}_{2 I_2(\bm{R})\ h_{ij}} \phi^i\phi^j + \underbrace{\left(\sum_a \lambda^a_i \lambda^a_j  \lambda^a_k\right)}_{A(\bm{R})\ d_{ijk}} \phi^i\phi^j \phi^k \right)\,.
\ea
\ee
Here, the quantities $I_2$ and $A$ are the second index and the triangle anomaly coefficient, respectively, and only depend on $\bm{R}$ and $G$. For the gauge groups and representations of interest they are summarised in table \ref{tab:IndexAnomaly}. Clearly, $A$ is only relevant for $G=SU(N>2)$ as $d_{ijk}$ otherwise vanishes.

Additionally, we generally have
\be \label{CCRep}
I_2(\bar{\bm{R}})=I_2(\bm{R})\,, \qquad A(\bar{\bm{R}})=-A(\bm{R})\,.
\ee
By comparison with \eqref{1-loop} we see that integrating out a matter multiplet by taking its mass to $\pm\infty$ shifts the parameters as 
\be \label{CSshift}
m_0\to m_0 \mp m_I I_2(\bm{R})\,, \qquad
k\to k \pm \half A(\bm{R})\,.
\ee
Note that the constant shift in the prepotential is non-physical, whereas we should view the new shifted gauge coupling as physical.

In fact we will start with the set of so-called marginal or KK-theories in 5d, which UV complete in 6d. The set of single gauge node theories we consider are listed in table \ref{tab:JKVZMarginals}, including their flavor symmetries \cite{Jefferson:2017ahm, Bhardwaj:2020gyu}. Decoupling matter from these theories we obtain a descendant tree, each of which describes another SCFT in the UV. The associated flavor symmetries for these descendants were determined in \cite{Apruzzi:2019opn}.

%
\begin{table}
\center
\begin{tabular}{|c|c|c|c|cc|}
\hline
$G$ & $N_{AS}$ & $N_F$ & $k$ & \multicolumn{2}{c|}{$\mathfrak{g}_F$}\\[3pt]
\hline\hline
\multirow{7}{*}{$SU(N)$}
& 0 & $2N+4$ & 0 & \multicolumn{2}{c|}{$\mathfrak{so}(4N+8)^{(1)}$} \\[3pt]
\cline{2-6}
& 1 & 8 & $\frac{N}{2}$ & \multicolumn{2}{c|}{$\mathfrak{e}_8 \oplus \mathfrak{su}(2)$}\\[3pt]
\cline{2-6}
& 1 & $N+6$ & 0 & \multicolumn{2}{c|}{$\mathfrak{su}(N+7)^{(1)} \oplus \mathfrak{u}(1)$}\\[3pt]
\cline{2-6}
& \multirow{2}{*}{2} & \multirow{2}{*}{8} & \multirow{2}{*}{0} & $N$ even: & $\mathfrak{e}_7^{(1)} \oplus \mathfrak{su}(2)^{(1)} \oplus \mathfrak{su}(2)^{(1)} \oplus \mathfrak{su}(2) $   \\[3pt]
& & &  & $N$ odd: & $\mathfrak{so}(16)^{(1)} \oplus \mathfrak{su}(2)^{(1)} \oplus \mathfrak{su}(2)$ \\[3pt]
\cline{2-6}
& \multirow{2}{*}{2} & \multirow{2}{*}{7} & \multirow{2}{*}{$\frac{3}{2}$} & $N$ even: & $\mathfrak{so}(16)^{(1)} \oplus \mathfrak{su}(2)$\\[3pt]
& & & & $N$ odd: & $\mathfrak{e}_7^{(1)} \oplus \mathfrak{su}(2) \oplus \mathfrak{su}(2)$\\[3pt]
\hline
\multirow{2}{*}{$Sp(N)$}
& 0 & $2N+6$ & $-$ & \multicolumn{2}{c|}{$\mathfrak{so}(4N+12)^{(1)}$} \\[3pt]
\cline{2-6}
& 1 & 8 & $-$& \multicolumn{2}{c|}{$\mathfrak{e}_8^{(1)} \oplus \mathfrak{su}(2)$} \\[3pt]
\hline
\end{tabular}
\caption{List of marginal 5d gauge theories, which whose descendant trees will considered in this paper. The gauge group is $G$ and $N_{R}$ denotes the number of matter hypermultiplets in the representation $R$. For $SU(N)$ we list the CS-levels  $k$. $\mathfrak{g}_F$ is the flavor symmetry algebra of the marginal theory. Here, the superscript $(1)$ denotes the affine extension.
\label{tab:JKVZMarginals}}
\end{table}
%


\subsection{Webs, HW-Moves and GTPs}
\label{sec:BasicsWebs}

5d $\CN=1$ SCFTs can be realized in type IIB string theory in terms of webs of $(p,q)$ 5- and 7-branes \cite{Aharony:1997ju}.
Here, we review the basic features of this description as well as their dual graphs, the generalized toric polygons (GTPs). 
Consider type IIB string theory with axio-dilaton $\tau$ and a web of 5-branes extended along the $(x_0,\dots,x_4)$-directions and at some angle $\theta$ in the $(x_5,x_6)$-plane. We adopt the convention that a $(p,q)$ 5-brane is a bound state of $p$ D5-branes and $q$ NS5-branes.
Rotations in the $(x_7,x_8,x_9)$-directions realize the $SU(2)_R$ R-symmetry of the 5d theory.
Each 5-brane is a $\half$-BPS object, preserving 16 supercharges. To be precise, a $(p,q)$ 5-brane at an angle $\theta$ in the $(x_5,x_6)$-direction imposes the condition \cite{Hanany:1996ie}
\be
\epsilon_2= \Gamma_{01234} \left(\cos\theta\ \Gamma_5+\sin\theta\ \Gamma_6\right)\left(\sin\beta\ \epsilon_1 +\cos\beta\ \epsilon_2\right)
\ee
on the supersymmetry parameters $\epsilon_{1,2}$ of type IIB, where
\be
\tan\beta=\frac{\Re(p+q\tau)}{\Im(p+q\tau)}\,.
\ee
Additionally, a pure NS5-brane imposes
\be
\epsilon_1=-\Gamma_{01234} \left(\cos\theta\ \Gamma_5+\sin\theta\ \Gamma_6\right)\epsilon_1\,.
\ee 
A general configuration of branes breaks supersymmetry completely. However, one can show that brane configurations satisfying
\be
\tan\theta=\tan\beta
\ee
preserve 8 supercharges. This condition implies that the angle of a $(p,q)$ 5-brane in the $(x_5,x_6)$-direction is completely determined by $(p,q)$ and the choice of $\tau$. We can use the $SL(2,\Z)$ invariance of type IIB to fix $\tau=\ii$, so that
\be
\tan\theta=\frac{p}{q}\,.
\ee
All brane configurations related by a global $SL(2,\Z)$ transformation are physically equivalent and all pictures are drawn at $\tau=\ii$. If there are no coincident branes the most general web is engineered out of triple-intersections called junctions. At each junction, charge conservation implies
\be
\sum_{i=1}^3\pm \left(p_i,q_i\right)=\left(0,0\right)\,,
\ee
where the signs reflect the orientation of each brane with respect to the junction. In other words, any allowed junction is locally $SL(2,\Z)$-equivalent to the standard configuration of a D5-, an NS5- and a (1,1)-brane:
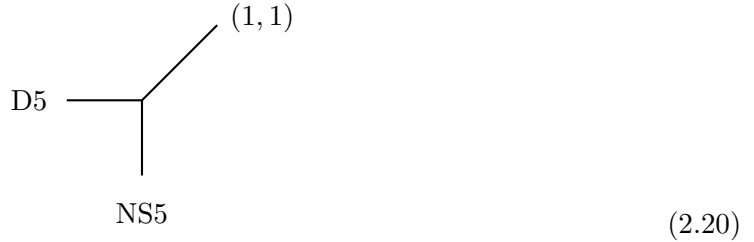
\be
\begin{tikzpicture}
\node at (-.5,0) {D5};
\node at (1,-1.5) {NS5};
\node at (2.6,1.1) {$(1,1)$};
\draw[thick] (0,0)--(1,0)--(2,1) {};
\draw[thick] (1,-1)--(1,0) {};
\end{tikzpicture}
\ee

A generic brane-web contains both finite (internal) and semi-infinite (external) 5-branes. 
It is straightforward to read off the ranks of the gauge and flavor group of the 5d theory from the web. We have \cite{Aharony:1997bh}
\be
\ba
r&=\text{\# local deformations}\,,\\
r_F&=\text{\# global deformations}=\text{\# external 5-branes}-3\,.
\ea
\ee
A deformation is a variation of the brane-web that leaves the type and relative position of the 5-branes invariant.
Local deformations also leave the position of the external 5-branes invariant, whereas global deformations change their positions. The position of an external 5-brane is specified by a single coordinate, corresponding to one deformation for each semi-infinite 5-brane. However, there are three unphysical global deformations comprising two global translations in the $(x_5,x_6)$-plane and the position of one of the external 5-branes, which is fixed by supersymmetry.

Any semi-infinite $(p,q)$ 5-brane can be made finite by letting it end on a corresponding $(p,q)$ 7-brane, extending in the $(x_0,\dots,x_4,x_7,x_8,x_9)$-directions. A 7-brane will be indicated by a circle and its $(p,q)$ charge will be implied by the angle of the 5-brane ending on it. The branch cut of a given $(p,q)$ 7-brane will be taken to point away from the 5-brane-web, along the direction $\tan \theta=p/q$, so as to not disrupt the 5-brane configuration. Furthermore, a $(p,q)$ 7-brane can be translated along this direction without affecting the 5d theory.

Consider a brane-web with two external 5-branes ending on 7-branes whose branch cuts (shown explicitly here as a dashed red line) intersect, e.g.
\be
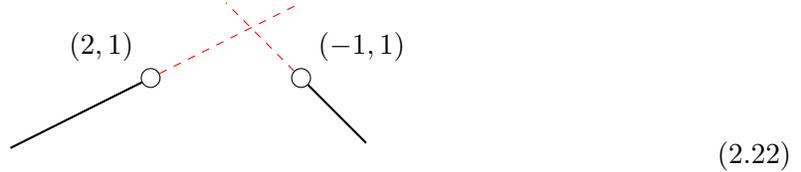
\label{monodromyintersect}
\begin{tikzpicture}
\node (1) {};
\node (2) at (5,0) {};
\node[D7,label=above left:{$(2,1)$}] (3) at (2,1) {};
\node[D7,label=above right:{$(-1,1)$}] (4) at (4,1) {};

\draw[thick] (1)--(3) {};
\draw[thick] (2) -- (4) {};
\draw[dashed, red] (3)--(4,2) {};
\draw[dashed, red] (4)--(3,2) {};
\end{tikzpicture}
\ee
Going to the strong coupling point of the 5d theory corresponds to taking the length of the internal 5-branes to zero. If a brane-web contains monodromy cuts that cross, we must exchange the order of the corresponding 7-branes, preserving both the total monodromy and charge conservation, before taking this limit. In the exchange, one 7-brane will change type because it crosses the branch cut of the other, whereas the type of the second 7-brane is preserved but the number of 5-branes that end on it changes to conserve charge. This is known as Hanany-Witten (HW) brane creation \cite{Hanany:1996ie}.
Specifically, the branch cut of a $(p,q)$ 7-brane induces a transformation
\be
M_{p,q}=\begin{pmatrix}
1-p q & p^2 \\ -q^2 & 1+p q
\end{pmatrix}\,.
\ee
The total monodromy is given by the counter-clockwise product of the $M_{p,q}$. Exchanging two 7-branes while preserving their total monodromy implies
\be \label{MonodromyExamples}
\ba
M_{p_1,q_1}M_{p_2,q_2}= \left\{ \begin{array}{l}
M_{p_2,q_2}M_{p_1',q_1'}\,,\quad \begin{pmatrix} p_1' \\ q_1' \end{pmatrix}= \begin{pmatrix} p_1 \\ q_1 \end{pmatrix} + \left(p_1 q_2-p_2 q_1\right)\begin{pmatrix} p_2 \\ q_2 \end{pmatrix}
\\
M_{p_2',q_2'}M_{p_1,q_1}\,,\quad \begin{pmatrix} p_2' \\ q_2' \end{pmatrix}= \begin{pmatrix} p_2 \\ q_2 \end{pmatrix} + \left(p_1 q_2-p_2 q_1\right)\begin{pmatrix} p_1 \\ q_1 \end{pmatrix}
\end{array} \right.\,.
\ea
\ee
In the configuration \eqref{monodromyintersect}, one has $(p_1 , q_1) = (-1,1)$ and $(p_2 , q_2) = (2,1)$, giving $(p'_1,q'_1)=(-7, -2)$ and $(p'_2,q'_2)=(5, -2)$. 

In order to guarantee charge conservation at the intersection, additional 5-branes are created when a 7-brane crosses a stack of 5-branes. The number of new 5-branes created depends on how many 5-branes are in the stack it crosses. In particular, exchanging two 7-branes as in the top line of \eqref{MonodromyExamples}, with $N_1$ 5-branes ending on the first 7-brane and $N_2$ on the second, the number of new 5-branes created is
\be
\Delta N_2=\left|p_1 q_2-p_2 q_1\right| N_1\,.
\ee
The number of 5-branes created in the move in the bottom line of \eqref{MonodromyExamples} is obtained by taking $N_1 \leftrightarrow N_2$. For example, the two possible HW moves for the configuration in \eqref{monodromyintersect} are
\be
\resizebox{\textwidth}{!}
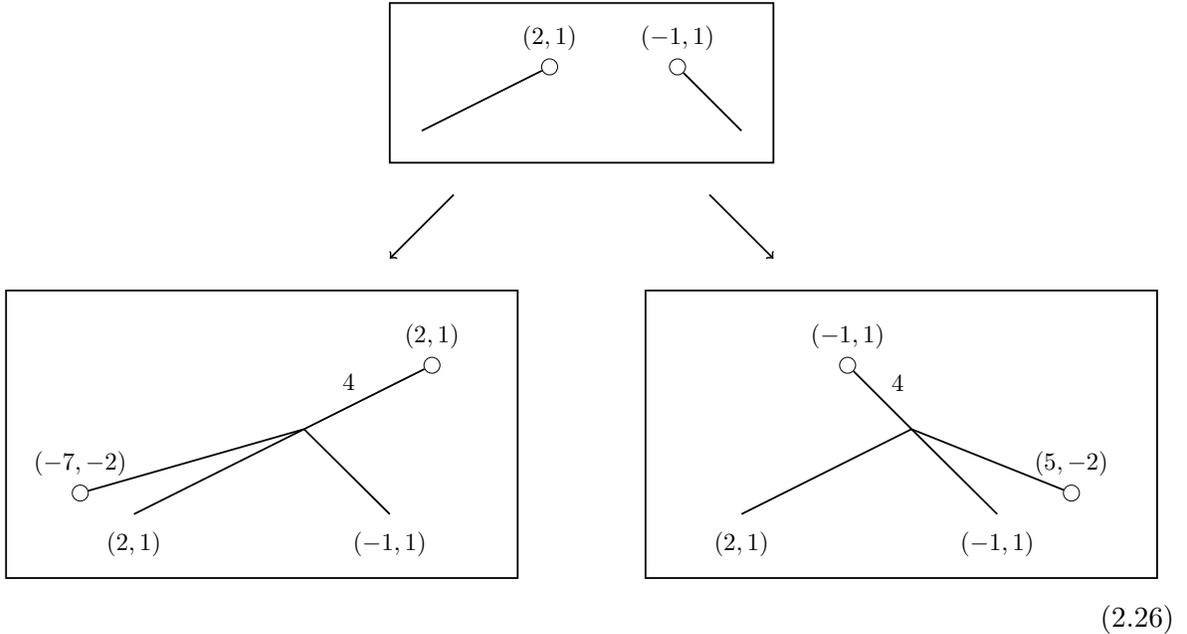
{%
{
\begin{tikzpicture}
\node (1) {};
\node (2) at (5,0) {};
\node[D7,label=above:{$(2,1)$}] (3) at (2,1) {};
\node[D7,label=above:{$(-1,1)$}] (4) at (4,1) {};

\node (5) at (.5,-1) {};
\node (6) at ($(5)+(-1,-1)$) {};
\node (7) at (4.5,-1) {};
\node (8) at ($(7)+(1,-1)$) {};

\node[label=below:{$(2,1)$}] (9) at (-4.5,-6) {};
\node[label=below:{$(-1,1)$}] (10) at ($(9)+(4,0)$) {};
\node (11) at ($(9)+(2.66,1.33)$) {};
\node[D7,,label=above:{$(-7,-2)$}] (12) at ($(11)+(-3.5,-1)$) {};
\node[D7,label=above:{$(2,1)$}] (13) at ($(11)+(2,1)$) {};

\node[label=below:{$(2,1)$}] (14) at (5,-6) {};
\node[label=below:{$(-1,1)$}] (15) at ($(14)+(4,0)$) {};
\node (16) at ($(14)+(2.66,1.33)$) {};
\node[D7,,label=above:{$(-1,1)$}] (17) at ($(16)+(-1,1)$) {};
\node[D7,label=above:{$(5,-2)$}] (18) at ($(16)+(2.5,-1)$) {};

\draw[thick] ($(1)$) -- (3) node[midway,below right=-.2] {};
\draw[thick] ($(2)$) -- (4) node[midway,below left=-.2] {};

\draw[thick,->] ($(5)$) -- ($(6)$) {};
\draw[thick,->] ($(7)$) -- ($(8)$) {};

\draw[thick] ($(9)$) -- ($(11)$) node[midway,below right=-.2] {} ;
\draw[thick] ($(10)$) -- ($(11)$)  node[midway,below left=-.2] {} ;
\draw[thick] (12) -- ($(11)$) node[midway,above left=-.2] {};
\draw[thick] (13) -- ($(11)$)  node[midway,above left] {4};

\draw[thick] ($(14)$) -- ($(16)$) node[midway,below right=-.2] {} ;
\draw[thick] ($(15)$) -- ($(16)$)  node[midway,below left=-.2] {} ;
\draw[thick] (17) -- ($(16)$) node[midway,above right] {4};
\draw[thick] (18) -- ($(16)$)  node[midway,below right=-.2] {};

\draw[thick] (-.5,-.5) rectangle (5.5,2);
\draw[thick] (-6.5,-7) rectangle (1.5,-2.5);
\draw[thick] (3.5,-7) rectangle (11.5,-2.5);

\end{tikzpicture}
}
}
\ee
Here, and in the following, we only write the number of coincident 5-branes if it is greater than one. As this example illustrates, performing a HW move usually leads to configurations where multiple 5-branes end on the same 7-brane, which gives rise to more general brane-webs. 

In general, the number of 5-branes that can end on a single 7-brane is constrained by the so-called {\it s-rule}. Given a web that satisfies the s-rule (e.g. one where each 7-brane has a single 5-brane ending on it), performing HW moves conserves this property, i.e. it gives another consistent supersymmetric brane-web.
The simplest formulation of the s-rule constrains the number of D5-branes $n_\text{D5}$ that connect a single D7-brane and a stack of $n_\text{NS5}$ NS5-branes to be $n_\text{D5} \leq n_\text{NS5}$.  
A brane-web is said to obey the s-rule if there exists a full resolution of the web such that each junction locally obeys the s-rule \cite{Benini:2009gi}. The $SL(2,\Z)$-invariant generalization of the s-rule for a junction of three branes $(p_i,q_i)$ is \cite{vanBeest:2020kou}
\be
\left|\det\begin{pmatrix} p_i & q_i \\ p_j & q_j \end{pmatrix}\right| \geq \text{gcd}(p_k,q_k)^2\,,
\ee
for every permutation $(i,j,k)$ of $(1,2,3)$.  

Finally, it is possible to add other types of branes to the brane-webs, such as orientifold planes \cite{Bergman:2015dpa,Zafrir:2015ftn,Hayashi:2015vhy}. In the case of $O7$-planes, these can be resolved in terms of $(p,q)$ 7-branes, and we will use the resolved webs.

A 5d SCFT realized as a brane-web in type IIB has a dual description in terms of a GTP, or dot-diagram \cite{Aharony:1997bh,Benini:2009gi}. In the toric case, the dual polygon determines a Calabi-Yau three-fold. Compactifying M-theory on this space will give rise to the same 5d SCFT. We briefly review the toric construction, before generalizing to GTPs and summarizing the duality with the brane-webs.

A complex toric singular non-compact Calabi-Yau three-fold $X_{\Sigma}$ is specified by a toric fan $\Sigma$, which is a collection of strongly convex rational polyhedral cones in a 3d vector space $N_{\mathbb{R}} = N \otimes \mathbb{R}$, where $N \simeq \mathbb{Z}^3$ is the cocharacter lattice of the complex torus $(\mathbb{C}^\ast)^3$. The Calabi-Yau condition imposes that the one-dimensional cones in the fan be generated by vectors in $N$ that all lie in a common affine hyperplane of $N_{\mathbb{R}}$, which we pick to be defined by the equation $z=1$, where $(x,y,z)$ are affine coordinates on $N_{\mathbb{R}}$. As a consequence, these vectors can be parametrized by their other two coordinates $(x,y) \in \mathbb{Z}^2$. The set of all such points define the vertices $\bm{v}_i \in \Z^2$ of the toric diagram. In a similar way, the two-dimensional cones in the fan are represented by edges $E_\alpha$, connecting a subset of the vertices in the toric diagram, and three-dimensional cones are represented by polygons. All cones in the fan are assumed to be convex, so the polygons in the toric diagrams also have to be convex. 
An edge $E_\alpha$ is a vector in $\Z^2$ on which lie at least two vertices $\bm{v}_{\alpha,i}\,, i=0,\dots, b_\alpha+1$ with $b_\alpha \geq 0$. Vertices that are the boundary of an edge are referred to as extreme vertices 
\be
\partial E_\alpha=\bm{v}_{\alpha,0}^\text{ex} \cup \bm{v}_{\alpha,b_\alpha+1}^\text{ex}\,.
\ee
We call the greatest common divisor of the components of an edge
\be 
\lambda_\alpha= \text{gcd}(E_\alpha)\,,
\ee
and define the reduced vector
\be 
L_\alpha=\frac{E_\alpha}{\lambda_\alpha}\,,
\ee
which is an elementary line segment along $E_\alpha$, i.e. connecting adjacent lattice points along the edge. 

Considering M-theory on a background $\mathbb{R}^{1,4} \times X_{\Sigma}$ gives at low energy a 5d $\mathcal{N}=1$ theory. This way, one can associate a 5d $\mathcal{N}=1$ theory to a given toric diagram. Blowing down all possible divisors gives a singular $X_{\Sigma}$, which is associated to an SCFT. 

The dual of a toric diagram can be interpreted as a brane-web. 
The map takes a vertex $\bm{v}_i$ in the polygon to a face (compact or non-compact) in the brane-web, and a line segment $L_\alpha$ to a 5-brane. Each external 5-brane is taken to end on a corresponding 7-brane.
An example of a toric polygon (realizing strongly coupled $SU(3)_\half+3\bm{F}$ or $SU(2)_\pi-SU(2)-[1]$) and the dual web is 
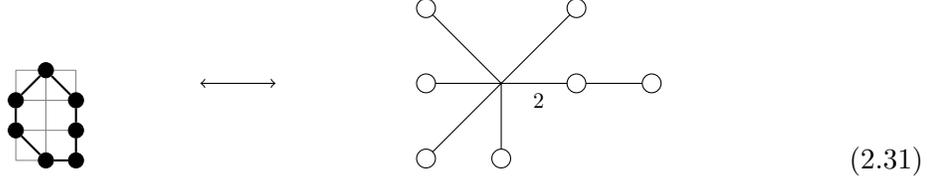
\be
\begin{tikzpicture}[x=.4cm,y=.4cm] 
\draw[step=.4cm,gray,very thin] (0,-1) grid (2,2);
\node[bd] (d1) at (0,0) {}; 
\node[bd] (d2) at (1,-1) {}; 
\node[bd] (d3) at (2,-1) {}; 
\node[bd] (d4) at (2,0) {};  
\node[bd] (d5) at (2,1) {}; 
\node[bd] (d6) at (1,2) {}; 
\node[bd] (d7) at (0,1) {}; 
\draw[ligne] (d1)--(d2)--(d3)--(d4)--(d5)--(d6)--(d7)--(d1); 
\end{tikzpicture} \qquad \qquad
\begin{tikzpicture}
\draw[<->] (-3,0)--(-2,0);
\draw[brane] (0,0)--(3,0) node[below,xshift=-1.5cm] {{\scalebox{.76}{$2$}}} ; 
\draw[brane] (0,-1)--(1,0)--(1,-1); 
\draw[brane] (0,1)--(1,0)--(2,1); 
\node[D7] at (0,0) {};
\node[D7] at (2,0) {};
\node[D7] at (3,0) {};
\node[D7] at (0,-1) {};
\node[D7] at (1,-1) {};
\node[D7] at (0,1) {};
\node[D7] at (2,1) {};
\end{tikzpicture}
\ee
with
\be 
\ba
E_\alpha&=((1,-1),(1,0),(0,2),(-1,1),(-1,-1),(0,-1))\,,\\
\lambda_\alpha&=(1,1,2,1,1,1)\,,\\
L_\alpha&=((1,-1),(0,1),(0,1),(-1,1),(-1,-1),(0,-1))\,.
\ea
\ee

However, as explained above, many brane-webs realizing SCFTs will have a 7-brane with more than one 5-brane ending on it. The GTPs were introduced in order to describe such configurations. The generalization from toric to GTP allows for lattice points belonging to the interior of an external edge that are not (conventional) vertices. In particular, a set of $n_5$ 5-branes ending on the same 7-brane in the web is mapped to $n_5$ consecutive line segments in the GTP separated by white dots, or ``empty'' vertices, as in \cite{Benini:2009gi}. 
The white dots define a partition of $\lambda_\alpha$ of an external edge $E_\alpha$ in the polygon as
\be \label{defmu}
\lambda_\alpha=\sum_{i=1}^{b_\alpha+1} \mu_{\alpha,i}\,, \qquad \mu_{\alpha,i}L_\alpha=\overline{\bm{v}_{\alpha,i}\bm{v}_{\alpha,i-1}}\,.
\ee
For a toric polygon these partitions are always given by $\{1^{b_\alpha+1}\}$. We order the partitions in descending magnitude
\be 
\mu_{\alpha,x} \geq \mu_{\alpha,x+1}\,,
\ee
where the index $x$ is introduced as an ordered version of $i$.
	An example of a GTP with white dots (realizing the SCFT with an $SU(3)_{5/2}+3\bm{F}$ gauge theory description) and its dual web is
	\be
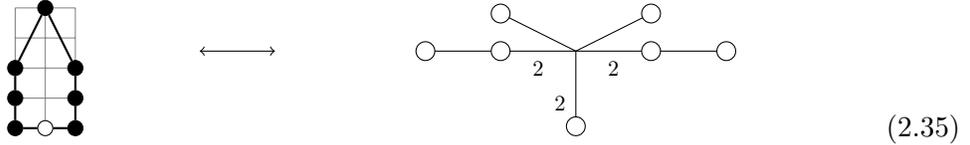
	\begin{tikzpicture}[x=.4cm,y=.4cm] 
		\draw[step=.4cm,gray,very thin] (0,-1) grid (2,3);
		\node[bd] (d1) at (0,-1) {}; 
		\node[wd] (d2) at (1,-1) {}; 
		\node[bd] (d3) at (2,-1) {}; 
		\node[bd] (d4) at (2,0) {};  
		\node[bd] (d5) at (2,1) {}; 
		\node[bd] (d6) at (1,3) {}; 
		\node[bd] (d7) at (0,1) {};
		\node[bd] (d8) at (0,0) {};  
		\draw[ligne] (d1)--(d2)--(d3)--(d4)--(d5)--(d6)--(d7)--(d8)--(d1); 
	\end{tikzpicture} \qquad \qquad
	\begin{tikzpicture}
		\draw[<->] (-3,0)--(-2,0);
		\draw[brane] (0,0)--(4,0) node[below,xshift=-1.5cm] {{\scalebox{.76}{$2$}}} node[below,xshift=-2.5cm] {{\scalebox{.76}{$2$}}}; 
		\draw[brane] (2,0)--(2,-1) node[left,yshift=.3cm] {{\scalebox{.76}{$2$}}} ;
		\draw[brane] (1,.5)--(2,0)--(3,.5);
		\node[D7] at (0,0) {}; 
		\node[D7] at (1,0) {};
		\node[D7] at (3,0) {};
		\node[D7] at (4,0) {};
		\node[D7] at (2,-1) {};
		\node[D7] at (1,.5) {};
		\node[D7] at (3,.5) {};
	\end{tikzpicture}
	\ee
	with
	\be 
	\ba
	E_\alpha&=((2,0),(0,2),(-1,2),(-1,-2),(0,-2))\,,\\
	\mu_\alpha&=(\{2\},\{1^2\},\{1\},\{1\},\{1^2\})\,,\\
	L_\alpha&=((1,0),(0,1),(-1,2),(-1,-2),(0,-1))\,.
	\ea
	\ee
In \cite{vanBeest:2020kou}, we defined the rank of a theory or (p)GTP $P$ as
\be
r=r_0-\half\sum_{\alpha}\sum_{x=1}^{b_\alpha+1} \mu_{\alpha,x} \left(\mu_{\alpha,x}-1\right)\,,
\ee
where $r_0$ is the number of internal nodes of $P$, or in terms of the area $A_P$ of $P$
\be \label{rankP}
r=A_P + 1 -\half \sum_{\alpha}\sum_{x=1}^{b_\alpha+1} \mu_{\alpha,x}^2 \,.
\ee
In particular, we postulated that a consistent 5d SCFT will have $r\geq 0$ (the r-rule) \cite{vanBeest:2020kou} (v2)\footnote{We thank Oren Bergman for discussions and for alerting us to the references \cite{DeWolfe:1998bi,Iqbal:1998xb}. In those papers a general rule, that agrees with the r-rule, was formulated for strings (in particular the case of three string junctions with 7-branes at each end, see (5.3) and (5.4) in \cite{Iqbal:1998xb}). This condition was also discussed in \cite{Bergman:2020myx} in the context of 5-brane-webs. }.

The GTP is a combinatorial object which, contrary to the brane-web, does not track the specific values of the (extended) Coulomb branch parameters, leading to simplifications of several operations.
In particular, decoupling of hyper-multiplet matter corresponds to a flop transition in the polygon, and we can implement the equivalent of HW moves in the GTP in a computationally simple and systematic fashion. This is why in the present paper we will determine the GTPs for the sub-marginal theories, and then perform all further analysis using the GTPs alone (never referring back to the webs).

\subsection{Magnetic Quivers, Flavor Symmetry, and Hasse Diagrams}
\label{sec:HB&MQs}

The Higgs branch of a 5d $\CN=1$ theory can be characterized in terms of its magnetic quiver (MQ) and Hasse diagram. 
The MQ is a graph, representing a 3d quiver gauge theory, whose Coulomb branch is 
 a symplectic singularity, which conjecturally is identified with the Higgs branch of the 5d theory. 
In this paper we focus on theories where the MQs are unitary gauge group quivers. The Higgs branch of 5d SCFTs has quantum corrections and the 
advantage of using the detour via MQs is that these instanton corrections in 5d can be understood from the exact Coulomb 
branch of the 3d $\CN=4$ MQ using the monopole formula (to compute its Hilbert series) \cite{Cremonesi:2013lqa, Bullimore:2015lsa,Nakajima:2015txa,Braverman:2016pwk,Nakajima:2017bdt}. The expectation is that the MQ will yield the complete Higgs branch for the 5d SCFTs. 

So far the MQs were derived indirectly either from the knowledge of the moduli space, or 
using brane-webs. The MQ can be determined also for 5d SCFTs realized as M-theory on canonical three-fold singularities by employing string dualities, and relating this to 4d SCFT in Type IIB obtained from the same canonical singularity \cite{Closset:2020scj}. Further compactifying to 3d and using mirror symmetry, i.e. T-duality, one finds the MQ. At present, this is however limited to hypersurface singularities, where the CB branch spectrum for the 4d SCFT can be computed. 

In \cite{vanBeest:2020kou} we proposed a third alternative, where we derived the MQ directly from a GTP on general grounds. Of course this is in principle dual to the web, however as we argued in \cite{vanBeest:2020kou}, this approach is far more elegant and concise. Furthermore, it eventually should provide a direct bridge to the geometric approach in M-theory on canonical singularities. 

Let us briefly summarize the algorithm to get from a GTP $P$ to the MQ of the associated 5d theory  \cite{vanBeest:2020kou}. The key concept is an edge coloring of the GTP, which determines a Minkowski sum decomposition into closed sub-polygons $S^c$, where $c=1,\dots,N$ labels the colors. Each closed sub-polygon is in turn a refined Minkowski sum of $m^c$ irreducible and minimal polygons (IMPs) $T^c$
\be 
S^c = \underbrace{T^c \oplus \dots \oplus T^c}_{m^c}\,,
\ee
where the refined Minkowski sum of two polygons $P_a \oplus P_b$ is defined such that the edges agree with the ones of $P_a + P_b$, i.e. the ordinary Minkowski sum, and the partitions are related by
\be
\mu_{\alpha,x}^{P_a \oplus P_b} = \mu_{\alpha,x}^{P_a} + \mu_{\alpha,x}^{P_b}\,.
\ee
That $T^c$ is irreducible means that it cannot be written as the refined Minkowski sum of two other polygons. Minimality of $T^c$ implies that one cannot remove any vertices (and replace them with a white dot) in $T^c$ without violating the s-rule. Since the $T^c$ constitute Higgsed sub-sectors of the theory, they must have non-negative rank, $r(T^c) \geq 0$. The building blocks of two distinct colored sub-polygons cannot be the same, so $T^c \neq T^d$ if $c \neq d$. Finally, the partitions of $P$ and those of the sub-polygons $S^c$ satisfy
\be 
\{\mu_\alpha\} \leq \sum_{c=1}^{N} \{\mu^c_\alpha\}\,,
\ee
where the inequality is the dominance partial order of partitions. A given polygon may have several valid colorings with each coloring associated to an MQ, realizing a cone in the 5d Higgs branch.
We obtain the MQs from a given polygon $P$ as follows:

\begin{enumerate}
\item Identify all valid colorings of $P$.
\item For each coloring, there are two kinds of nodes in the magnetic quiver:
\subitem {\it Color nodes:} Each colored sub-polygon $S^c$ maps to a node in the MQ with label $m^c$.
\subitem {\it Tail nodes:} Each edge $E_\alpha$ gives rise to a sequence of nodes of length $b_\alpha$ in the MQ with labels 
\be 
m_{\alpha,x}=\sum_{y=1}^x \left( -\mu_{\alpha,y}+\sum_{c=1}^{N} \mu^c_{\alpha,y} \right)\,.
\ee
\item The intersection of two color nodes, associated to sub-polygons $S^c$ and $S^d$ in $P$, receive a contribution from their mixed volume (MV) and a contribution from the two colors sharing an edge in $P$. The number of edges between color nodes $c$ and $d$ is
\be 
k^{c,d}=\frac{1}{m^c m^d} \left( MV(S^c,S^d)-\sum_\alpha \sum_{x=1}^{b_\alpha} \mu_{\alpha,x}^c \mu_{\alpha,x}^d\right) \,.
\ee
Tail nodes pertaining to the same edge in $P$ intersect their nearest neighbor once
\be 
k_{\alpha,x;\beta,y}=\delta_{\alpha,\beta} \delta_{\vert x-y \vert,1}\,.
\ee
The number of edges connecting a color node and tail node in the MQ is
\be 
k_{\alpha,x}^c=\frac{1}{m^c} \left( \mu_{\alpha,x}^c-\mu_{\alpha,x+1}^c \right)\,.
\ee
\item The number of loops $\ell$ on a given node is given by
\be
\ell^c=1+\frac{k^{cc}}{2}\,, \qquad \ell_{\alpha,x}=0\,,
\ee
where $k^{cc}=2(r(T^c)-1)$ is the self-intersection number.
\end{enumerate}

From the magnetic quiver it is possible to compute the flavor symmetry algebra of the 5d SCFT. Given a MQ with nodes $m_I$, number of loops $\ell_I$ and number of edges $k_{IJ}$, we define the balance of a node as
\be
\beta_I = -2m_I + 2 \ell_I (m_I-1)+ \sum_{J \neq I} k_{IJ} m_J\,.
\ee
Naturally, there are three different types of nodes,
\be\label{balancies}
\ba
\begin{tabular} {c c c}
$\beta_I=0$ & balanced & \begin{tikzpicture}[x=.8cm,y=.8cm] \node at (0,0) [gauge] {}; \end{tikzpicture}\\
$\beta_I>0$ & overbalanced & \begin{tikzpicture}[x=.8cm,y=.8cm] \node at (0,0) [gaugeb] {}; \end{tikzpicture}\\
$\beta_I<0$ & underbalanced & \begin{tikzpicture}[x=.8cm,y=.8cm] \node at (0,0) [gauger] {}; \end{tikzpicture}\\
\end{tabular}
\ea
\ee
which we indicate with different colours. A MQ with an underbalanced node is said to be ``bad'' or ``ugly'', see \cite{Gaiotto:2008ak}. This means that there are BPS monopole operators violating or saturating the unitary bound respectively. For ugly quivers, it is known that one can use a Seiberg duality to flow to an SCFT, together with a set of free hypermultiplets.
All magnetic quivers discussed in this paper are either good or ugly, as they describe the Higgs branch of a physical SCFT, i.e. they always describe a hyper-K\"ahler cone. For ugly quivers we can use the following local Seiberg duality to balance the nodes:
\be \label{SimoneDuality}
U(n_I) + (2n_I-\left|\beta_I\right|) \bm{F} \longleftrightarrow U(n_I-\left|\beta_I\right|) + (2n_I-\left|\beta_I\right|) \bm{F} + \left|\beta_I\right| \bm{H}\,,
\ee
where $\bm{H}$ indicates free hypermultiplets and $\beta_I<0$.\footnote{In \cite{Gaiotto:2008ak} it was argued that this duality can only be applied for $\beta_I=-1$. However, evidence suggests that in the case of ugly quivers it also holds in general. For bad quivers the duality \eqref{SimoneDuality} can not be applied. Thus, in principle one first needs to check the dimensions of the monopole operators. It would indeed be interesting to derive the general duality from first principle.} The process of repeatedly applying this duality always terminates for ugly quivers. We regard the endpoint of this process as the MQ of the theory and do not usually include the free hypermultiplets. Then, the balanced nodes form the Dynkin diagram of a finite Lie algebra. In many cases the balanced nodes directly gives the non-abelian part of the enhanced flavor symmetry of the theory.
However, we find examples in which the flavor symmetry is larger than expected from the balanced nodes. To determine the flavor symmetry in general one has to compute the Hilbert series. 

The Hasse diagram is a graphical representation of the foliation structure of the Higgs branch. It describes a partially ordered set of symplectic leaves, with each leaf corresponding to a physical phase of the 5d SCFT. The transition from one leaf to another (respecting the partial order) is described by the transverse slice between the two leaves, which again corresponds to a symplectic singularity. The Hasse diagram can be either obtained from the standard quiver subtraction \cite{Cabrera:2018ann,Bourget:2019aer,Bourget:2020mez}, or by the method proposed in \cite{vanBeest:2020kou}. 
In our analysis we encounter magnetic quivers which correspond to non-simply-laced flavor symmetries, even though the quivers themselves may or may not be simply laced. This is discussed in detail in section \ref{sec:NSLQuivers}.


\section{Construction of GTPs and Edge-Moves}
\label{sec:SUNGTPs}

 In this paper we will construct the GTPs for a large class of 5d SCFTs. 
 The general strategy is to compute the (p)GTP for the sub-marginal theory from the dual web (and to then determine the descendants using polygon-only operations). 
 Generically, the resulting diagram is not convex, which in the web-picture means that in order to take the strong-coupling limit, one has to first apply
 a sequence of Hanany-Witten moves. As the brane-practitioner will know, this can be a rather cumbersome enterprise. By mapping the problem to the dual 
 graphs, these moves are realized in terms of monodromy-preserving edge-moves, which are much simpler to identify and work with. We will now explain how to determine these.

\subsection{Generalized Toric Polygons and Webs}

In this section we present the pGTPs of $SU(N)$ gauge theories with fundamental and anti-symmetric matter that admit a UV completion. Our strategy is to start with the weakly coupled $(p,q)$ 5-brane webs which are reviewed in appendix \ref{app:SUNWebs}. From this, we can easily determine the pGTP of the gauge theory. These pGTPs, specified uniquely by the set of black vertices $V_b$, are summarised in table \ref{tab:ToricPolygons}. However, if the polygon is non-convex, it is not possible to remove the internal edge representing the gauge coupling. To be able to pass to strong coupling, 
we need to perform the GTP equivalent of the Hanany-Witten moves, i.e. monodromy- and charge-preserving edge-moves.

\begin{table}
\center
    \begin{tabular}{|c|c|} \hline 
    Theory & pGTP \\ \hline 
        \multirow{2}{*}{$SU(N)_k + N_F \bm{F}$} & $\Big\{ (0,0) ; (1,i)_{a \leq i \leq N_R + a} ; (0,N) ;
        $\\& $ 
        (-1,i)_{N-k-\frac{N_L-N_R}{2}-a \geq i \geq  N-k-\frac{N_L+N_R}{2}-a} \Big\}$ \\
        \hline
       \multirow{2}{*}{$SU(2n)_k +  1 \bm{AS} + N_F \bm{F}$} & $\Big\{ (0,-n) ; (1,i)_{k - \frac{N_F+2}{2}  \leq i \leq k + \frac{N_F-2}{2} } ;
        $\\& $ 
	 (0,n) ; (-1,n-1); (-n,0) \Big\}$ \\
	 \hline
        \multirow{2}{*}{$SU(2n+1)_{k}  +  1 \bm{AS} + N_F \bm{F}$} & $\Big\{ (0,-n-1) ; (1,i)_{k - \frac{N_F+5}{2}  \leq i \leq k + \frac{N_F-5}{2} } ;
                $\\& $
           (0,n) ; (-1,n); (-n-1,0) \Big\}$ \\ 
           \hline
        \multirow{3}{*}{$SU(2n)_{k}  +  2 \bm{AS} + N_F \bm{F}$} & $\Big\{ (0,-n) ; (1,-n) ; (n,n\ i)_{-1  \leq i \leq N_R-1 } ;
         $\\& $
          (0,n) ; (-1,n+k+\frac{N_L-N_R}{2});
           $\\& $
            (-n,n\ i)_{k+\frac{N_L-N_R}{2}+1  \geq i \geq k-\frac{N_L+N_R}{2}+1 } \Big\}$ \\
            \hline
         \multirow{4}{*}{$SU(2n+1)_{k}  +  2 \bm{AS} + N_F \bm{F}$} & $\Big\{ (0,-n-1) ; (n,-n-1) ; (n,n\ i)_{-1  \leq i \leq N_R-1 } ; 
         $\\& $
          (0,n) ; (-n,n\ i)_{k+\frac{N_L-N_R}{2}+1  \geq i \geq k-\frac{N_L+N_R}{2}+1 };
         $\\& $
           (-n,n(k-\frac{N_L+N_R}{2}+1)-1) \Big\}$ \\ \hline 
    \end{tabular}
    \caption{Construction of toric polygons (not necessarily convex) for the theories studied in this paper. For each theory, we give a list of pairs of integers which should be interpreted as the list of coordinates of the black dots defining the polygons in counter clockwise order; the white dots can then be added on the points of the interior of the edges with integer coefficients. The free parameter $a$ can take any integer value, parametrising the $SL(2,\Z)$-frame. The polygons presented here can be deduced from the brane webs constructed in \cite{Zafrir:2015rga}.}
    \label{tab:ToricPolygons}
    \end{table}

Let us first fix the notation. We let $P$ be a pre-GTP (pGTP), i.e. a GTP where we relax the convexity requirement. For simplicity we ignore interior edges, such as rulings, for now but will return to this issue later. The pGTP is characterised by the list of its black vertices $\bm{v}_i$ with $i \in \mathbb{Z}_{n_V}$, labelling them in counter clockwise order. Equivalently, $P$ can be characterised by the list of vectors $[\bm{u}_i]_{i \in \mathbb{Z}_{n_V}}$ defined by 
\be
\bm{u}_i = \bm{v}_{i} - \bm{v}_{i-1} \, . 
\ee
Note that such a list of vectors of $\mathbb{Z}^2$ defines a polygon only if they add up to zero. 
The polygon is convex if and only if two consecutive edges have direction vectors $\bm{u}_i$ and $\bm{u}_{i+1}$ forming a positively oriented basis of the plane. It is useful to introduce shorthand for the determinants and greatest common divisors
\be
D_{i,j} =  \det\left(\bm{u}_{i},\bm{u}_{j}\right)\,,\qquad D_i = D_{i,i+1}\,,  \qquad \mu_i =\mathrm{gcd}(\bm{u}_i)   \, . 
\ee
In other words,  
\be
P \textrm{ convex } \Longleftrightarrow  \ D_{i} \geq 0\quad  \forall i \in \mathbb{Z}_{n_V} \,.
\ee
Let 
\be
\mathfrak{C} := \{ i \mid D_{i} < 0  \}
\ee
be the set of labels, which indicate the non-convexities of $P$. Clearly, the polygon is convex if and only if $\mathfrak{C} = \emptyset$.

\subsection{From Non-Convex to Convex Polygons: Edge-moves}

In this section, we assume $\mathfrak{C} \neq \emptyset$, and our goal is to find a GTP, which is convex and physically equivalent to $P$. 
Inspired by the brane-web, we exhibit two basic, related operations which, upon being iterated in an appropriate way, can result in such a GTP:

\begin{enumerate}

\item{Monodromies:} 

Let $i \in \mathfrak{C}$. We define two operations on the polygon $P$, which affect the two vectors $\bm{u}_{i}$ and $\bm{u}_{i+1}$ in the following way, leaving the other vectors unchanged: 
\begin{eqnarray}
\label{edgemoverules}
\mathfrak{M}_i^+ &:& \left[ \dots , \bm{u}_{i-1} , \bm{u}_{i} , \bm{u}_{i+1} , \bm{u}_{i+2} , \dots \right] \mapsto \\ 
& & \qquad  \left[ \dots , \bm{u}_{i-1} ,  \bm{u}_{i+1} \left(1- \frac{D_i}{\mu_{i+1}^2} \right),  \bm{u}_{i} + \bm{u}_{i+1} \left( \frac{D_i}{\mu_{i+1}^2} \right), \bm{u}_{i+2} , \dots \right] \nonumber \\ 
\mathfrak{M}_i^- &:& \left[ \dots , \bm{u}_{i-1} , \bm{u}_{i} , \bm{u}_{i+1} , \bm{u}_{i+2} , \dots \right] \mapsto \\ 
& & \qquad  \left[ \dots , \bm{u}_{i-1} ,  \bm{u}_{i+1}  + \bm{u}_{i} \left(\frac{D_i}{\mu_i^2} \right),  \bm{u}_{i}  \left( 1-\frac{D_i}{\mu_i^2} \right), \bm{u}_{i+2} , \dots \right] \nonumber 
\end{eqnarray}
One can check that after having applied $\mathfrak{M}_i^{\pm}$, the polygon is convex at $i$, as 
\be
\det \left[  \bm{u}_{i+1} \left(1- \frac{D_i}{\mu_{i+1}^2} \right),  \bm{u}_{i} + \bm{u}_{i+1} \left( \frac{D_i}{\mu_{i+1}^2} \right) \right] =  - D_i \left(1- \frac{D_i}{\mu_{i+1}^2} \right) > 0 \, ,  
\ee
and similarly for $\mathfrak{M}_i^-$. However, it is now possible that $i \pm 1 \in \mathfrak{C}$. 

\item {Pruning:}

Let $i\in \mathfrak{C}$. Let $k_\pm$ be the smallest positive integers such that
\be
\ba
&D_{i+1,i+k_+ +1}>0\quad &\wedge \quad &D_{i+1,i+k_+ +2}\leq 0 \quad &\wedge \quad &i+1,\dots,i+k_+ \notin \mathfrak{C}\\
&D_{i-k_-,i}>0\quad &\wedge \quad &D_{i-k_- -1,i} \leq 0 \quad &\wedge \quad &i-1,\dots,i-k_- \notin \mathfrak{C}\,.
\ea
\ee
If the respective integer $k_\pm$ exists we define the operations
\begin{eqnarray}
\mathfrak{P}_i^+ &:& \left[ \dots , \bm{u}_{i} , \bm{u}_{i+1} , \bm{u}_{i+2} , \dots,  \bm{u}_{i+k_+ +1}, \bm{u}_{i+k_+ +2}, \dots \right] \mapsto \\ 
& & \qquad  \left[ \dots , \bm{u}_{i} ,  \bm{u}_{i+2} + \frac{D_{i+1,i+2}}{\mu_{i+1}^2} \bm{u}_{i+1}, \dots ,  \bm{u}_{i+k_+ +1} +  \frac{\left|D_{i+1,i+k_+ +1}\right|}{\mu_{i+1}^2} \bm{u}_{i+1},\right. \nonumber\\
& & \qquad \qquad \left. \bm{u}_{i+1}\left(1-\sum_{j=1}^{k_+} \frac{D_{i+1,i+j+1}}{\mu_{i+1}^2}\right) , \bm{u}_{i+k_+ +2} \dots \right] \nonumber \\ 
\mathfrak{P}_i^- &:& \left[ \dots , \bm{u}_{i-k_- -1} , \bm{u}_{i-k_-} , \dots,  \bm{u}_{i-1}, \bm{u}_{i}, \bm{u}_{i+1}, \dots \right] \mapsto \\ 
& & \qquad  \left[ \dots ,\bm{u}_{i-k_- -1} , \bm{u}_i \left(1-\sum_{j=1}^{k_-} \frac{D_{i,i-j}}{\mu_{i}^2}\right),  \bm{u}_{i-k_-} +\frac{D_{i,i-k_-}}{\mu_{i}^2} \bm{u}_{i}, \dots ,\right. \nonumber\\
& & \qquad \qquad \left.   \bm{u}_{i-1} + \frac{D_{i,i-1}}{\mu_{i}^2} \bm{u}_{i}, \bm{u}_{i+1} \dots \right]\,. \nonumber 
\end{eqnarray}
This does not necessarily remove the non-convexity at $i$, but may increase $D_i$.

\end{enumerate}
We can represent any monodromy- and charge-preserving transformation on $P$ as a (finite) word using these operations as letters. We will present detailed examples in section \ref{sec:SUNasf}.

Let us now consider internal edges, which in our context represent finite parameters along the extended Coulomb branch, i.e, gauge couplings $m_0$ or hypermultiplet masses $m_f$. Under an edge-move $\bm{u} \mapsto \mathfrak{M}_i^\pm \bm{u}$, the original internal edges are unchanged and $\bm{u}_i$ and $\bm{u}_{i+1}$ become internal lines in the transformed polygon.
We can also prune a polygon with internal edges, however, the transformation is very impractical to implement, and will therefore not be used in the following.

\section{Decoupling}
\label{sec:Decoupling}

5d SCFTs can be related by decoupling matter hypermultiplets. In the context of geometric models realized as singular Calabi-Yau spaces, this is achieved by flop transitions, which take matter curves of normal bundle degree $(-1,-1)$ and flop them out of the compact surfaces, that realize the Cartan subgroups of the gauge group. In this way in 5d, one can construct decoupling trees, which start with a marginal theory, which itself UV completes in 6d, and apply decoupling limits \cite{Apruzzi:2019opn}. 
Applied in the context of toric models, these flops have a simple realization: the $(-1,-1)$ curves are particular  curves, which are connections of an internal vertex (compact divisors) to an external one -- subject to the normal bundle condition. This decoupling was applied in \cite{Eckhard:2020jyr} to toric models and the construction of decoupling trees and relation to so-called CFDs was discussed. Here we will extend this to pGTPs and GTPs. 

The first important point to note is that the decoupling in the pGTP is straight forward, as we will explain in this section. We can trace the decoupling process through the monodromy transformations, which map the pGTP to a convex GTP. However, we do not find a simple decoupling rule on GTPs: the main complication is that applying the decoupling to a GTP usually results in a pGTP, i.e. convexity is not retained in this process. We will therefore construct our decoupling trees before applying the edge-moves, and then map them to GTPs afterwards. 

There are two cases of interest to consider: those that are purely made of black dots, convex or not, and those that do not have such a representation, e.g. the situation with $\mathbf{AS}$ matter. We now discuss decoupling in all these instances for the SCFTs, and also the decoupling for IR magnetic quivers, which has a much more straight forward realization in terms of quiver subtractions.

\subsection{Decoupling Fundamental Matter}

\begin{figure}
	\centering
\begin{tikzpicture}
\begin{scope}[x=.5cm,y=.5cm] 
\draw[step=.5cm,gray,very thin] (0,-1) grid (2,2);
\node[bd] (d1) at (0,0) {}; 
\node[bd] (d2) at (1,-1) {}; 
\node[bd] (d3) at (2,-1) {}; 
\node[bd] (d4) at (2,0) {};  
\node[bd] (d5) at (2,1) {}; 
\node[bd] (d6) at (1,2) {}; 
\node[bd] (d7) at (0,1) {}; 
\node[bd] (c1) at (1,0) {} ; 
\node[bd] (c2) at (1,1) {} ; 
\draw[ligne] (d1)--(d2)--(d3)--(d4)--(d5)--(d6)--(d7)--(d1); 
\draw[ligne] (d1)--(d4); 
\draw[ligne] (d2)--(c1); 
\draw[ligne] (d7)--(c1)--(d5); 
\draw[ligne,red] (c1)--(d3);

 \draw[->] (4,.5)--(6,.5);
\draw[step=.5cm,gray,very thin] (8,-1) grid (10,2);
\node[bd] (d1) at (8,0) {}; 
\node[bd] (d2) at (9,-1) {}; 
\node[bd] (d3) at (10,-1) {}; 
\node[bd] (d4) at (10,0) {};  
\node[bd] (d5) at (10,1) {}; 
\node[bd] (d6) at (9,2) {}; 
\node[bd] (d7) at (8,1) {}; 
\node[bd] (c1) at (9,0) {} ; 
\node[bd] (c2) at (9,1) {} ; 
\draw[ligne] (d1)--(d2)--(d3)--(d4)--(d5)--(d6)--(d7)--(d1); 
\draw[ligne] (d1)--(d4); 
\draw[ligne] (d2)--(c1); 
\draw[ligne] (d7)--(c1)--(d5); 
\draw[ligne,red] (d2)--(d4);
\end{scope}

\draw[brane] (-2,-2)--(-1,-3)--(0,-2); 
\draw[brane] (-1,-3)--(-2,-4)--(-3,-4); 
\draw[brane] (-1,-3)--(0,-4)--(1,-4);
\draw[brane] (-2,-4)--(-2,-5)--(-3,-6); 
\draw[brane] (0,-4)--(0,-4.5)--(1,-4.5); 
\draw[brane] (-2,-5)--(-.5,-5)--(-.5,-6); 
\draw[brane,red] (-.5,-5)--(0,-4.5); 
\node[D7] at (-2,-2) {};
\node[D7] at (0,-2) {};
\node[D7] at (-3,-4) {};
\node[D7] at (1,-4) {};
\node[D7] at (-3,-6) {};
\node[D7] at (1,-4.5) {};
\node[D7] at (-.5,-6) {};
\draw[->] (2,-4)--(3,-4);
\draw[brane] (5,-2)--(6,-3)--(7,-2); 
\draw[brane] (6,-3)--(5,-4)--(4,-4); 
\draw[brane] (6,-3)--(7,-4)--(8,-4);
\draw[brane] (5,-4)--(5,-5)--(4,-6); 
\draw[brane] (7,-4)--(7,-5); 
\draw[brane] (5,-5)--(7,-5); 
\draw[brane,red] (7,-5)--(7.5,-5.5);
\draw[brane] (7.5,-6.5)--(7.5,-5.5)--(8.5,-5.5);  
\node[D7] at (5,-2) {};
\node[D7] at (7,-2) {};
\node[D7] at (4,-4) {};
\node[D7] at (8,-4) {};
\node[D7] at (4,-6) {};
\node[D7] at (7.5,-6.5) {};
\node[D7] at (8.5,-5.5) {};

\begin{scope}[x=.5cm,y=.5cm] 
\draw[step=.5cm,gray,very thin] (0,-19) grid (2,-16);
\node[bd] (d1) at (0,-18) {}; 
\node[bd] (d2) at (1,-18) {}; 
\node[bd] (d3) at (2,-19) {}; 
\node[bd] (d4) at (2,-18) {};  
\node[bd] (d5) at (2,-17) {}; 
\node[bd] (d6) at (1,-16) {}; 
\node[bd] (d7) at (0,-16) {}; 
\node[bd] (d8) at (0,-17) {}; 
\node[bd] (c1) at (1,-17) {} ; 
\draw[ligne] (d1)--(d2)--(d3)--(d4)--(d5)--(d6)--(d7)--(d8)--(d1); 
\draw[ligne] (d6)--(c1)--(d2);
\draw[ligne] (d8)--(c1)--(d4); 
\draw[ligne,red] (c1)--(d3);

 \draw[->] (4,-17.5)--(6,-17.5);
\draw[step=.5cm,gray,very thin] (8,-17) grid (10,-16);
\node[bd] (d1) at (8,-18) {}; 
\node[bd] (d2) at (9,-18) {}; 
\node[bd] (d3) at (10,-19) {}; 
\node[bd] (d4) at (10,-18) {};  
\node[bd] (d5) at (10,-17) {}; 
\node[bd] (d6) at (9,-16) {}; 
\node[bd] (d7) at (8,-16) {}; 
\node[bd] (d8) at (8,-17) {}; 
\node[bd] (c1) at (9,-17) {} ; 
\draw[ligne] (d1)--(d2)--(d3)--(d4)--(d5)--(d6)--(d7)--(d8)--(d1); 
\draw[ligne] (d6)--(c1)--(d2);
\draw[ligne] (d8)--(c1)--(d4); 
\draw[ligne,red] (d2)--(d4);
\end{scope}

\draw[brane] (-3,-11)--(-3,-14); 
\draw[brane] (-4,-12)--(1,-12); 
\draw[brane] (-4,-13)--(-1.5,-13)--(-2.5,-14);
\draw[brane] (1,-11)--(-.5,-12.5)--(.5,-12.5); 
\draw[brane,red] (-1.5,-13)--(-.5,-12.5); 
\node[D7] at (-3,-11) {};
\node[D7] at (-3,-14) {};
\node[D7] at (-4,-12) {};
\node[D7] at (-4,-13) {};
\node[D7] at (1,-12) {};
\node[D7] at (-2.5,-14) {};
\node[D7] at (.5,-12.5) {};
\node[D7] at (1,-11) {};

\draw[->] (2,-13)--(3,-13);

\draw[brane] (5,-11)--(5,-14); 
\draw[brane] (4,-12)--(9,-12); 
\draw[brane] (4,-13)--(7,-13)--(9,-11); 
\draw[brane,red] (7,-13)--(7,-14); 
\draw[brane] (6,-15)--(7,-14)--(8,-14); 
\node[D7] at (5,-11) {};
\node[D7] at (5,-14) {};
\node[D7] at (4,-12) {};
\node[D7] at (4,-13) {};
\node[D7] at (9,-12) {};
\node[D7] at (8,-14) {};
\node[D7] at (9,-11) {};
\node[D7] at (6,-15) {};
\end{tikzpicture}
\caption{Decoupling in a toric (top) and pGTP (bottom), as well as their counterparts in the brane-web. In each of the figures the left hand side shows the configuration before the flop, with the matter hypermultiplet "curve" shown in red, and the right hand side shows the configuration, after the decoupling/flop. 
\label{fig:DecoupleToric}}
\end{figure}
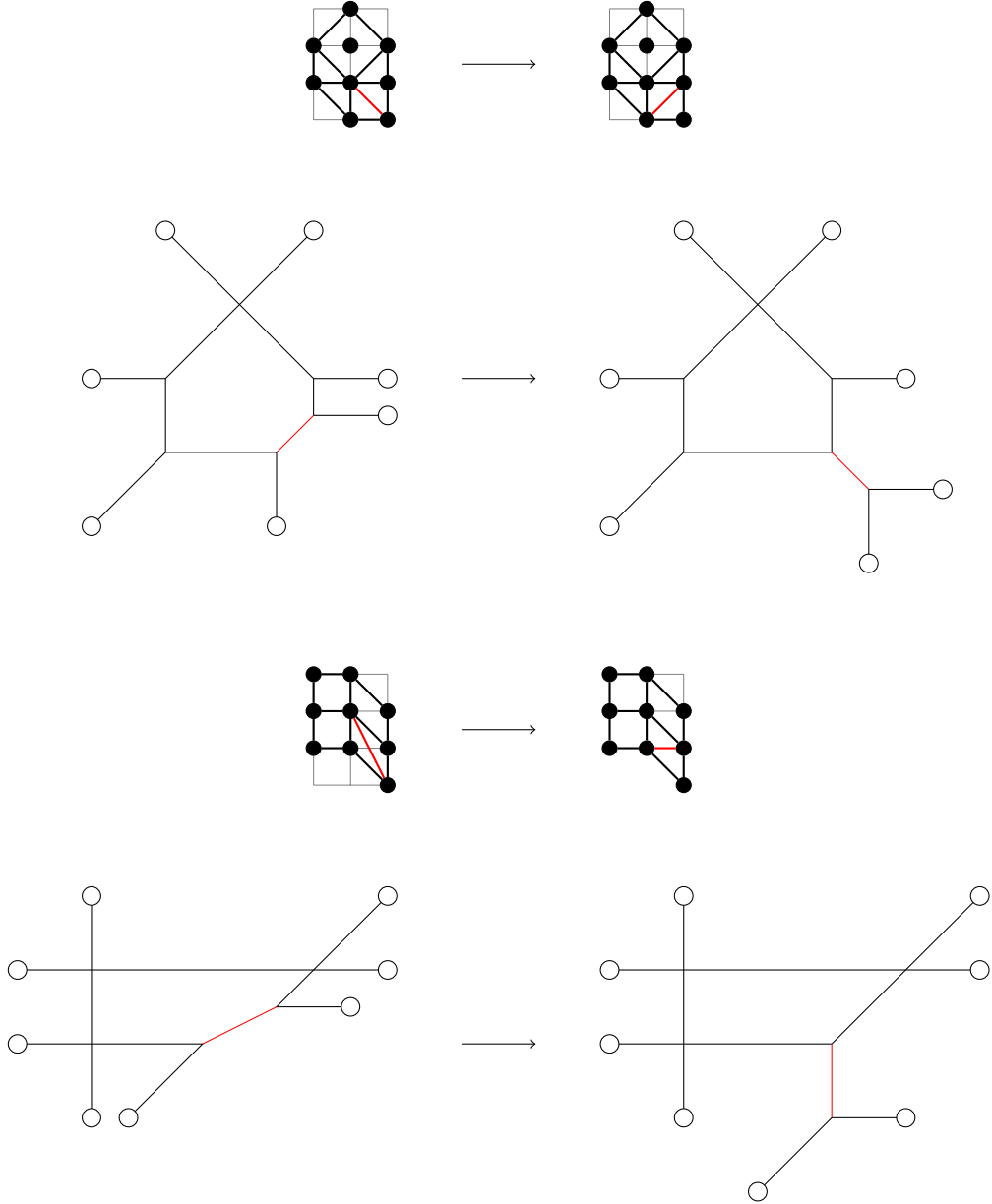

Let us first recall briefly the toric situation. In this case a hyper-multiplet curve is decoupled by a flop, which contracts the curve and flops it outside of the compact surfaces -- in practice this means that one of the external vertices (corresponding to non-compact divisors) is no longer connected to an internal vertex. An example is shown in figure \ref{fig:DecoupleToric}. 

Similarly for the case of pGTPs a fundamental matter multiplet will correspond locally to a curve connecting an internal to an external corner vertex. Decoupling this we follow the same procedure as in the toric diagram (even though of course the geometric analogy does not exist here). Figure \ref{fig:DecoupleToric} shows an example.

\subsection{Decoupling Anti-Symmetric Matter}

The decoupling of anti-symmetric hypermultiplets proceeds in a similar way. The theories which we consider have one or two $\mathbf{AS}$ of an $SU(N)$ gauge theory. 
First in the (p)GTP we identify the ruling, which are internal black dots, corresponding to the Cartan subgroups of the gauge group. To decouple an $\mathbf{AS}$, a sub-division of the (p)GTP needs to be achieved, such that there is no connection between the vertices of the ruling and the external node corresponding to the $\mathbf{AS}$ matter. 
In addition this decoupling needs to satisfy all the rules of GTPs (the r-rule in particular).

\begin{figure}
	\centering 
	\begin{equation*}
		\raisebox{-.5\height}{ \scalebox{.99}{  
				\begin{tikzpicture}[x=.5cm,y=.5cm] 
					\draw[step=.5cm,gray,very thin] (-4,-4) grid (1,4);
					\draw[ligne] (-4,0)--(0,-4); 
					\draw[ligne] (0,-4)--(1,-1); 
					\draw[ligne] (1,-1)--(0,4); 
					\draw[ligne] (0,4)--(-1,3); 
					\draw[ligne] (-1,3)--(-4,0); 
					\node[bd] at (0,-4) {}; 
					\node[bd] at (1,-1) {}; 
					\node[bd] at (0,4) {}; 
					\node[bd] at (-1,3) {}; 
					\node[bd] at (-4,0) {}; 
					\node[wd] at (-2,2) {}; 
					\node[wd] at (-3,1) {}; 
					\node[wd] at (-3,-1) {}; 
					\node[wd] at (-2,-2) {}; 
					\node[wd] at (-1,-3) {}; 
					\draw[decoration={brace,mirror,raise=5pt},decorate]  (-5,4) -- node[left=8pt] {$1$} (-5,3.2);
					\draw[decoration={brace,mirror,raise=5pt},decorate]  (-5,2.8) -- node[left=8pt] {$n-1$} (-5,.2);
					\draw[decoration={brace,mirror,raise=5pt},decorate]  (-5,-.2) -- node[left=8pt] {$n$} (-5,-4);
					\draw[decoration={brace,raise=5pt},decorate]  (2,4) -- node[right=8pt] {$n+1$} (2,-0.8);
					\draw[decoration={brace,raise=5pt},decorate]  (2,-1.2) -- node[right=8pt] {$n-1$} (2,-4);
					\node at (-1.5,-5) {$SU(2n)_0 + 1 \mathbf{AS}$}; 
		\end{tikzpicture}}}
	\end{equation*}
	\begin{equation*}
		\raisebox{-.5\height}{ \scalebox{.99}{  
				\begin{tikzpicture}[x=.5cm,y=.5cm] 
					\draw[step=.5cm,gray,very thin] (-5,-5) grid (1,4);
					\draw[ligne] (-5,0)--(0,-5); 
					\draw[ligne] (0,-5)--(1,-2); 
					\draw[ligne] (1,-2)--(0,4); 
					\draw[ligne] (0,4)--(-1,4); 
					\draw[ligne] (-1,4)--(-5,0); 
					\node[bd] at (0,-5) {}; 
					\node[bd] at (1,-2) {}; 
					\node[bd] at (0,4) {}; 
					\node[bd] at (-1,4) {}; 
					\node[bd] at (-5,0) {}; 
					\node[wd] at (-2,3) {}; 
					\node[wd] at (-3,2) {}; 
					\node[wd] at (-4,1) {}; 
					\node[wd] at (-4,-1) {}; 
					\node[wd] at (-3,-2) {}; 
					\node[wd] at (-2,-3) {}; 
					\node[wd] at (-1,-4) {}; 
					\draw[decoration={brace,mirror,raise=5pt},decorate]  (-6,3.8) -- node[left=8pt] {$n$} (-6,.2);
					\draw[decoration={brace,mirror,raise=5pt},decorate]  (-6,-.2) -- node[left=8pt] {$n+1$} (-6,-5);
					\draw[decoration={brace,raise=5pt},decorate]  (2,4) -- node[right=8pt] {$n+2$} (2,-1.8);
					\draw[decoration={brace,raise=5pt},decorate]  (2,-2.2) -- node[right=8pt] {$n-1$} (2,-5);
					\node at (-2,-6) {$SU(2n+1)_{1/2} + 1 \mathbf{AS}$}; 
		\end{tikzpicture}}} \qquad 
		\raisebox{-.5\height}{ \scalebox{.99}{  
				\begin{tikzpicture}[x=.5cm,y=.5cm] 
					\draw[step=.5cm,gray,very thin] (-5,-5) grid (1,4);
					\draw[ligne] (-5,0)--(0,-5); 
					\draw[ligne] (0,-5)--(1,-3); 
					\draw[ligne] (1,-3)--(0,4); 
					\draw[ligne] (0,4)--(-1,4); 
					\draw[ligne] (-1,4)--(-5,0); 
					\node[bd] at (0,-5) {}; 
					\node[bd] at (1,-3) {}; 
					\node[bd] at (0,4) {}; 
					\node[bd] at (-1,4) {}; 
					\node[bd] at (-5,0) {}; 
					\node[wd] at (-2,3) {}; 
					\node[wd] at (-3,2) {}; 
					\node[wd] at (-4,1) {}; 
					\node[wd] at (-4,-1) {}; 
					\node[wd] at (-3,-2) {}; 
					\node[wd] at (-2,-3) {}; 
					\node[wd] at (-1,-4) {}; 
					\draw[decoration={brace,mirror,raise=5pt},decorate]  (-6,3.8) -- node[left=8pt] {$n$} (-6,.2);
					\draw[decoration={brace,mirror,raise=5pt},decorate]  (-6,-.2) -- node[left=8pt] {$n+1$} (-6,-5);
					\draw[decoration={brace,raise=5pt},decorate]  (2,4) -- node[right=8pt] {$n+3$} (2,-2.8);
					\node at (-2,-6) {$SU(2n+1)_{-1/2} + 1 \mathbf{AS}$}; 
		\end{tikzpicture}}}
	\end{equation*}
	\caption{Toric polygons for $SU(N)_{k} + 1 \mathbf{AS}$ for the lowest possible values of the Chern-Simons level, $|k|<1$. We have to distinguish between the cases where $N=2n$ is even and $N=2n+1$ is odd. The polygons presented here can be deduced from the brane webs constructed in \cite{Zafrir:2015rga}. The integers next to the braces indicate the vertical length of the corresponding segment; for the drawings we picked $n=4$. The Chern-Simons level can be increased / decreased by one unit by moving the rightmost dot upwards / downwards by one unit of distance.  }
	\label{figpureAS}
\end{figure}
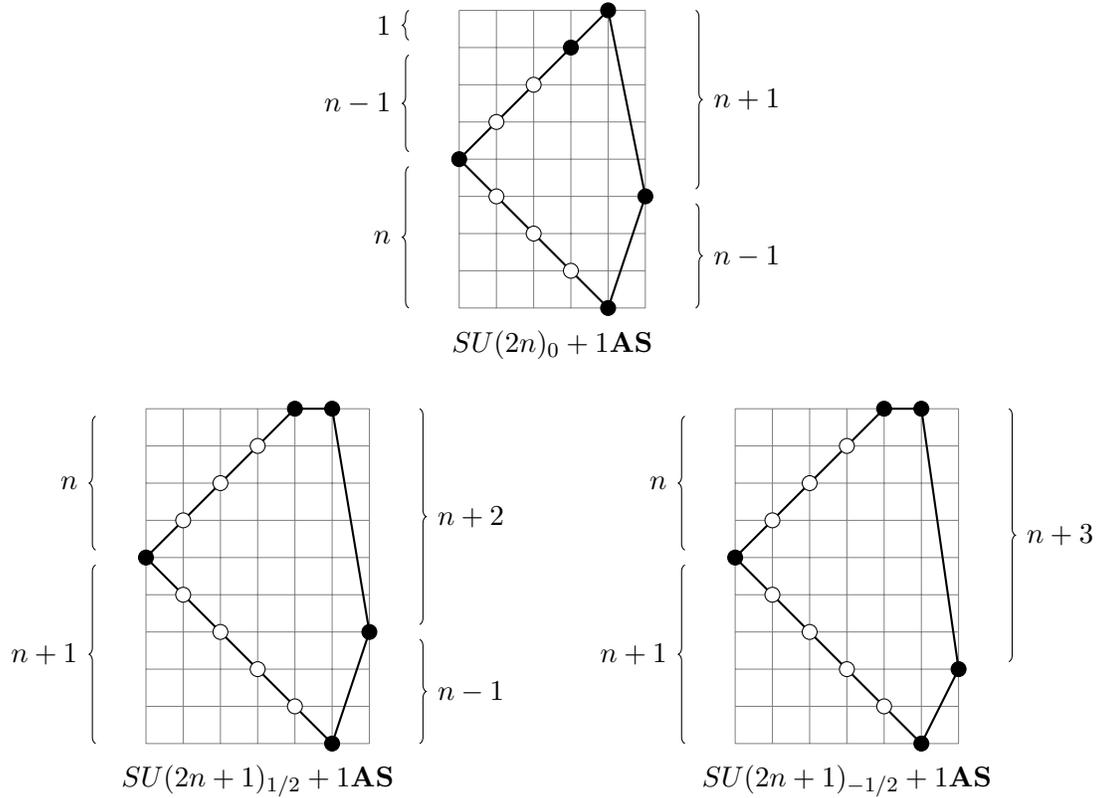

Let us illustrate the decoupling of an anti-symmetric hypermultiplet. It is particularly interesting to keep track of the Chern-Simons level in the process. Starting from the first polygon in figure \ref{figpureAS}, which is $SU(2n)_0 + 1 \mathbf{AS}$, the decoupling of the $\mathbf{AS}$ hyper is given by 
 \begin{equation}
        \raisebox{-.5\height}{ \scalebox{.99}{  
\begin{tikzpicture}[x=.5cm,y=.5cm] 
\draw[step=.5cm,gray,very thin] (-5,-4) grid (1,4);
\draw[ligne] (-4,0)--(0,-4); 
\draw[ligne] (0,-4)--(1,-1); 
\draw[ligne] (1,-1)--(0,4); 
\draw[ligne] (0,4)--(-1,3); 
\draw[ligne] (-1,3)--(-4,0); 
\draw[ligne,red] (-1,3)--(0,-4); 
\draw[ligne, thin] (0,4)--(0,-4) ;
\node[bd] at (0,-4) {}; 
\node[bd] at (1,-1) {}; 
\node[bd] at (0,4) {}; 
\node[bd] at (-1,3) {}; 
\node[bd] at (-4,0) {}; 
\node[wd] at (-2,2) {}; 
\node[wd] at (-3,1) {}; 
\node[wd] at (-3,-1) {}; 
\node[wd] at (-2,-2) {}; 
\node[wd] at (-1,-3) {}; 
\draw[decoration={brace,mirror,raise=5pt},decorate]  (-5,4) -- node[left=8pt] {$1$} (-5,3.2);
\draw[decoration={brace,mirror,raise=5pt},decorate]  (-5,2.8) -- node[left=8pt] {$n-1$} (-5,.2);
\draw[decoration={brace,mirror,raise=5pt},decorate]  (-5,-.2) -- node[left=8pt] {$n$} (-5,-4);
\draw[decoration={brace,raise=5pt},decorate]  (2,4) -- node[right=8pt] {$n+1$} (2,-0.8);
\draw[decoration={brace,raise=5pt},decorate]  (2,-1.2) -- node[right=8pt] {$n-1$} (2,-4);
\end{tikzpicture}}}
    \end{equation}
 The sub-division is shown in terms of the red line, and the ruling is indicated in black.    
After decoupling, we obtain a polygon which corresponds to the pure theory $SU(2n)_{n-2}$. Therefore the $\mathbf{AS}$ hyper contributes to $n-2$ units of Chern-Simons level in $SU(2n)$ theories. This is in perfect agreement with the expectation (\ref{CSshift}), according to which the Chern-Simons level is shifted by $\frac{1}{2}A (\bm{AS}) = n-2$ upon decoupling of the antisymmetric hyper. 
    
We can do the same for the other two polygons of figure \ref{figpureAS}, and obtain 
    \begin{equation}
    \raisebox{-.5\height}{ \scalebox{.99}{  
\begin{tikzpicture}[x=.5cm,y=.5cm] 
\draw[step=.5cm,gray,very thin] (-5,-5) grid (1,4);
\draw[ligne] (-5,0)--(0,-5); 
\draw[ligne] (0,-5)--(1,-2); 
\draw[ligne] (1,-2)--(0,4); 
\draw[ligne] (0,4)--(-1,4); 
\draw[ligne] (-1,4)--(-5,0); 
\draw[ligne,red] (-1,4)--(0,-5); 
\draw[ligne, thin] (0,4)--(0,-5) ;
\node[bd] at (0,-5) {}; 
\node[bd] at (1,-2) {}; 
\node[bd] at (0,4) {}; 
\node[bd] at (-1,4) {}; 
\node[bd] at (-5,0) {}; 
\node[wd] at (-2,3) {}; 
\node[wd] at (-3,2) {}; 
\node[wd] at (-4,1) {}; 
\node[wd] at (-4,-1) {}; 
\node[wd] at (-3,-2) {}; 
\node[wd] at (-2,-3) {}; 
\node[wd] at (-1,-4) {}; 
\draw[decoration={brace,mirror,raise=5pt},decorate]  (-6,3.8) -- node[left=8pt] {$n$} (-6,.2);
\draw[decoration={brace,mirror,raise=5pt},decorate]  (-6,-.2) -- node[left=8pt] {$n+1$} (-6,-5);
\draw[decoration={brace,raise=5pt},decorate]  (2,4) -- node[right=8pt] {$n+2$} (2,-1.8);
\draw[decoration={brace,raise=5pt},decorate]  (2,-2.2) -- node[right=8pt] {$n-1$} (2,-5);
\node at (-2,-6) {$SU(2n+1)_{1/2} + 1 \mathbf{AS}$}; 
\end{tikzpicture}}} \qquad 
        \raisebox{-.5\height}{ \scalebox{.99}{  
\begin{tikzpicture}[x=.5cm,y=.5cm] 
\draw[step=.5cm,gray,very thin] (-5,-5) grid (2,4);
\draw[ligne] (-5,0)--(0,-5); 
\draw[ligne] (0,-5)--(1,-3); 
\draw[ligne] (1,-3)--(0,4); 
\draw[ligne] (0,4)--(-1,4); 
\draw[ligne] (-1,4)--(-5,0); 
\draw[ligne,red] (-1,4)--(0,-5); 
\draw[ligne, thin] (0,4)--(0,-5) ;
\node[bd] at (0,-5) {}; 
\node[bd] at (1,-3) {}; 
\node[bd] at (0,4) {}; 
\node[bd] at (-1,4) {}; 
\node[bd] at (-5,0) {}; 
\node[wd] at (-2,3) {}; 
\node[wd] at (-3,2) {}; 
\node[wd] at (-4,1) {}; 
\node[wd] at (-4,-1) {}; 
\node[wd] at (-3,-2) {}; 
\node[wd] at (-2,-3) {}; 
\node[wd] at (-1,-4) {}; 
\draw[decoration={brace,mirror,raise=5pt},decorate]  (-6,3.8) -- node[left=8pt] {$n$} (-6,.2);
\draw[decoration={brace,mirror,raise=5pt},decorate]  (-6,-.2) -- node[left=8pt] {$n+1$} (-6,-5);
\draw[decoration={brace,raise=5pt},decorate]  (2,4) -- node[right=8pt] {$n+3$} (2,-2.8);
\draw[decoration={brace,raise=5pt},decorate]  (2,-3.2) -- node[right=8pt] {$n-2$} (2,-5);
\node at (-2,-6) {$SU(2n+1)_{-1/2} + 1 \mathbf{AS}$}; 
\end{tikzpicture}}}
    \end{equation}
In the first case, the level after decoupling is $n-1$ while in the second case it is $n-2$. This means that the $\mathbf{AS}$ hyper contributes $\frac{1}{2}-(n-1)=-\frac{1}{2}-(n-2)=\frac{3}{2}-n$ units of Chern-Simons level in $SU(2n+1)$ theories. Again this is in perfect agreement with (\ref{CSshift}), according to which the Chern-Simons level is shifted by $\frac{1}{2}A (\bm{AS}) = n- \frac{3}{2}$ upon decoupling of the antisymmetric hyper.

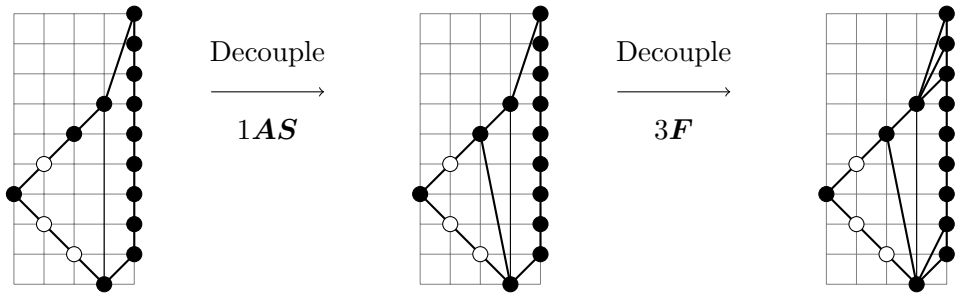
\begin{figure}
\begin{equation*}
\raisebox{-.5\height}{ \scalebox{1}{  
\begin{tikzpicture}[x=.4cm,y=.4cm] 
\draw[step=.4cm,gray,very thin] (-3,-3) grid (1,6);
\draw[ligne] (-3,0)--(0,-3); 
\draw[ligne] (0,-3)--(1,-2); 
\draw[ligne] (1,-2)--(1,-1); 
\draw[ligne] (1,-1)--(1,0); 
\draw[ligne] (1,0)--(1,1); 
\draw[ligne] (1,1)--(1,2); 
\draw[ligne] (1,2)--(1,3); 
\draw[ligne] (1,3)--(1,4); 
\draw[ligne] (1,4)--(1,5); 
\draw[ligne] (1,5)--(1,6); 
\draw[ligne] (1,6)--(0,3); 
\draw[ligne] (0,3)--(-1,2); 
\draw[ligne] (-1,2)--(-3,0); 
\draw[ligne, thin] (0,3)--(0,-3) ;
\node[bd] at (0,-3) {}; 
\node[bd] at (1,-2) {}; 
\node[bd] at (1,-1) {}; 
\node[bd] at (1,0) {}; 
\node[bd] at (1,1) {}; 
\node[bd] at (1,2) {}; 
\node[bd] at (1,3) {}; 
\node[bd] at (1,4) {}; 
\node[bd] at (1,5) {}; 
\node[bd] at (1,6) {}; 
\node[bd] at (0,3) {}; 
\node[bd] at (-1,2) {}; 
\node[bd] at (-3,0) {}; 
\node[wd] at (-2,1) {}; 
\node[wd] at (-2,-1) {}; 
\node[wd] at (-1,-2) {}; 
\node at (0,-3.5) {};
\node at (0,6.5) {};
\end{tikzpicture}}}
\qquad
\begin{tikzpicture}[x=.5cm,y=.5cm] 
\draw[->] (0,0)--(3,0);
\node[] at (1.5,1) {Decouple};
\node[] at (1.5,-1) {$1 \bm{AS}$};
\end{tikzpicture}
\qquad
\raisebox{-.5\height}{ \scalebox{1}{  
\begin{tikzpicture}[x=.4cm,y=.4cm] 
\draw[step=.4cm,gray,very thin] (-3,-3) grid (1,6);
\draw[ligne] (-3,0)--(0,-3); 
\draw[ligne] (0,-3)--(1,-2); 
\draw[ligne] (1,-2)--(1,-1); 
\draw[ligne] (1,-1)--(1,0); 
\draw[ligne] (1,0)--(1,1); 
\draw[ligne] (1,1)--(1,2); 
\draw[ligne] (1,2)--(1,3); 
\draw[ligne] (1,3)--(1,4); 
\draw[ligne] (1,4)--(1,5); 
\draw[ligne] (1,5)--(1,6); 
\draw[ligne] (1,6)--(0,3); 
\draw[ligne] (0,3)--(-1,2); 
\draw[ligne] (-1,2)--(-3,0); 
\draw[ligne] (-1,2)--(0,-3); 
\draw[ligne, thin] (0,3)--(0,-3) ;
\node[bd] at (0,-3) {}; 
\node[bd] at (1,-2) {}; 
\node[bd] at (1,-1) {}; 
\node[bd] at (1,0) {}; 
\node[bd] at (1,1) {}; 
\node[bd] at (1,2) {}; 
\node[bd] at (1,3) {}; 
\node[bd] at (1,4) {}; 
\node[bd] at (1,5) {}; 
\node[bd] at (1,6) {}; 
\node[bd] at (0,3) {}; 
\node[bd] at (-1,2) {}; 
\node[bd] at (-3,0) {}; 
\node[wd] at (-2,1) {}; 
\node[wd] at (-2,-1) {}; 
\node[wd] at (-1,-2) {}; 
\node at (0,-3.5) {};
\node at (0,6.5) {};
\end{tikzpicture}}}
\qquad
\begin{tikzpicture}[x=.5cm,y=.5cm] 
\draw[->] (0,0)--(3,0);
\node[] at (1.5,1) {Decouple};
\node[] at (1.5,-1) {$3 \bm{F}$};
\end{tikzpicture}
\qquad
\raisebox{-.5\height}{ \scalebox{1}{  
\begin{tikzpicture}[x=.4cm,y=.4cm] 
\draw[step=.4cm,gray,very thin] (-3,-3) grid (1,6);
\draw[ligne] (-3,0)--(0,-3); 
\draw[ligne] (0,-3)--(1,-2); 
\draw[ligne] (1,-2)--(1,-1); 
\draw[ligne] (1,-1)--(1,0); 
\draw[ligne] (1,0)--(1,1); 
\draw[ligne] (1,1)--(1,2); 
\draw[ligne] (1,2)--(1,3); 
\draw[ligne] (1,3)--(1,4); 
\draw[ligne] (1,4)--(1,5); 
\draw[ligne] (1,5)--(1,6); 
\draw[ligne] (1,6)--(0,3); 
\draw[ligne] (1,5)--(0,3); 
\draw[ligne] (1,4)--(0,3); 
\draw[ligne] (1,-1)--(0,-3); 
\draw[ligne] (0,3)--(-1,2); 
\draw[ligne] (-1,2)--(-3,0); 
\draw[ligne] (-1,2)--(0,-3); 
\draw[ligne, thin] (0,3)--(0,-3) ;
\node[bd] at (0,-3) {}; 
\node[bd] at (1,-2) {}; 
\node[bd] at (1,-1) {}; 
\node[bd] at (1,0) {}; 
\node[bd] at (1,1) {}; 
\node[bd] at (1,2) {}; 
\node[bd] at (1,3) {}; 
\node[bd] at (1,4) {}; 
\node[bd] at (1,5) {}; 
\node[bd] at (1,6) {}; 
\node[bd] at (0,3) {}; 
\node[bd] at (-1,2) {}; 
\node[bd] at (-3,0) {}; 
\node[wd] at (-2,1) {}; 
\node[wd] at (-2,-1) {}; 
\node[wd] at (-1,-2) {}; 
\node at (0,-3.5) {};
\node at (0,6.5) {};
\end{tikzpicture}}}
\end{equation*}
\caption[Example of decoupling anti-symmetric and fundamental matter]{Example of decoupling matter. We start from $SU(2n)_n+1\bm{AS}+ 8 \bm{F}$ and after decoupling an anti-symmetric we obtain $SU(2n)_{2n-2} + 8 \bm{F}$. We then decouple three fundamentals (one with positive and two with negative mass) to reach $SU(2n)_{2n-\frac{5}{2}} + 5 \bm{F}$. }
\label{fig:ExampleDecoupling}
\end{figure}

In all cases, we see that the $\mathbf{AS}$ hyper as constructed in these polygons contribute $\frac{N-4}{2}$ units of Chern-Simons level in $SU(N)$ theories, in accordance with \eqref{CSshift} and table \ref{tab:IndexAnomaly}. Note that for $SU(3)$ theories, this is indeed $-\frac{1}{2}$, as expected from the fact that $\mathbf{AS}\equiv \bar{\mathbf{F}}$ in that case. In particular, we see that 
\begin{itemize}
    \item The marginal theory $SU(N)_{N/2} + 1 \mathbf{AS} + 8 \mathbf{F}$ gives after decoupling the $\mathbf{AS}$ the non-marginal theory $SU(N)_{N-2} + 8 \mathbf{F}$, which can also be seen as the marginal theory $SU(N)_{0} + (2N+4) \mathbf{F}$ where we have decoupled $N-4$ fundamentals with level $\frac{1}{2}$. 
    \item The marginal theory $SU(N)_{0} + 1 \mathbf{AS} + (N+6) \mathbf{F}$ gives after decoupling the $\mathbf{AS}$ the non-marginal theory $SU(N)_{\frac{N}{2}-2} + (N+6) \mathbf{F}$, which can also be seen as the marginal theory $SU(N)_{0} + (2N+4) \mathbf{F}$ where we have decoupled $N-3$ fundamentals with level $\frac{1}{2}$ and one fundamental with level $- \frac{1}{2}$. 
\end{itemize}
The toric diagram interpretation guarantees that these statements are valid at the SCFT (strong coupling) point. 
One can also decouple several hypers at the same time using the same procedure. This is illustrated in figure \ref{fig:ExampleDecoupling}.

\subsection{Decoupling for IR Magnetic Quivers}
\label{sec:DecouplingMQs}

So far we discussed the decoupling for a generic pGTP, having in mind GTPs that correspond to strongly-coupled UV fixed points. 
In this section we implement decoupling directly in the IR magnetic quivers. A word of caution is in place here: the IR magnetic quivers are blind to several important characteristics of a theory, e.g. the CS-couplings. This means that one cannot (at least not without additional information) infer the UV from the IR magnetic quiver. Nevertheless, the decoupling is a relevant concept also in the IR, and we will see that it reduces in may cases simply to a quiver subtraction. 

We consider the cases 
\be
SU(N)_{k} + N_F \mbf{F}\,,\qquad Sp(N) + N_F \mbf{F} \,.
\ee
Decoupling essentially corresponds to subtracting a hypermultiplet from the magnetic quiver. 
The magnetic quivers for these theories are given by 
\be
\ba
 &
\begin{tikzpicture}[x=.8cm,y=.8cm]
\node (g0) at (-3, 0) {$SU(N) + N_F \mbf{F}:\qquad$};
\node (g1) at (0,0) [gauge,label=below:{1}] {};
\node (g2) at (1,0) [gauge,label=below:{2}] {};
\node (g3) at (2,-.2)  {$\cdots$};
\node (g4) at (3,0) [gauge,label=below:{$N$}] {};
\node (g5) at (4,0) [gauge,label=below:{$N$}] {};
\node (g6) at (5,-0.2)  {$\cdots$};
\node (g7) at (6,0) [gauge,label=below:{$N$}] {};
\node (g8) at (7,0) [gauge,label=below:{$N$}] {};
\node (g9) at (8,-0.2)  {$\cdots$};
\node (g10) at (9,0) [gauge,label=below:{$2$}] {};
\node (g11) at (10,0) [gauge,label=below:{$1$}] {};
\node (g12) at (3,1) [gaugeb,label=right:{$1$}] {};
\node (g13) at (7,1) [gaugeb,label=right:{$1$}] {};
\draw (g1)--(g2)--(g4)--(g5)--(g7)--(g8)--(g10)--(g11);
\draw (g4)--(g12);
\draw (g8)--(g13);
\draw [decorate,decoration={brace,mirror, amplitude=10pt}] (0,-1) -- (10,-1);
\node (g14) at (5,-1.8) {$N_F-1$};
\end{tikzpicture}
\cr 
&
\begin{tikzpicture}[x=.8cm,y=.8cm]
\node (g0) at (-3, 0)  {$Sp(N) + N_F \mbf{F}:\qquad $};
\node (g1) at (0,0) [gauge,label=below:{1}] {};
\node (g2) at (1,0) [gauge,label=below:{2}] {};
\node (g3) at (2,-.2)  {$\cdots$};
\node (g4) at (3,0) [gauge,label=below:{$2 N$}] {};
\node (g5) at (4,0) [gauge,label=below:{$2 N$}] {};
\node (g6) at (5,-0.2)  {$\cdots$};
\node (g7) at (6,0) [gauge,label=below:{$2 N$}] {};
\node (g8) at (7,0) [gauge,label=below:{$2 N$}] {};
\node (g9) at (8,0) [gauge,label=below:{$N$}] {};
\node (g12) at (3,1) [gaugeb,label=right:{$1$}] {};
\node (g13) at (7,1) [gauge,label=right:{$N$}] {};
\draw (g1)--(g2)--(g4)--(g5)--(g7)--(g8)--(g9);
\draw (g4)--(g12);
\draw (g8)--(g13);
\draw [decorate,decoration={brace,mirror, amplitude=10pt}] (0,-1) -- (8,-1);
\node (g14) at (4,-1.8) {$N_F-1$};
\end{tikzpicture}
\ea\ee
To pass from these magnetic quivers to the ones with one fundamental decoupled is equivalent to the quiver subtraction by 
\be
\ba
 &
\begin{tikzpicture}[x=.8cm,y=.8cm]
\node (g0) at (0, 0) {Quiver subtraction for  $SU(N) + N_F \mbf{F}:\qquad$};
\node (g8) at (7,0) [gauge,label=below:{$1$}] {};
\node (g9) at (8,-0.2)  {$\cdots$};
\node (g10) at (9,0) [gauge,label=below:{$1$}] {};
\node (g11) at (10,0) [gauge,label=below:{$1$}] {};
\node (g13) at (7,1) [gauge,label=right:{$1$}] {};
\draw (g8)--(g10)--(g11);
\draw (g8)--(g13);
\draw [decorate,decoration={brace,mirror, amplitude=10pt}] (7,-1) -- (10,-1);
\node (g14) at (8,-1.8) {$N$};
\end{tikzpicture}
\cr 
&
\begin{tikzpicture}[x=.8cm,y=.8cm]
\node (g0) at (-5, 0)  {Quiver subtraction for  $Sp(N) + N_F \mbf{F}:\qquad $};
\node (g1) at (0,0) [gauge,label=below:{1}] {};
\node (g2) at (1,0) [gauge,label=below:{1}] {};
\node (g3) at (2,-.2)  {$\cdots$};
\node (g4) at (3,0) [gauge,label=below:{$1$}] {};
\node (g12) at (3,1) [gauge,label=right:{$1$}] {};
\draw (g1)--(g2)--(g4);
\draw (g4)--(g12);
\draw [decorate,decoration={brace,mirror, amplitude=10pt}] (0,-1) -- (3,-1);
\node (g14) at (1.5,-1.8) {$2N$};
\end{tikzpicture}
\ea\ee
The interpretation of each of these diagrams is of course that they correspond to a free hyper in the fundamental. Subtracting these using the rules in 
\cite{Cabrera:2018ann}, gives rise to the magnetic quiver of the theory with one less flavor.

%

This approach generalizes readily, e.g. to the 5d quiver theories $SU(2)-SU(2)$, where decoupling in the IR-MQ is just subtracting an $\su(4)$.


\section[UV Magnetic Quivers for \texorpdfstring{$SU(N)$}{SU(N)} with \texorpdfstring{$AS$}{AS} and \texorpdfstring{$F$}{F} via GTPs]
{UV Magnetic Quivers for \boldmath{$SU(N)$} with \boldmath{$AS$} and \boldmath{$F$} via GTPs}
\label{sec:SUNasf}

For general $SU(N)$ with fundamental and anti-symmetric matter there are five marginal theories introduced in \cite{Jefferson:2018irk}.  For each of them we consider the submarginal theory obtained by decoupling a single fundamental hypermultiplet. We give the (p)GTP before and after edge-moves and compute the magnetic quiver for the SCFT. Decoupling further matter was discussed in section \ref{sec:Decoupling} and the magnetic quivers of the descendant trees are given in appendix \ref{sec:MQTables}.

\subsection[\texorpdfstring{$SU(N)_\half + (2N+3) F$}{SU(N)1/2 + (2N+3) F}]{\boldmath{$SU(N)_\half + (2N+3) F$}}
\label{sec:SQCD}

\paragraph{GTP for Tree Tops.}

First, we study the submarginal theory associated to the circle reduction of $(D_{N+2},D_{N+2})$-conformal matter. This theory is the parent of all 5d SCFTs with an SQCD description in the IR.
The polygon for $SU(N)_\half + (2N+3) \bm{F}$ is given by
\be
\begin{tikzpicture}[x=.4cm,y=.4cm]
\draw[step=.4cm,gray,very thin] (0,0) grid (2,4);

\draw[ligne,black] (0,0)--(1,1)--(2,1)--(2,4)--(1,3)--(0,4)--(0,0);
\draw[black] (1,1)--(1,3);

\node[bd] at (0,0) {};
\node[bd] at (1,1) {};
\node[bd] at (2,1) {};
\node[bd] at (2,2) {};
\node[bd] at (2,3) {};
\node[bd] at (2,4) {};
\node[bd] at (1,3) {};
\node[bd] at (0,4) {};
\node[bd] at (0,3) {};
\node[bd] at (0,2) {};
\node[bd] at (0,1) {};

\node[] at (4,2.5) {$N+1$};
\node[] at (-2,2) {$N+2$};
\end{tikzpicture}
\ee
We draw the pGTP for $N=3$, but the generalization to any $N$ is obtained by increasing the number of vertical edges as shown. The ruling indicating the $SU(N)$ gauge theory description is given by the thin internal line. 
The edge-moves necessary to make this polygon convex were discussed in \cite{Cabrera:2018jxt} in the web description. However, let us exemplify them on the polygon and in our language.
Fix $\bm{v}_0$ to be the lower left corner. Then, the polygon is given by the vectors
\be
\bm{u}=\left[(1,1),(1,0),\underbrace{(0,1)}_{N+1},(-1,-1),(-1,1),\underbrace{(0,-1)}_{N+2}\right]\,.
\ee
The non-convex edges are $\mathfrak{C}=\{1,N+4\}$. First, let us perform an edge-move $\mathfrak{M}^-_{N+4}$. Using the rules laid out in \eqref{edgemoverules}, the polygon becomes
\be
\begin{tikzpicture}[x=.4cm,y=.4cm]
\node[] at (-18,2) {$\bm{u} \mapsto \mathfrak{M}^-_{N+4}\bm{u}=\left[(1,1),(1,0),\underbrace{(0,1)}_{N+1},(1,3),(-3,-3),\underbrace{(0,-1)}_{N+2}\right]\,,$};

\draw[step=.4cm,gray,very thin] (0,0) grid (3,7);

\draw[ligne,black] (0,0)--(1,1)--(2,1)--(2,4)--(3,7)--(0,4)--(0,0);
\draw[black] (1,1)--(1,3)--(2,4);
\draw[black] (1,3)--(0,4);

\node[bd] at (0,0) {};
\node[bd] at (1,1) {};
\node[bd] at (2,1) {};
\node[bd] at (2,2) {};
\node[bd] at (2,3) {};
\node[bd] at (2,4) {};
\node[bd] at (3,7) {};
\node[wd] at (2,6) {};
\node[wd] at (1,5) {};
\node[bd] at (0,4) {};
\node[bd] at (0,3) {};
\node[bd] at (0,2) {};
\node[bd] at (0,1) {};

\node[] at (4,2.5) {$N+1$};
\node[] at (-2,2) {$N+2$};
\end{tikzpicture}
\ee
As we can see there are now white vertices, because in the resulting polygon $\mathfrak{M}^-_{N+4}\bm{u}$ we have $\mu_{N+5}=\{3\}$. Next, we move the $(1,3)$ edge through all the $(0,1)$ edges by applying $\mathfrak{M}_{3}^+ \cdots \mathfrak{M}_{N+3}^+$. Then, the vectors are given by
\be 
\mathfrak{M}_3^+ \cdots \mathfrak{M}_{N+3}^+ \mathfrak{M}^-_{N+4}\bm{u}=\left[(1,1),(1,0),(N+2,3N+6),\underbrace{(-1,-2)}_{N+1},(-3,-3),\underbrace{(0,-1)}_{N+2}\right]\,,
\ee
and the polygon is 
\be
\begin{tikzpicture}[x=.4cm,y=.4cm]
\draw[step=.4cm,gray,very thin] (0,0) grid (6,13);

\draw[ligne,black] (0,0)--(1,1)--(2,1)--(6,13)--(5,11)--(4,9)--(3,7)--(0,4)--(0,0);
\draw[black] (1,1)--(1,3)--(2,4)--(2,1);
\draw[black] (1,3)--(0,4);
\draw[black] (2,4)--(3,7);

\node[bd] at (0,0) {};
\node[bd] at (1,1) {};
\node[bd] at (2,1) {};
\node[wd] at (3,4) {};
\node[wd] at (4,7) {};
\node[wd] at (5,10) {};
\node[bd] at (6,13) {};
\node[bd] at (5,11) {};
\node[bd] at (4,9) {};
\node[bd] at (3,7) {};
\node[wd] at (2,6) {};
\node[wd] at (1,5) {};
\node[bd] at (0,4) {};
\node[bd] at (0,3) {};
\node[bd] at (0,2) {};
\node[bd] at (0,1) {};

\node[] at (2.5,10) {$N+1$};
\node[] at (6,7) {$N+2$};
\node[] at (-2,2) {$N+2$};
\end{tikzpicture}
\ee
Now, we remove the non-convexity at $i=1$. We apply the edge-move $\mathfrak{M}^+_1$ which yields the vectors
\be
\mathfrak{M}^+_1 \mathfrak{M}_3^+ \cdots \mathfrak{M}_{N+3}^+ \mathfrak{M}^-_{N+4}\bm{u}=\left[(2,0),(0,1),(N+2,3N+6),\underbrace{(-1,-2)}_{N+1},(-3,-3),\underbrace{(0,-1)}_{N+2}\right]\,,
\ee
and polygon
\be
\begin{tikzpicture}[x=.4cm,y=.4cm]
\draw[step=.4cm,gray,very thin] (0,0) grid (6,13);

\draw[ligne,black] (0,0)--(2,0)--(2,1)--(6,13)--(5,11)--(4,9)--(3,7)--(0,4)--(0,0);
\draw[black] (0,0)--(1,1)--(1,3)--(2,4)--(2,1)--(1,1);
\draw[black] (1,3)--(0,4);
\draw[black] (2,4)--(3,7);

\node[bd] at (0,0) {};
\node[wd] at (1,0) {};
\node[bd] at (2,0) {};
\node[bd] at (2,1) {};
\node[wd] at (3,4) {};
\node[wd] at (4,7) {};
\node[wd] at (5,10) {};
\node[bd] at (6,13) {};
\node[bd] at (5,11) {};
\node[bd] at (4,9) {};
\node[bd] at (3,7) {};
\node[wd] at (2,6) {};
\node[wd] at (1,5) {};
\node[bd] at (0,4) {};
\node[bd] at (0,3) {};
\node[bd] at (0,2) {};
\node[bd] at (0,1) {};

\node[] at (2.5,10) {$N+1$};
\node[] at (6,7) {$N+2$};
\node[] at (-2,2) {$N+2$};
\end{tikzpicture}
\ee
We apply the operation $\mathfrak{M}^-_2$ to arrive at
\be
\mathfrak{M}^-_2 \mathfrak{M}^+_1 \mathfrak{M}_3^+ \cdots \mathfrak{M}_{N+3}^+ \mathfrak{M}^-_{N+4}\bm{u}=\left[(2,0),(N+2,2N+4),(0,N+3),\underbrace{(-1,-2)}_{N+1},(-3,-3),\underbrace{(0,-1)}_{N+2}\right]\,,
\ee
\be
\begin{tikzpicture}[x=.4cm,y=.4cm]
\draw[step=.4cm,gray,very thin] (0,0) grid (6,13);

\draw[ligne,black] (0,0)--(2,0)--(6,8)--(6,13)--(5,11)--(4,9)--(3,7)--(0,4)--(0,0);
\draw[black] (0,0)--(1,1)--(1,3)--(2,4)--(2,1)--(1,1);
\draw[black] (1,3)--(0,4);
\draw[black] (2,4)--(3,7);
\draw[black] (2,0)--(2,1)--(6,13);

\node[bd] at (0,0) {};
\node[wd] at (1,0) {};
\node[bd] at (2,0) {};
\node[wd] at (3,2) {};
\node[wd] at (4,4) {};
\node[wd] at (5,6) {};
\node[bd] at (6,8) {};
\node[wd] at (6,9) {};
\node[wd] at (6,10) {};
\node[wd] at (6,11) {};
\node[wd] at (6,12) {};
\node[bd] at (6,13) {};
\node[bd] at (5,11) {};
\node[bd] at (4,9) {};
\node[bd] at (3,7) {};
\node[wd] at (2,6) {};
\node[wd] at (1,5) {};
\node[bd] at (0,4) {};
\node[bd] at (0,3) {};
\node[bd] at (0,2) {};
\node[bd] at (0,1) {};

\node[] at (6.5,4) {$N+2$};
\node[] at (8,10.5) {$N+3$};
\node[] at (2.5,10) {$N+1$};
\node[] at (-2,2) {$N+2$};
\end{tikzpicture}
\ee
Finally, we move the $(-3,-3)$ edge through the $(-1,-2)$ edges, and obtain the vectors 
\be
\mathfrak{T} \bm{u}=\left[(2,0),(N+2,2N+4),(0,N+3),(-N-4,-N-4),\underbrace{(0,-1)}_{2N+3}\right]\,,
\ee
where we have written the entire sequence of transformations in a compact form as
\be \label{MonoSQCD}
\mathfrak{T}= \mathfrak{M}_4^+ \cdots \mathfrak{M}_{N+4}^+ \mathfrak{M}^-_2 \mathfrak{M}^+_1 \mathfrak{M}_3^+ \cdots \mathfrak{M}_{N+3}^+ \mathfrak{M}^-_{N+4}\,.
\ee
The convex polygon is
\be \label{GTPSQCD}
\begin{tikzpicture}[x=.4cm,y=.4cm]
\draw[step=.4cm,gray,very thin] (0,0) grid (6,13);

\draw[ligne,black] (0,0)--(2,0)--(6,8)--(6,13)--(0,7)--(0,0);

\node[bd] at (0,0) {};
\node[wd] at (1,0) {};
\node[bd] at (2,0) {};
\node[wd] at (3,2) {};
\node[wd] at (4,4) {};
\node[wd] at (5,6) {};
\node[bd] at (6,8) {};
\node[wd] at (6,9) {};
\node[wd] at (6,10) {};
\node[wd] at (6,11) {};
\node[wd] at (6,12) {};
\node[bd] at (6,13) {};
\node[wd] at (5,12) {};
\node[wd] at (4,11) {};
\node[wd] at (3,10) {};
\node[wd] at (2,9) {};
\node[wd] at (1,8) {};
\node[bd] at (0,7) {};
\node[bd] at (0,6) {};
\node[bd] at (0,5) {};
\node[bd] at (0,4) {};
\node[bd] at (0,3) {};
\node[bd] at (0,2) {};
\node[bd] at (0,1) {};

\node[] at (6.5,4) {$N+2$};
\node[] at (8,10.5) {$N+3$};
\node[] at (1,10) {$N+4$};
\node[] at (-2.5,3.5) {$2N+3$};
\end{tikzpicture}
\ee
We have removed the interior edges, since we are interested in the SCFT corresponding to this polygon. It is easy to check that $\mathfrak{C}=\emptyset$ for $\mathfrak{T}\bm{u}$. 

\paragraph{Magnetic Quivers and Decoupling}

In section \ref{sec:BasicsWebs} it was discussed how to compute the magnetic quiver from such a GTP. In this case there is a single coloring of \eqref{GTPSQCD}, i.e. a single magnetic quiver. As is well-known, this is given by
\be
\label{MQ1stDescDkDk}
\begin{tikzpicture}[x=.8cm,y=.8cm]
\node (g1) at (0,0) [gauge,label=below:{1}] {};
\node (g2) at (1,0) {$\cdots$};
\node (g3) at (2,0) [gauge,label=below:{$2N$+2}] {};
\node (g4) at (3.5,0) [gauge,label=below:{$N$+2}] {};
\node (g5) at (4.5,0) [gaugeb,label=below:{2}] {};
\node (g6) at (2,1) [gauge,label=right:{$N$+1}] {};
\draw (g1)--(g2)--(g3)--(g4)--(g5);
\draw (g3)--(g6);
\end{tikzpicture}
\ee
from which one can read off the $\mathfrak{so}(4N+8)$ flavor symmetry.

%

%
%

\subsection[\texorpdfstring{$SU(N)_\frac{N+1}{2} + 1 AS + 7 F$}{SU(N)(N+1)/2 + 1 AS + 7 F}]{\boldmath{$SU(N)_\frac{N+1}{2} + 1 AS + 7 F$}}
\label{sec:HigherRkEString}

\paragraph{GTP for Tree Tops.}

Next, let us consider the higher rank E-string. Its descendants include the rank $r=(N-1)$ $E_{N_F+1}$ theories, which have two weakly coupled gauge theory descriptions \footnote{Note, that there are also descriptions in terms of quiver gauge theories, which we will not discuss here.}
\be
SU(N)_{\frac{N+8-N_F}{2}} + 1 \bm{AS} + N_F \bm{F} \quad \leftrightarrow \quad Sp(N-1) + 1 \bm{AS} + N_F \bm{F}\,, \qquad N_F=0,\dots,7\,.
\ee
The sub-marginal theory  has seven flavors and is known as the higher rank $E_8$ theory.

\paragraph{$Sp(N-1)$ Realization.}
Let us first consider the $Sp(N-1)$ gauge theory realization, which was discussed in \cite{Bergman:2015dpa,Zafrir:2015rga}. It turns out that the magnetic quiver is far easier to compute in this description. 
The polygon for the rank $r$ $E_8$-theory before edge-moves are given by (drawn for $r=3$)
\be
\begin{tikzpicture}[x=.4cm,y=.4cm]
\draw[step=.4cm,gray,very thin] (0,0) grid (6,12);

\draw[ligne,black] (0,0)--(3,3)--(6,3)--(6,12)--(3,9)--(0,12)--(0,0);
\draw[black] (3,3)--(3,9);

\node[bd] at (0,0) {};
\node[wd] at (1,1) {};
\node[wd] at (2,2) {};
\node[bd] at (3,3) {};
\node[wd] at (4,3) {};
\node[wd] at (5,3) {};
\node[bd] at (6,3) {};
\node[wd] at (6,4) {};
\node[wd] at (6,5) {};
\node[bd] at (6,6) {};
\node[wd] at (6,7) {};
\node[wd] at (6,8) {};
\node[bd] at (6,9) {};
\node[wd] at (6,10) {};
\node[wd] at (6,11) {};
\node[bd] at (6,12) {};
\node[wd] at (5,11) {};
\node[wd] at (4,10) {};
\node[bd] at (3,9) {};
\node[wd] at (2,10) {};
\node[wd] at (1,11) {};
\node[bd] at (0,12) {};
\node[wd] at (0,11) {};
\node[wd] at (0,10) {};
\node[bd] at (0,9) {};
\node[wd] at (0,8) {};
\node[wd] at (0,7) {};
\node[bd] at (0,6) {};
\node[wd] at (0,5) {};
\node[wd] at (0,4) {};
\node[bd] at (0,3) {};
\node[wd] at (0,2) {};
\node[bd] at (0,1) {};

\node at (2.5,0.5) {\small $r$};
\node at (4.5,2) {\small $r$};
\node at (7,4.5) {\small $r$};
\node at (7,7.5) {\small $r$};
\node at (7,10.5) {\small $r$};
\node at (3.5,11.5) {\small $r$};
\node at (2.5,11.5) {\small $r$};
\node at (-1,10.5) {\small $r$};
\node at (-1,7.5) {\small $r$};
\node at (-1,4.5) {\small $r$};
\node at (-1.5,2) {\small $r-1$};

\node at (-1.2,0) {$\bm{v}_0$};

\end{tikzpicture}
\ee
We need to apply the same set of edge-moves as in \eqref{MonoSQCD}, namely
\be 
\mathfrak{T}= \mathfrak{M}_4^+ \mathfrak{M}_5^+ \mathfrak{M}_6^+ \mathfrak{M}^-_2 \mathfrak{M}^+_1 \mathfrak{M}_3^+ \mathfrak{M}_4^+ \mathfrak{M}_5^+ \mathfrak{M}^-_6\,,
\ee
with the origin $\bm{v}_0$ in the lower left corner. The vectors of the initial and convex polygons are given by
\be
\ba 
\bm{u}&=\left[(r,r),(r,0),\underbrace{(0,r)}_3,(-r,-r),(-r,r),\underbrace{(0,-r)}_3,(0,-r+1),(0,-1) \right]\,,\\
\mathfrak{T} \bm{u}&=\left[(2r,0),(4r,8r),(0,5r),(-6r,-6r),\underbrace{(0,-r)}_6,(0,-r+1),(0,-1)\right].
\ea
\ee
After a simple $SL(2,\Z)$ transformation the final GTP is given by (drawn for $r=2$)
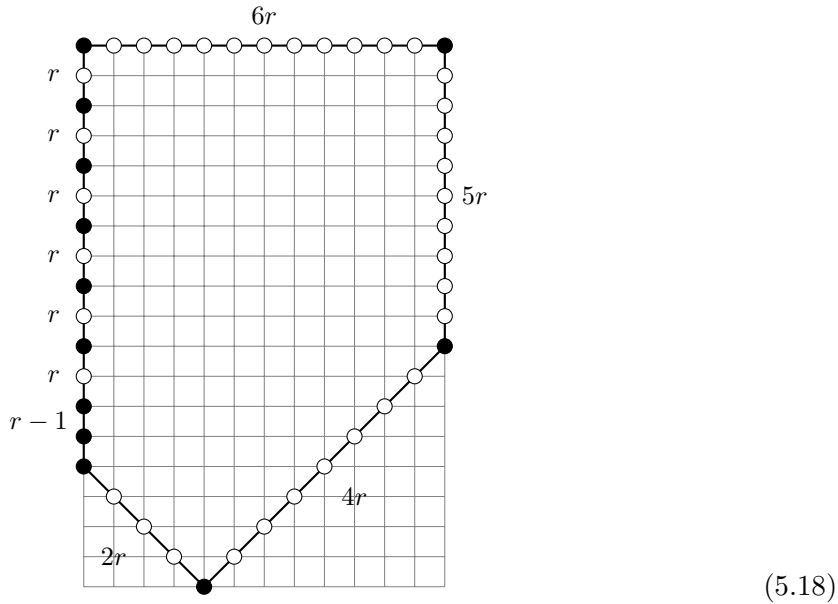
\be
\begin{tikzpicture}[x=.4cm,y=.4cm]

\draw[step=.4cm,gray,very thin] (0,-4) grid (12,14);

\draw[ligne,black] (0,0)--(4,-4)--(12,4)--(12,14)--(0,14)--(0,0);

\node[bd] at (0,0) {};
 \foreach \x in {1,2,3}
    \node[wd] at (\x,-\x) {};
    
\node[bd] at (4,-4) {};
 \foreach \x in {1,2,...,7}
    \node[wd] at (4+\x,-4+\x) {};
    
\node[bd] at (12,4) {};
 \foreach \x in {1,2,...,9}
    \node[wd] at (12,4+\x) {};
    
\node[bd] at (12,14) {};
 \foreach \x in {1,2,...,11}
    \node[wd] at (12-\x,14) {};

 \foreach \x in {1,3,...,11}
    \node[wd] at (0,14-\x) {};
 \foreach \x in {0,2,...,12}
    \node[bd] at (0,14-\x) {};
    
    \node[bd] at (0,1) {};

\node at (1,-3) {\small $2r$};
\node at (9,-1) {\small $4r$};
\node at (13,9) {\small $5r$};
\node at (6,15) {\small $6r$};

 \foreach \x in {1,3,...,11}
    \node at (-1,14-\x) {\small $r$};
    
\node at (-1.5,1.5) {\small $r-1$};
\end{tikzpicture}
\ee
From this we deduce that the magnetic quiver for the rank $(N-1)$ $E_8$ theory is given by the affine Dynkin diagram of $\mathfrak{e}_{8}$, with each label multiplied by $(N-1)$, and an additional $\mathfrak{u}(1)$ node attached to the affine node, i.e.,
\be 
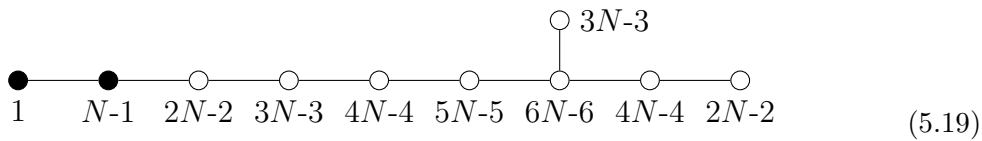
\label{MQE8Theory}
\begin{tikzpicture}[x=.8cm,y=.8cm]
\node (h1) at (0,0) [gaugeb,label=below:\large{1}] {};
\node (h2) at (1.5,0) [gaugeb,label=below:\large{$N$-1}] {};
\node (h3) at (3,0) [gauge,label=below:\large{$2N$-2}] {};
\node (h4) at (4.5,0) [gauge,label=below:\large{$3N$-3}] {};
\node (h5) at (6,0) [gauge,label=below:\large{$4N$-4}] {};
\node (h6) at (7.5,0) [gauge,label=below:\large{$5N$-5}] {};
\node (h7) at (9,0) [gauge,label=below:\large{$6N$-6}] {};
\node (h8) at (10.5,0) [gauge,label=below:\large{$4N$-4}] {};
\node (h9) at (12,0) [gauge,label=below:\large{$2N$-2}] {};
\node (h10) at (9,1) [gauge,label=right:\large{$3N$-3}] {};
\draw (h1)--(h2)--(h3)--(h4)--(h5)--(h6)--(h7)--(h8)--(h9);
\draw (h7)--(h10);
\end{tikzpicture}
\ee
We will see later that the logic of this higher rank generalization holds for all higher rank $E_{N_F+1}$ theories. From the balanced nodes one reads off an $\e_{N_F+1} \oplus \uu(1)$ flavor symmetry. However, for $N>1$, the flavor symmetry is conjectured to enhance to $\e_{N_F+1} \oplus \su(2)$. The appearance of this additional $\su(2)$ symmetry is somewhat mysterious both from the field theoretic and the geometric point of view. We will comment on this enhancement in section \ref{sec:NSLQuivers}.

\paragraph{$SU(N)$ Realization.}
Let us now reproduce this in the $SU(N)$ description, which illustrates the issues we will encounter in the remainder of this section.
First, let us check how these theories are connected to the $(D_{N+2},D_{N+2})$ conformal matter discussed in section \ref{sec:SQCD}. Integrating out the anti-symmetric shifts the Chern-Simons level as
\be
k\to k+\frac{N-4}{2}\,,
\ee
i.e. the $E_{N_F+1}$ theory descends to an $SU(N)_{N+2-\frac{N_F}{2}}+N_F\bm{F}$.

The brane-web was discussed in \cite{Bergman:2015dpa,Jefferson:2018irk}. In general these are constructed by the inclusion of $O7^-$-planes and differ qualitatively depending on whether $N$ is even or odd. The corresponding pGTPs for the $SU(N)_{\frac{N+1}{2}} + 1\bm{AS} + 7 \bm{F}$ are given by (drawn for $n=3$)
\be \label{pGTPEString}
\begin{tikzpicture}[x=.3cm,y=.3cm]

\draw[step=.3cm,gray,very thin] (0,-3) grid (4,6);

\draw[ligne,black] (0,0)--(3,3)--(4,6)--(4,-1)--(3,-3)--(0,0);
\draw[black] (3,-3)--(3,3);

\node[bd] at (0,0) {};
\node[wd] at (1,-1) {};
\node[wd] at (2,-2) {};
\node[bd] at (3,-3) {};
\node[bd] at (4,-1) {};
\node[bd] at (4,0) {};
\node[bd] at (4,1) {};
\node[bd] at (4,2) {};
\node[bd] at (4,3) {};
\node[bd] at (4,4) {};
\node[bd] at (4,5) {};
\node[bd] at (4,6) {};
\node[bd] at (3,3) {};
\node[bd] at (2,2) {};
\node[wd] at (1,1) {};

\node at (0.5,-2.5) {\small $n$};
\node at (0,2) {\small $n$-1};

\node at (-1.2,0) {$\bm{v}_0$};

\draw[decoration={calligraphic brace,amplitude=5pt,mirror}, decorate, line width=0.7pt] (4.5,-3)--(4.5,-1);

\node at (6.5,-2) {\small $2n$-4};

\end{tikzpicture}
\qquad
\qquad
\begin{tikzpicture}[x=.3cm,y=.3cm]

\draw[step=.3cm,gray,very thin] (0,-4) grid (5,5);

\draw[ligne,black] (0,0)--(3,3)--(4,3)--(5,5)--(5,-2)--(4,-4)--(0,0);
\draw[black] (4,-4)--(4,3);

\node[bd] at (0,0) {};
\node[wd] at (1,-1) {};
\node[wd] at (2,-2) {};
\node[wd] at (3,-3) {};
\node[bd] at (4,-4) {};
\node[bd] at (5,-2) {};
\node[bd] at (5,-1) {};
\node[bd] at (5,0) {};
\node[bd] at (5,1) {};
\node[bd] at (5,2) {};
\node[bd] at (5,3) {};
\node[bd] at (5,4) {};
\node[bd] at (5,5) {};
\node[bd] at (4,3) {};
\node[bd] at (3,3) {};
\node[wd] at (2,2) {};
\node[wd] at (1,1) {};

\node at (1,-3) {\small $n$+1};
\node at (0.5,2.5) {\small $n$};

\node at (-1.2,0) {$\bm{v}_0$};

\draw[decoration={calligraphic brace,amplitude=5pt,mirror}, decorate, line width=0.7pt] (5.5,-4)--(5.5,-2);

\node at (7.5,-3) {\small $2n$-4};

\end{tikzpicture}
\ee
for $N=2n$ even on the left and $N=2n+1$ odd on the right. These correspond to
\be
\ba
N&=2n: \quad &&\bm{u}=\left[(n,-n),(1,2n-4),\underbrace{(0,1)}_{7},(-1,-3),(-1,-1),(-n+1,-n+1) \right]\\
N&=2n+1: \quad &&\bm{u}=\left[(n+1,-n-1),(1,2n-4),\underbrace{(0,1)}_{7},(-1,-2),(-1,0),(-n,-n) \right]
\ea
\ee
Unlike in the other cases there is no closed form for the monodromies necessary to turn this into a GTP. However, one can show that such a monodromy exists in all cases. In appendix \ref{app:HigherRankEString} we discuss this in more detail and especially consider the case $N_F=6$. One can indeed check that, after applying the local Seiberg dualities \eqref{SimoneDuality}, the resulting magnetic quivers agree with \eqref{MQE8Theory} for $N_F=7$, and more generally are given by $N-1$ times the affine Dynkin diagram of $\e_{N_F+1}$ with a $\uu(1)$ attached to the affine node.

\subsection[\texorpdfstring{$SU(N)_\half + 1 AS + (N+5)F$}{SU(N)1/2 + 1 AS + (N+5) F}]{\boldmath{$SU(N)_\half + 1 AS + (N+5)F$}}
\label{sec:1AS(N+5)F}

\paragraph{GTP for Tree Tops.}

The next submarginal single gauge node SCFT has an $SU(N)_\half + 1 \bm{AS} + (N+5) \bm{F}$ description.
Also in this case the polygon depends on whether $N$ is even or odd, so we will discuss them in turn. The polygons before edge-moves are (drawn for $n=3$)
\be
\begin{tikzpicture}[x=.3cm,y=.3cm]

\draw[step=.3cm,gray,very thin] (0,-6) grid (4,5);

\draw[ligne,black] (0,0)--(3,3)--(4,5)--(4,-6)--(3,-3)--(0,0);
\draw[black] (3,3)--(3,-3);

\node[bd] at (4,-6) {};
\node[bd] at (3,-3) {};
\node[wd] at (2,-2) {};
\node[wd] at (1,-1) {};
\node[bd] at (0,0) {};
\node[wd] at (1,1) {};
\node[bd] at (2,2) {};
\node[bd] at (3,3) {};
\node[bd] at (4,5) {};
\node[bd] at (4,4) {};
\node[bd] at (4,3) {};
\node[bd] at (4,2) {};
\node[bd] at (4,1) {};
\node[bd] at (4,0) {};
\node[bd] at (4,-1) {};
\node[bd] at (4,-2) {};
\node[bd] at (4,-3) {};
\node[bd] at (4,-4) {};
\node[bd] at (4,-5) {};

\node at (0.5,-2.5) {\small $n$};
\node at (0,2) {\small $n$-1};
\node at (6,-0.5) {\small $2n$+5};

\node at (-1.2,0) {$\bm{v}_0$};

\end{tikzpicture}
\qquad
\qquad
\begin{tikzpicture}[x=.3cm,y=.3cm]

\draw[step=.3cm,gray,very thin] (0,-8) grid (5,4);

\draw[ligne,black] (0,0)--(3,3)--(4,3)--(5,4)--(5,-8)--(4,-4)--(0,0);
\draw[black] (4,3)--(4,-4);

\node[bd] at (5,-8) {};
\node[bd] at (4,-4) {};
\node[wd] at (3,-3) {};
\node[wd] at (2,-2) {};
\node[wd] at (1,-1) {};
\node[bd] at (0,0) {};
\node[wd] at (1,1) {};
\node[wd] at (2,2) {};
\node[bd] at (3,3) {};
\node[bd] at (4,3) {};
\node[bd] at (5,4) {};
\node[bd] at (5,3) {};
\node[bd] at (5,2) {};
\node[bd] at (5,1) {};
\node[bd] at (5,0) {};
\node[bd] at (5,-1) {};
\node[bd] at (5,-2) {};
\node[bd] at (5,-3) {};
\node[bd] at (5,-4) {};
\node[bd] at (5,-5) {};
\node[bd] at (5,-6) {};
\node[bd] at (5,-7) {};

\node at (1,-3) {\small $n$+1};
\node at (0.5,2.5) {\small $n$};
\node at (7,-2) {\small $2n$+6};

\node at (-1.2,0) {$\bm{v}_0$};

\end{tikzpicture}
\ee
where $N=2n$ and $N=2n+1$, respectively. The vectors of the pGTPs are
\be
\ba
N&=2n: \qquad &&\bm{u}=\left[(n,-n),(1,-3),\underbrace{(0,1)}_{2n+5},(-1,-2),(-1,-1),(-n+1,-n+1) \right]\\
N&=2n+1: \qquad &&\bm{u}=\left[(n+1,-n-1),(1,-4),\underbrace{(0,1)}_{2n+6},(-1,-1),(-1,0),(-n,-n) \right]
\ea
\ee
The sequences of edge-moves that make the polygons convex are given by
\be
\ba
N&=2n: \qquad &&\mathfrak{T}=\mathfrak{M}_{-1}^- \mathfrak{M}_{0}^- \mathfrak{M}_{1}^- \mathfrak{M}_{-1}^+ \mathfrak{M}_{-2}^-\\
N&=2n+1: \qquad &&\mathfrak{T}=\mathfrak{M}_{-2}^- \mathfrak{M}_{-3}^- \mathfrak{M}_{0}^- \mathfrak{M}_{1}^- 
\ea
\ee
The resulting GTPs are (drawn for $n=3$)
\be
\begin{tikzpicture}[x=.3cm,y=.3cm]

\draw[step=.3cm,gray,very thin] (0,-5) grid (5,8);

\draw[ligne,black] (0,0)--(0,3)--(5,8)--(5,-5)--(0,0) ;

\node[bd] at (0,0) {};
\node[wd] at (0,1) {};
\node[wd] at (0,2) {};
\node[bd] at (0,3) {};
\node[wd] at (1,4) {};
\node[wd] at (2,5) {};
\node[wd] at (3,6) {};
\node[wd] at (4,7) {};
\node[bd] at (5,8) {};
\node[bd] at (5,7) {};
\node[bd] at (5,6) {};
\node[bd] at (5,5) {};
\node[bd] at (5,4) {};
\node[bd] at (5,3) {};
\node[bd] at (5,2) {};
\node[bd] at (5,1) {};
\node[bd] at (5,0) {};
\node[bd] at (5,-1) {};
\node[bd] at (5,-2) {};
\node[bd] at (5,-3) {};
\node[bd] at (5,-4) {};
\node[bd] at (5,-5) {};
\node[wd] at (4,-4) {};
\node[wd] at (3,-3) {};
\node[wd] at (2,-2) {};
\node[wd] at (1,-1) {};

\node at (1.5,6.5) {\small $n$+2};
\node at (1.5,-3.5) {\small $n$+2};
\node at (7,1.5) {\small $2n$+7};

\end{tikzpicture}
\qquad
\qquad
\begin{tikzpicture}[x=.3cm,y=.3cm]

\draw[step=.3cm,gray,very thin] (0,-7) grid (7,7);

\draw[ligne,black] (0,0)--(7,7)--(7,-7)--(0,0) ;

\node[bd] at (0,0) {};
\node[wd] at (1,1) {};
\node[wd] at (2,2) {};
\node[wd] at (3,3) {};
\node[bd] at (4,4) {};
\node[wd] at (5,5) {};
\node[wd] at (6,6) {};
\node[bd] at (7,7) {};
\node[bd] at (7,6) {};
\node[bd] at (7,5) {};
\node[bd] at (7,4) {};
\node[bd] at (7,3) {};
\node[bd] at (7,2) {};
\node[bd] at (7,1) {};
\node[bd] at (7,0) {};
\node[bd] at (7,-1) {};
\node[bd] at (7,-2) {};
\node[bd] at (7,-3) {};
\node[bd] at (7,-4) {};
\node[bd] at (7,-5) {};
\node[bd] at (7,-6) {};
\node[bd] at (7,-7) {};
\node[wd] at (6,-6) {};
\node[wd] at (5,-5) {};
\node[wd] at (4,-4) {};
\node[wd] at (3,-3) {};
\node[wd] at (2,-2) {};
\node[wd] at (1,-1) {};

\node at (1,3) {\small $n$+1};
\node at (2.5,-4.5) {\small $n$+4};
\node at (9,0) {\small $2n$+8};

\end{tikzpicture}
\ee
given by the vectors
\be
\ba
N&=2n: \qquad &&\mathfrak{T}\bm{u}=\left[(n+2,-n-2),\underbrace{(0,1)}_{2n+7},(-n-2,-n-2),(0,-3) \right]\\
N&=2n+1: \qquad &&\mathfrak{T}\bm{u}=\left[(n+4,-n-4),\underbrace{(0,1)}_{2n+8},(-3,-3),(-n-1,-n-1) \right]
\ea
\ee

\paragraph{Magnetic Quivers and Decoupling.}

From this we can compute the magnetic quivers for general $N$
\be
\begin{tikzpicture}[x=.8cm,y=.8cm]
\node (g1) at (0,0) [gauge,label=below:{1}] {};
\node (g2) at (1,0) {$\cdots$};
\node (g3) at (2,0) [gauge,label=below:{$2n$+5}] {};
\node (g4) at (3.5,0) [gauge,label=below:{$n$+4}] {};
\node (g5) at (4.5,0) [gaugeb,label=below:{3}] {};
\node (g6) at (2,1) [gaugeb,label=right:{$n$+2}] {};
\draw (g1)--(g2)--(g3)--(g4)--(g5);
\draw (g3)--(g6);

\node (h1) at (7,0) [gauge,label=below:{1}] {};
\node (h2) at (8,0) {$\cdots$};
\node (h3) at (9,0) [gauge,label=below:{$2n$+6}] {};
\node (h4) at (10.5,0) [gaugeb,label=below:{$n$+4}] {};
\node (h5) at (11.5,0) [gaugeb,label=below:{3}] {};
\node (h6) at (9,1) [gauge,label=right:{$n$+3}] {};
\draw (h1)--(h2)--(h3)--(h4)--(h5);
\draw (h3)--(h6);
\end{tikzpicture}
\ee
for $N=2n$ and $N=2n+1$ respectively.
From the balanced nodes we see that generally the theory has flavor group $\su(N+7)\oplus \uu(1)$, which is enhanced for small $N$. Concretely, the symmetry enhances to $\su(12) \oplus \su(2)$ for $N=5$ and $\su(12)$ for $N=4$. For $N=3$ the theory is $SU(3)_\half + 9 \bm{F}$ with $\so(20)$ flavor symmetry, whereas the $N=2$ case is the rank one $E_8$ theory.\\

{\example $SU(4)_k + 1 \mathbf{AS} + N_F \mathbf{F}$

Let us illustrate the decoupling of matter in the polygons associated to $SU(4)_k + 1 \mathbf{AS} + N_F \mathbf{F}$ theories. We consider the theories with $0 \leq N_F \leq 9$ with $k=0$ or $k=\frac{1}{2}$ depending on the parity of $N_F$. 
We first identify the ruling associated to the gauge theory phase. The decoupling of fundamental matter is then implemented by flopping a curve connecting an internal vertex on the ruling to a corner vertex. The relevant fundamental matter curves will be indicated in red in the following. For $N_F=9$, the pGTP before and after the flop of the matter curve is 
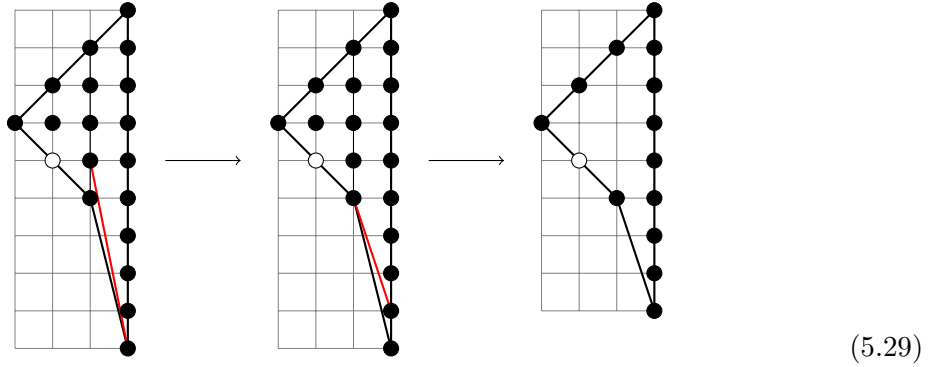
\be 
	\begin{tikzpicture}[x=.5cm,y=.5cm] 
	\draw[step=.5cm,gray,very thin] (-2,-4) grid (1,5);		\draw[ligne] (1,5)--(-2,2)--(0,0)--(1,-4)--(1,5); 
	\draw[ligne,red] (0,1)--(1,-4); 
	\draw[ligne,thin] (0,0)--(0,4);
		\foreach \x in {0,1,...,9}
			\node[bd] at (1,-4+\x) {}; 	
		\foreach \x in {0,1,2}
			\node[bd] at (0,1+\x) {}; 
	\node[bd] at (-1,2) {};  
		\foreach \x in {0,1,2}
			\node[bd] at (-\x,4-\x) {};
	\node[bd] at (0,0) {}; 
	\node[wd] at (-1,1) {}; 

	\draw[->] (2,1)--(4,1);

	\draw[step=.5cm,gray,very thin] (5,-4) grid (8,5);
	\draw[ligne] (8,5)--(5,2)--(7,0)--(8,-4)--(8,5); 
	\draw[ligne,red] (7,0)--(8,-3); 		\draw[ligne,thin] (7,0)--(7,4);
		\foreach \x in {0,1,...,9}
			\node[bd] at (8,-4+\x) {}; 	
		\foreach \x in {0,1,2}
			\node[bd] at (7,1+\x) {}; 
	\node[bd] at (6,2) {};  
		\foreach \x in {0,1,2}
			\node[bd] at (7-\x,4-\x) {};
	\node[bd] at (7,0) {}; 
	\node[wd] at (6,1) {}; 

	\draw[->] (9,1)--(11,1);
	
	\draw[step=.5cm,gray,very thin] (12,-3) grid (15,5);
	\draw[ligne] (15,5)--(12,2)--(14,0)--(15,-3)--(15,5); 
		\foreach \x in {0,1,...,8}
			\node[bd] at (15,-3+\x) {}; 	
		\foreach \x in {0,1,2}
			\node[bd] at (14-\x,4-\x) {};
	\node[bd] at (14,0) {}; 
	\node[wd] at (13,1) {}; 
\end{tikzpicture}
\ee
After the flop, the corner vertex is no longer connected to the internal gauge vertices, so this theory is represented by the right-most polygon, which is exactly the pGTP of $SU(4)_{0} + 1 \mathbf{AS} + 8 \mathbf{F}$. 
This pGTP also has a fundamental matter curve connecting its bottom right vertex to an internal gauge node, which we can again flop to obtain the pGTP of $SU(4)_{1/2} + 1 \mathbf{AS} + 7 \mathbf{F}$:
\be 
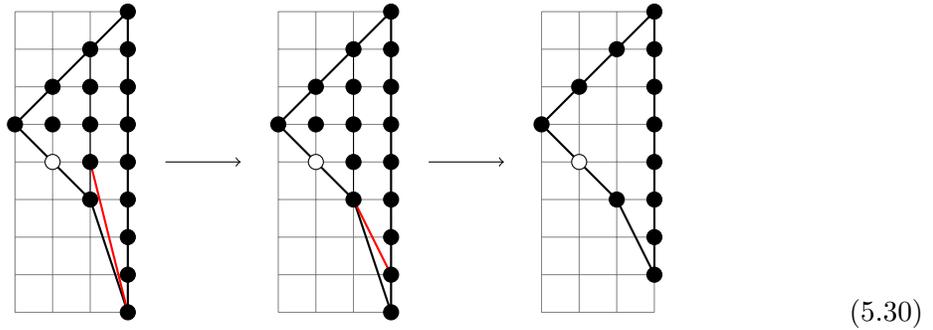
\begin{tikzpicture}[x=.5cm,y=.5cm] 
	\draw[step=.5cm,gray,very thin] (-2,-3) grid (1,5);
	\draw[ligne] (1,5)--(-2,2); 
	\draw[ligne] (-2,2)--(0,0); 
	\draw[ligne] (0,0)--(1,-3); 
	\draw[ligne] (1,-3)--(1,5); 	
	\draw[ligne,red] (0,1)--(1,-3);	
	\draw[ligne,thin] (0,0)--(0,4);  
	\foreach \x in {0,1,...,8}
	\node[bd] at (1,-3+\x) {}; 
	\foreach \x in {0,1,2}
	\node[bd] at (0,1+\x) {}; 
	\node[bd] at (-1,2) {};  	
	\foreach \x in {0,1,2}
	\node[bd] at (-\x,4-\x) {};
	\node[bd] at (0,0) {}; 
	\node[wd] at (-1,1) {};

	\draw[->] (2,1)--(4,1);
	
	\draw[step=.5cm,gray,very thin] (5,-3) grid (8,5);
	\draw[ligne] (8,5)--(5,2); 
	\draw[ligne] (5,2)--(7,0); 
	\draw[ligne] (7,0)--(8,-3); 
	\draw[ligne] (8,-3)--(8,5); 	
	\draw[ligne,red] (7,0)--(8,-2);	
	\draw[ligne,thin] (7,0)--(7,4);  
	\foreach \x in {0,1,...,8}
	\node[bd] at (8,-3+\x) {}; 
	\foreach \x in {0,1,2}
	\node[bd] at (7,1+\x) {}; 
	\node[bd] at (6,2) {};  	
	\foreach \x in {0,1,2}
	\node[bd] at (7-\x,4-\x) {};
	\node[bd] at (7,0) {}; 
	\node[wd] at (6,1) {}; 
	
	\draw[->] (9,1)--(11,1);
	
	\draw[step=.5cm,gray,very thin] (12,-3) grid (15,5);
	\draw[ligne] (15,5)--(12,2); 
	\draw[ligne] (12,2)--(14,0); 
	\draw[ligne] (14,0)--(15,-2); 
	\draw[ligne] (15,-2)--(15,5); 		
	\foreach \x in {0,1,...,7}
	\node[bd] at (15,-2+\x) {}; 
	\foreach \x in {0,1,2}
	\node[bd] at (14-\x,4-\x) {};
	\node[bd] at (14,0) {}; 
	\node[wd] at (13,1) {}; 
\end{tikzpicture}
\ee
Strictly speaking, the two previous flops correspond to decoupling fundamental matter multiplets by giving them a mass with the same sign. For this reason, the change in CS level is actually $k=\half \rightarrow 0 \rightarrow -\half$. However, since the Higgs branch only depends on the absolute value of the CS level, the previous flops are indeed equivalent to the sequence $k=\half \rightarrow 0 \rightarrow \half$. To stay in the $k=0, \half$ regime, we now flop a matter curve connecting to the top right corner vertex, which corresponds to giving a fundamental a mass with the opposite sign of the previous two decoupled matter multiplets. Flopping this curve, we thus arrive at the polygon for $SU(4)_{0} + 1 \mathbf{AS} + 6 \mathbf{F}$:
\be 
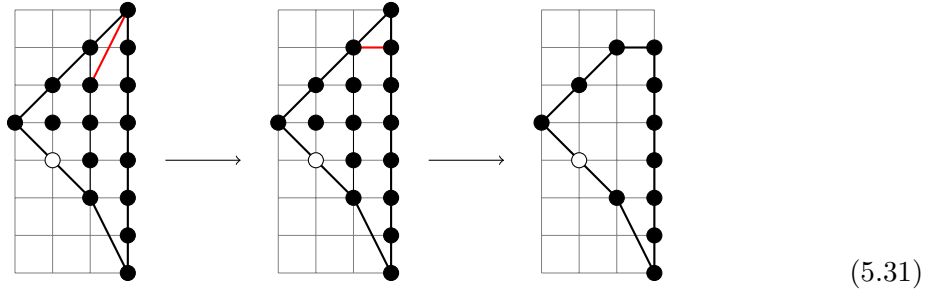
\begin{tikzpicture}[x=.5cm,y=.5cm] 
	\draw[step=.5cm,gray,very thin] (-2,-2) grid (1,5);
	\draw[ligne] (1,5)--(-2,2); 
	\draw[ligne] (-2,2)--(0,0); 
	\draw[ligne] (0,0)--(1,-2); 
	\draw[ligne] (1,-2)--(1,5); 	
	\draw[ligne,red] (0,3)--(1,5);	
	\draw[ligne,thin] (0,0)--(0,4);  
	\foreach \x in {0,1,...,7}
	\node[bd] at (1,-2+\x) {}; 
	\foreach \x in {0,1,2}
	\node[bd] at (0,1+\x) {}; 
	\node[bd] at (-1,2) {};  	
	\foreach \x in {0,1,2}
	\node[bd] at (-\x,4-\x) {};
	\node[bd] at (0,0) {}; 
	\node[wd] at (-1,1) {};

	\draw[->] (2,1)--(4,1);
	
	\draw[step=.5cm,gray,very thin] (5,-2) grid (8,5);
	\draw[ligne] (8,5)--(5,2); 
	\draw[ligne] (5,2)--(7,0); 
	\draw[ligne] (7,0)--(8,-2); 
	\draw[ligne] (8,-2)--(8,5); 	
	\draw[ligne,red] (7,4)--(8,4);	
	\draw[ligne,thin] (7,0)--(7,4);  
	\foreach \x in {0,1,...,7}
	\node[bd] at (8,-2+\x) {}; 
	\foreach \x in {0,1,2}
	\node[bd] at (7,1+\x) {}; 
	\node[bd] at (6,2) {};  	
	\foreach \x in {0,1,2}
	\node[bd] at (7-\x,4-\x) {};
	\node[bd] at (7,0) {}; 
	\node[wd] at (6,1) {}; 
	
	\draw[->] (9,1)--(11,1);
	
	\draw[step=.5cm,gray,very thin] (12,-2) grid (15,5);
	\draw[ligne] (14,4)--(12,2); 
	\draw[ligne] (12,2)--(14,0); 
	\draw[ligne] (14,0)--(15,-2); 
	\draw[ligne] (15,-2)--(15,4); 
	\draw[ligne] (15,4)--(14,4); 			
	\foreach \x in {0,1,...,6}
	\node[bd] at (15,-2+\x) {}; 
	\foreach \x in {0,1,2}
	\node[bd] at (14-\x,4-\x) {};
	\node[bd] at (14,0) {}; 
	\node[wd] at (13,1) {}; 
\end{tikzpicture}
\ee
We can continue in this way until all fundamental matter is decoupled.
For $N_F=5,\dots,1,0$, one finds
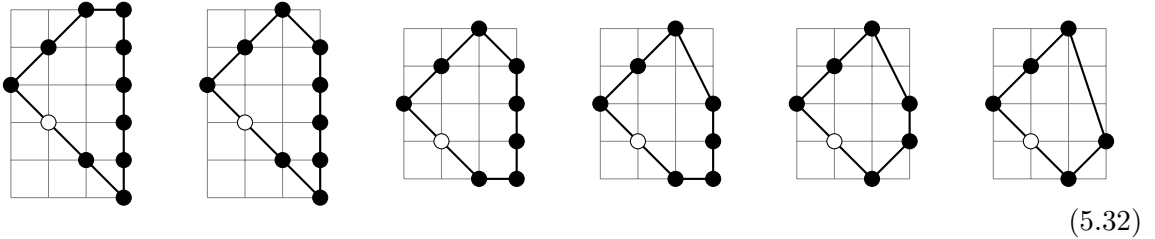
\begin{equation}
\raisebox{-.5\height}{ 
	\begin{tikzpicture}[x=.5cm,y=.5cm] 
		\draw[step=.5cm,gray,very thin] (-2,-1) grid (1,4);
		\draw[ligne] (1,4)--(0,4); 
		\draw[ligne] (0,4)--(-1,3); 
		\draw[ligne] (-1,3)--(-2,2); 
		\draw[ligne] (-2,2)--(0,0); 
		\draw[ligne] (0,0)--(1,-1); 
		\draw[ligne] (1,-1)--(1,0); 
		\draw[ligne] (1,0)--(1,1); 
		\draw[ligne] (1,1)--(1,2); 
		\draw[ligne] (1,2)--(1,3); 
		\draw[ligne] (1,3)--(1,4); 
		\node[bd] at (0,4) {}; 
		\node[bd] at (-1,3) {}; 
		\node[bd] at (-2,2) {}; 
		\node[bd] at (0,0) {}; 
		\node[bd] at (1,-1) {}; 
		\node[bd] at (1,0) {}; 
		\node[bd] at (1,1) {}; 
		\node[bd] at (1,2) {}; 
		\node[bd] at (1,3) {}; 
		\node[bd] at (1,4) {}; 
		\node[wd] at (-1,1) {}; 
\end{tikzpicture}} \qquad	\raisebox{-.5\height}{ 
\begin{tikzpicture}[x=.5cm,y=.5cm] 
\draw[step=.5cm,gray,very thin] (-2,-1) grid (1,4);
\draw[ligne] (1,3)--(0,4); 
\draw[ligne] (0,4)--(-1,3); 
\draw[ligne] (-1,3)--(-2,2); 
\draw[ligne] (-2,2)--(0,0); 
\draw[ligne] (0,0)--(1,-1); 
\draw[ligne] (1,-1)--(1,0); 
\draw[ligne] (1,0)--(1,1); 
\draw[ligne] (1,1)--(1,2); 
\draw[ligne] (1,2)--(1,3); 
\node[bd] at (0,4) {}; 
\node[bd] at (-1,3) {}; 
\node[bd] at (-2,2) {}; 
\node[bd] at (0,0) {}; 
\node[bd] at (1,-1) {}; 
\node[bd] at (1,0) {}; 
\node[bd] at (1,1) {}; 
\node[bd] at (1,2) {}; 
\node[bd] at (1,3) {}; 
\node[wd] at (-1,1) {}; 
\end{tikzpicture}} \qquad \raisebox{-.5\height}{ 
\begin{tikzpicture}[x=.5cm,y=.5cm] 
\draw[step=.5cm,gray,very thin] (-2,0) grid (1,4);
\draw[ligne] (1,3)--(0,4); 
\draw[ligne] (0,4)--(-1,3); 
\draw[ligne] (-1,3)--(-2,2); 
\draw[ligne] (-2,2)--(0,0); 
\draw[ligne] (0,0)--(1,0); 
\draw[ligne] (1,0)--(1,1); 
\draw[ligne] (1,1)--(1,2); 
\draw[ligne] (1,2)--(1,3); 
\node[bd] at (0,4) {}; 
\node[bd] at (-1,3) {}; 
\node[bd] at (-2,2) {}; 
\node[bd] at (0,0) {}; 
\node[bd] at (1,0) {}; 
\node[bd] at (1,1) {}; 
\node[bd] at (1,2) {}; 
\node[bd] at (1,3) {}; 
\node[wd] at (-1,1) {}; 
\end{tikzpicture}}  \qquad \raisebox{-.5\height}{ 
\begin{tikzpicture}[x=.5cm,y=.5cm] 
\draw[step=.5cm,gray,very thin] (-2,0) grid (1,4);
\draw[ligne] (1,2)--(0,4); 
\draw[ligne] (0,4)--(-1,3); 
\draw[ligne] (-1,3)--(-2,2); 
\draw[ligne] (-2,2)--(0,0); 
\draw[ligne] (0,0)--(1,0); 
\draw[ligne] (1,0)--(1,1); 
\draw[ligne] (1,1)--(1,2); 
\node[bd] at (0,4) {}; 
\node[bd] at (-1,3) {}; 
\node[bd] at (-2,2) {}; 
\node[bd] at (0,0) {}; 
\node[bd] at (1,0) {}; 
\node[bd] at (1,1) {}; 
\node[bd] at (1,2) {}; 
\node[wd] at (-1,1) {}; 
\end{tikzpicture}} \qquad
\raisebox{-.5\height}{ 
\begin{tikzpicture}[x=.5cm,y=.5cm] 
\draw[step=.5cm,gray,very thin] (-2,0) grid (1,4);
\draw[ligne] (1,2)--(0,4); 
\draw[ligne] (0,4)--(-1,3); 
\draw[ligne] (-1,3)--(-2,2); 
\draw[ligne] (-2,2)--(0,0); 
\draw[ligne] (0,0)--(1,1); 
\draw[ligne] (1,1)--(1,2); 
\node[bd] at (0,4) {}; 
\node[bd] at (-1,3) {}; 
\node[bd] at (-2,2) {}; 
\node[bd] at (0,0) {}; 
\node[bd] at (1,1) {}; 
\node[bd] at (1,2) {}; 
\node[wd] at (-1,1) {}; 
\end{tikzpicture}} \qquad  \raisebox{-.5\height}{ 
\begin{tikzpicture}[x=.5cm,y=.5cm] 
	\draw[step=.5cm,gray,very thin] (-2,0) grid (1,4);
	\draw[ligne] (1,1)--(0,4); 
	\draw[ligne] (0,4)--(-2,2); 
	\draw[ligne] (-2,2)--(0,0); 
	\draw[ligne] (0,0)--(1,1); 
	\node[bd] at (0,4) {}; 
	\node[bd] at (-1,3) {}; 
	\node[bd] at (-2,2) {}; 
	\node[bd] at (0,0) {}; 
	\node[bd] at (1,1) {}; 
	\node[wd] at (-1,1) {}; 
\end{tikzpicture}} 
\end{equation}
Clearly the last polygon has no more fundamental matter curves that can be flopped.
These pGTPs are convex so the magnetic quiver for the corresponding SCFT can be obtained directly from these polygons, using the method summarized in section \ref{sec:HB&MQs}. On the contrary for $N_F=6,7,8,9$ the pGTPs are not convex, and edge-moves have to be performed. The resulting GTPs as well as their UV magnetic quivers are shown in table \ref{table:GTPExample}. Note that the edge-moves can be done in different ways, giving distinct GTPs. 
However, all resolutions lead to the same UV magnetic quivers. 

One can also at any point choose to decouple the anti-symmetric matter, thereby making contact with the $SU(4)$ SQDC theories. If we take the left-most vertex as the origin, the relevant matter curve connects the vertex at $(1,1)$ to the bottom vertex of the ruling at $(2,-2)$. Clearly this curve exists in all the above polygons. E.g. for $SU(4)_{1/2} + 1 \mathbf{AS} + 7 \mathbf{F}$ the relevant curve is
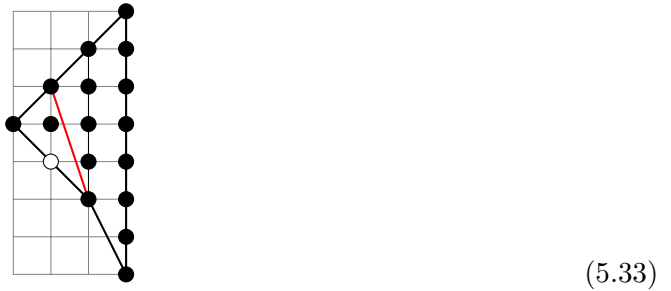
\be 
\begin{tikzpicture}[x=.5cm,y=.5cm] 
%
%
	
	\draw[step=.5cm,gray,very thin] (5,-2) grid (8,5);
	\draw[ligne] (8,5)--(5,2); 
	\draw[ligne] (5,2)--(7,0); 
	\draw[ligne] (7,0)--(8,-2); 
	\draw[ligne] (8,-2)--(8,5); 	
	\draw[ligne,red] (6,3)--(7,0);	
	\draw[ligne,thin] (7,0)--(7,4);  
	\foreach \x in {0,1,...,7}
	\node[bd] at (8,-2+\x) {}; 
	\foreach \x in {0,1,2}
	\node[bd] at (7,1+\x) {}; 
	\node[bd] at (6,2) {};  	
	\foreach \x in {0,1,2}
	\node[bd] at (7-\x,4-\x) {};
	\node[bd] at (7,0) {}; 
	\node[wd] at (6,1) {}; 
	
%
\end{tikzpicture}
\ee

\begin{table}[]
    \centering
    \begin{tabular}{c|c|c|c} \toprule 
      $N_F$ &  pGTP & GTP  & Magnetic quiver and  Flavor Symmetry  \\ \midrule 
      6 &\raisebox{-.5\height}{ 
\begin{tikzpicture}[x=.5cm,y=.5cm] 
	\draw[step=.5cm,gray,very thin] (-2,-2) grid (1,4);
\draw[ligne] (1,4)--(0,4); 
\draw[ligne] (0,4)--(-1,3); 
\draw[ligne] (-1,3)--(-2,2); 
\draw[ligne] (-2,2)--(0,0); 
\draw[ligne] (0,0)--(1,-2); 
\draw[ligne] (1,-2)--(1,-1); 
\draw[ligne] (1,-1)--(1,0); 
\draw[ligne] (1,0)--(1,1); 
\draw[ligne] (1,1)--(1,2); 
\draw[ligne] (1,2)--(1,3); 
\draw[ligne] (1,3)--(1,4); 
\node[bd] at (0,4) {}; 
\node[bd] at (-1,3) {}; 
\node[bd] at (-2,2) {}; 
\node[bd] at (0,0) {}; 
\node[bd] at (1,-2) {}; 
\node[bd] at (1,-1) {}; 
\node[bd] at (1,0) {}; 
\node[bd] at (1,1) {}; 
\node[bd] at (1,2) {}; 
\node[bd] at (1,3) {}; 
\node[bd] at (1,4) {}; 
\node[wd] at (-1,1) {}; 
\end{tikzpicture}} & \raisebox{-.5\height}{ 
\begin{tikzpicture}[x=.5cm,y=.5cm] 
	\draw[step=.5cm,gray,very thin] (-2,-2) grid (1,4);
\draw[ligne] (1,4)--(0,4); 
\draw[ligne] (0,4)--(-1,3); 
\draw[ligne] (-1,3)--(-2,2); 
\draw[ligne] (-2,2)--(-2,1); 
\draw[ligne] (-2,1)--(1,-2); 
\draw[ligne] (1,-2)--(1,-1); 
\draw[ligne] (1,-1)--(1,0); 
\draw[ligne] (1,0)--(1,1); 
\draw[ligne] (1,1)--(1,2); 
\draw[ligne] (1,2)--(1,3); 
\draw[ligne] (1,3)--(1,4); 
\node[bd] at (0,4) {}; 
\node[bd] at (-1,3) {}; 
\node[bd] at (-2,2) {}; 
\node[bd] at (-2,1) {}; 
\node[bd] at (1,-2) {}; 
\node[bd] at (1,-1) {}; 
\node[bd] at (1,0) {}; 
\node[bd] at (1,1) {}; 
\node[bd] at (1,2) {}; 
\node[bd] at (1,3) {}; 
\node[bd] at (1,4) {}; 
\node[wd] at (-1,0) {}; 
\node[wd] at (0,-1) {}; 
\end{tikzpicture}} &
\begin{tabular}{c}
     \raisebox{-.5\height}{\begin{tikzpicture}[x=.8cm,y=.8cm]
\node (g1) at (1,0) [gauge,label=below:{\scalebox{.76}{$1$}}] {};
\node (g2) at (2,0) [gauge,label=below:{\scalebox{.76}{$2$}}] {};
\node (g3) at (3,0) [gauge,label=below:{\scalebox{.76}{$3$}}] {};
\node (g4) at (4,0) [gauge,label=below:{\scalebox{.76}{$4$}}] {};
\node (g5) at (5,0) [gauge,label=below:{\scalebox{.76}{$3$}}] {};
\node (g6) at (6,0) [gaugeb,label=below:{\scalebox{.76}{$1$}}] {};
\node (g7) at (5,1) [gaugeb,label=above:{\scalebox{.76}{$1$}}] {};
\node (g8) at (4,1) [gaugeb,label=above:{\scalebox{.76}{$2$}}] {};
\node (g9) at (3,1) [gauge,label=above:{\scalebox{.76}{$1$}}] {};
\draw (g1)--(g2)--(g3)--(g4)--(g5)--(g6);
\draw (g4)--(g8)--(g9);
\draw (g5)--(g7);
\end{tikzpicture}}  \\ \\
   $ \su(6) \oplus \su(2) \oplus \uu(1) \oplus \uu(1)$
\end{tabular}
 \\ \midrule 
      7 &\raisebox{-.5\height}{ 
\begin{tikzpicture}[x=.5cm,y=.5cm] 
	\draw[step=.5cm,gray,very thin] (-2,-2) grid (1,5);
\draw[ligne] (1,5)--(0,4); 
\draw[ligne] (0,4)--(-1,3); 
\draw[ligne] (-1,3)--(-2,2); 
\draw[ligne] (-2,2)--(0,0); 
\draw[ligne] (0,0)--(1,-2); 
\draw[ligne] (1,-2)--(1,-1); 
\draw[ligne] (1,-1)--(1,0); 
\draw[ligne] (1,0)--(1,1); 
\draw[ligne] (1,1)--(1,2); 
\draw[ligne] (1,2)--(1,3); 
\draw[ligne] (1,3)--(1,4); 
\draw[ligne] (1,4)--(1,5); 
\node[bd] at (0,4) {}; 
\node[bd] at (-1,3) {}; 
\node[bd] at (-2,2) {}; 
\node[bd] at (0,0) {}; 
\node[bd] at (1,-2) {}; 
\node[bd] at (1,-1) {}; 
\node[bd] at (1,0) {}; 
\node[bd] at (1,1) {}; 
\node[bd] at (1,2) {}; 
\node[bd] at (1,3) {}; 
\node[bd] at (1,4) {}; 
\node[bd] at (1,5) {}; 
\node[wd] at (-1,1) {}; 
\end{tikzpicture}} &  \raisebox{-.5\height}{ 
\begin{tikzpicture}[x=.5cm,y=.5cm] 
	\draw[step=.5cm,gray,very thin] (-2,-2) grid (1,5);
\draw[ligne] (1,5)--(0,4); 
\draw[ligne] (0,4)--(-1,3); 
\draw[ligne] (-1,3)--(-2,2); 
\draw[ligne] (-2,2)--(-2,1); 
\draw[ligne] (-2,1)--(1,-2); 
\draw[ligne] (1,-2)--(1,-1); 
\draw[ligne] (1,-1)--(1,0); 
\draw[ligne] (1,0)--(1,1); 
\draw[ligne] (1,1)--(1,2); 
\draw[ligne] (1,2)--(1,3); 
\draw[ligne] (1,3)--(1,4); 
\draw[ligne] (1,4)--(1,5); 
\node[bd] at (0,4) {}; 
\node[bd] at (-1,3) {}; 
\node[bd] at (-2,2) {}; 
\node[bd] at (-2,1) {}; 
\node[bd] at (1,-2) {}; 
\node[bd] at (1,-1) {}; 
\node[bd] at (1,0) {}; 
\node[bd] at (1,1) {}; 
\node[bd] at (1,2) {}; 
\node[bd] at (1,3) {}; 
\node[bd] at (1,4) {}; 
\node[bd] at (1,5) {}; 
\node[wd] at (-1,0) {}; 
\node[wd] at (0,-1) {}; 
\end{tikzpicture}} & 
\begin{tabular}{c}
\raisebox{-.5\height}{\begin{tikzpicture}[x=.8cm,y=.8cm]
\node (g1) at (1,0) [gauge,label=below:{\scalebox{.76}{$1$}}] {};
\node (g2) at (2,0) [gauge,label=below:{\scalebox{.76}{$2$}}] {};
\node (g3) at (3,0) [gauge,label=below:{\scalebox{.76}{$3$}}] {};
\node (g4) at (4,0) [gauge,label=below:{\scalebox{.76}{$4$}}] {};
\node (g5) at (5,0) [gauge,label=below:{\scalebox{.76}{$5$}}] {};
\node (g6) at (6,0) [gaugeb,label=below:{\scalebox{.76}{$3$}}] {};
\node (g7) at (7,0) [gauge,label=below:{\scalebox{.76}{$2$}}] {};
\node (g8) at (8,0) [gauge,label=below:{\scalebox{.76}{$1$}}] {};
\node (g9) at (5,1) [gauge,label=left:{\scalebox{.76}{$3$}}] {};
\node (g10) at (6,1) [gaugeb,label=right:{\scalebox{.76}{$1$}}] {};
\draw (g1)--(g2)--(g3)--(g4)--(g5)--(g6)--(g7)--(g8);
\draw (g5)--(g9)--(g10);
\end{tikzpicture} }  \\ \\ 
    $\su(7) \oplus \su(3) \oplus \uu(1)$
\end{tabular}   
\\ \midrule 
      8 &\raisebox{-.5\height}{ 
\begin{tikzpicture}[x=.5cm,y=.5cm] 
	\draw[step=.5cm,gray,very thin] (-2,-3) grid (1,5);
\draw[ligne] (1,5)--(0,4); 
\draw[ligne] (0,4)--(-1,3); 
\draw[ligne] (-1,3)--(-2,2); 
\draw[ligne] (-2,2)--(0,0); 
\draw[ligne] (0,0)--(1,-3); 
\draw[ligne] (1,-3)--(1,-2); 
\draw[ligne] (1,-2)--(1,-1); 
\draw[ligne] (1,-1)--(1,0); 
\draw[ligne] (1,0)--(1,1); 
\draw[ligne] (1,1)--(1,2); 
\draw[ligne] (1,2)--(1,3); 
\draw[ligne] (1,3)--(1,4); 
\draw[ligne] (1,4)--(1,5); 
\node[bd] at (0,4) {}; 
\node[bd] at (-1,3) {}; 
\node[bd] at (-2,2) {}; 
\node[bd] at (0,0) {}; 
\node[bd] at (1,-3) {}; 
\node[bd] at (1,-2) {}; 
\node[bd] at (1,-1) {}; 
\node[bd] at (1,0) {}; 
\node[bd] at (1,1) {}; 
\node[bd] at (1,2) {}; 
\node[bd] at (1,3) {}; 
\node[bd] at (1,4) {}; 
\node[bd] at (1,5) {}; 
\node[wd] at (-1,1) {}; 
\end{tikzpicture}} & \raisebox{-.5\height}{ 
\begin{tikzpicture}[x=.5cm,y=.5cm] 
	\draw[step=.5cm,gray,very thin] (-3,-3) grid (1,5);
\draw[ligne] (1,5)--(0,4); 
\draw[ligne] (0,4)--(-1,3); 
\draw[ligne] (-1,3)--(-2,2); 
\draw[ligne] (-2,2)--(-3,1); 
\draw[ligne] (-3,1)--(1,-3); 
\draw[ligne] (1,-3)--(1,-2); 
\draw[ligne] (1,-2)--(1,-1); 
\draw[ligne] (1,-1)--(1,0); 
\draw[ligne] (1,0)--(1,1); 
\draw[ligne] (1,1)--(1,2); 
\draw[ligne] (1,2)--(1,3); 
\draw[ligne] (1,3)--(1,4); 
\draw[ligne] (1,4)--(1,5); 
\node[bd] at (0,4) {}; 
\node[bd] at (-1,3) {}; 
\node[bd] at (-2,2) {}; 
\node[bd] at (-3,1) {}; 
\node[bd] at (1,-3) {}; 
\node[bd] at (1,-2) {}; 
\node[bd] at (1,-1) {}; 
\node[bd] at (1,0) {}; 
\node[bd] at (1,1) {}; 
\node[bd] at (1,2) {}; 
\node[bd] at (1,3) {}; 
\node[bd] at (1,4) {}; 
\node[bd] at (1,5) {}; 
\node[wd] at (-2,0) {}; 
\node[wd] at (-1,-1) {}; 
\node[wd] at (0,-2) {}; 
\end{tikzpicture}} &
\begin{tabular}{c}
\raisebox{-.5\height}{\begin{tikzpicture}[x=.8cm,y=.8cm]
\node (g1) at (1,0) [gauge,label=below:{\scalebox{.76}{$1$}}] {};
\node (g2) at (2,0) [gauge,label=below:{\scalebox{.76}{$2$}}] {};
\node (g3) at (3,0) [gauge,label=below:{\scalebox{.76}{$3$}}] {};
\node (g4) at (4,0) [gauge,label=below:{\scalebox{.76}{$4$}}] {};
\node (g5) at (5,0) [gauge,label=below:{\scalebox{.76}{$5$}}] {};
\node (g6) at (6,0) [gauge,label=below:{\scalebox{.76}{$6$}}] {};
\node (g7) at (7,0) [gaugeb,label=below:{\scalebox{.76}{$4$}}] {};
\node (g8) at (8,0) [gauge,label=below:{\scalebox{.76}{$3$}}] {};
\node (g9) at (9,0) [gauge,label=below:{\scalebox{.76}{$2$}}] {};
\node (g10) at (10,0) [gauge,label=below:{\scalebox{.76}{$1$}}] {};
\node (g11) at (6,1) [gauge,label=right:{\scalebox{.76}{$3$}}] {};
\draw (g1)--(g2)--(g3)--(g4)--(g5)--(g6)--(g7)--(g8)--(g9)--(g10);
\draw (g6)--(g11);
	\end{tikzpicture} } \\ \\ 
	$\su(8) \oplus \su(4)$
\end{tabular}  \\ \midrule 
     9 &   \raisebox{-.5\height}{ 
\begin{tikzpicture}[x=.5cm,y=.5cm] 
	\draw[step=.5cm,gray,very thin] (-2,-4) grid (1,5);
\draw[ligne] (1,5)--(0,4); 
\draw[ligne] (0,4)--(-1,3); 
\draw[ligne] (-1,3)--(-2,2); 
\draw[ligne] (-2,2)--(0,0); 
\draw[ligne] (0,0)--(1,-4); 
\draw[ligne] (1,-4)--(1,-3); 
\draw[ligne] (1,-3)--(1,-2); 
\draw[ligne] (1,-2)--(1,-1); 
\draw[ligne] (1,-1)--(1,0); 
\draw[ligne] (1,0)--(1,1); 
\draw[ligne] (1,1)--(1,2); 
\draw[ligne] (1,2)--(1,3); 
\draw[ligne] (1,3)--(1,4); 
\draw[ligne] (1,4)--(1,5); 
\node[bd] at (0,4) {}; 
\node[bd] at (-1,3) {}; 
\node[bd] at (-2,2) {}; 
\node[bd] at (0,0) {}; 
\node[bd] at (1,-4) {}; 
\node[bd] at (1,-3) {}; 
\node[bd] at (1,-2) {}; 
\node[bd] at (1,-1) {}; 
\node[bd] at (1,0) {}; 
\node[bd] at (1,1) {}; 
\node[bd] at (1,2) {}; 
\node[bd] at (1,3) {}; 
\node[bd] at (1,4) {}; 
\node[bd] at (1,5) {}; 
\node[wd] at (-1,1) {}; 
\end{tikzpicture}} & \raisebox{-.5\height}{ 
\begin{tikzpicture}[x=.5cm,y=.5cm] 
	\draw[step=.5cm,gray,very thin] (-3,-3) grid (1,8);
\draw[ligne] (1,6)--(1,7); 
\draw[ligne] (1,7)--(1,8); 
\draw[ligne] (1,8)--(-3,4); 
\draw[ligne] (-3,4)--(-3,1); 
\draw[ligne] (-3,1)--(1,-3); 
\draw[ligne] (1,-3)--(1,-2); 
\draw[ligne] (1,-2)--(1,-1); 
\draw[ligne] (1,-1)--(1,0); 
\draw[ligne] (1,0)--(1,1); 
\draw[ligne] (1,1)--(1,2); 
\draw[ligne] (1,2)--(1,3); 
\draw[ligne] (1,3)--(1,4); 
\draw[ligne] (1,4)--(1,5); 
\draw[ligne] (1,5)--(1,6); 
\node[bd] at (1,7) {}; 
\node[bd] at (1,8) {}; 
\node[bd] at (-3,4) {}; 
\node[bd] at (-3,1) {}; 
\node[bd] at (1,-3) {}; 
\node[bd] at (1,-2) {}; 
\node[bd] at (1,-1) {}; 
\node[bd] at (1,0) {}; 
\node[bd] at (1,1) {}; 
\node[bd] at (1,2) {}; 
\node[bd] at (1,3) {}; 
\node[bd] at (1,4) {}; 
\node[bd] at (1,5) {}; 
\node[bd] at (1,6) {}; 
\node[wd] at (0,7) {}; 
\node[wd] at (-1,6) {}; 
\node[wd] at (-2,5) {}; 
\node[wd] at (-3,3) {}; 
\node[wd] at (-3,2) {}; 
\node[wd] at (-2,0) {}; 
\node[wd] at (-1,-1) {}; 
\node[wd] at (0,-2) {}; 
\end{tikzpicture}} & 
\begin{tabular}{c}
\raisebox{-.5\height}{\begin{tikzpicture}[x=.8cm,y=.8cm]
\node (g1) at (1,0) [gauge,label=below:{\scalebox{.76}{$1$}}] {};
\node (g2) at (2,0) [gauge,label=below:{\scalebox{.76}{$2$}}] {};
\node (g3) at (3,0) [gauge,label=below:{\scalebox{.76}{$3$}}] {};
\node (g4) at (4,0) [gauge,label=below:{\scalebox{.76}{$4$}}] {};
\node (g5) at (5,0) [gauge,label=below:{\scalebox{.76}{$5$}}] {};
\node (g6) at (6,0) [gauge,label=below:{\scalebox{.76}{$6$}}] {};
\node (g7) at (7,0) [gauge,label=below:{\scalebox{.76}{$7$}}] {};
\node (g8) at (8,0) [gauge,label=below:{\scalebox{.76}{$8$}}] {};
\node (g9) at (9,0) [gauge,label=below:{\scalebox{.76}{$9$}}] {};
\node (g10) at (10,0) [gauge,label=below:{\scalebox{.76}{$6$}}] {};
\node (g11) at (11,0) [gauge,label=below:{\scalebox{.76}{$3$}}] {};
\node (g12) at (9,1) [gaugeb,label=right:{\scalebox{.76}{$4$}}] {};
\draw (g1)--(g2)--(g3)--(g4)--(g5)--(g6)--(g7)--(g8)--(g9)--(g10)--(g11);
\draw (g9)--(g12);
\end{tikzpicture} } \\ \\ 
 $\su(12)$
\end{tabular} 
  \\ \bottomrule 
    \end{tabular}
    \caption{pGTPs, GTPs, MQs and flavor symmetries for  $SU(4)_k + 1 \mathbf{AS} + N_F \mathbf{F}$ for $N_F=6,7,8,9$ and $k=0$ or $k=\half$. }
    \label{table:GTPExample}
\end{table}

\subsection[\texorpdfstring{$SU(N)_\half + 2 AS + 7F$}{SU(N)1/2 + 2 AS + 7 F}]{\boldmath{$SU(N)_\half + 2 AS + 7F$}}

\label{sec:2AS7F}

\paragraph{GTP for Tree Tops.}

Next, we are interested in the strongly coupled $SU(N)_{\half} + 2 \bm{AS} + 7 \bm{F}$ theory. The brane-web for the gauge theory description was determined in \cite{Zafrir:2015rga} and dualizes to the polygons (drawn for $n=3$)
\be
\begin{tikzpicture}[x=.3cm,y=.3cm]

\draw[step=.3cm,gray,very thin] (0,0) grid (6,12);

\draw[ligne,black] (0,0)--(3,3)--(6,3)--(6,12)--(3,9)--(0,12)--(0,0);
\draw[black] (3,3)--(3,9);

\node[bd] at (0,0) {};
\node[wd] at (1,1) {};
\node[wd] at (2,2) {};
\node[bd] at (3,3) {};
\node[bd] at (4,3) {};
\node[wd] at (5,3) {};
\node[bd] at (6,3) {};
\node[wd] at (6,4) {};
\node[wd] at (6,5) {};
\node[bd] at (6,6) {};
\node[wd] at (6,7) {};
\node[wd] at (6,8) {};
\node[bd] at (6,9) {};
\node[wd] at (6,10) {};
\node[wd] at (6,11) {};
\node[bd] at (6,12) {};
\node[wd] at (5,11) {};
\node[wd] at (4,10) {};
\node[bd] at (3,9) {};
\node[bd] at (2,10) {};
\node[wd] at (1,11) {};
\node[bd] at (0,12) {};
\node[wd] at (0,11) {};
\node[wd] at (0,10) {};
\node[bd] at (0,9) {};
\node[wd] at (0,8) {};
\node[wd] at (0,7) {};
\node[bd] at (0,6) {};
\node[wd] at (0,5) {};
\node[wd] at (0,4) {};
\node[bd] at (0,3) {};
\node[wd] at (0,2) {};
\node[wd] at (0,1) {};

\node at (2.5,0.5) {\small $n$};
\node at (5,2) {\small $n$-1};
\node at (7,4.5) {\small $n$};
\node at (7,7.5) {\small $n$};
\node at (7,10.5) {\small $n$};
\node at (4,11.5) {\small $n$};
\node at (1.5,12) {\small $n$-1};
\node at (-1,10.5) {\small $n$};
\node at (-1,7.5) {\small $n$};
\node at (-1,4.5) {\small $n$};
\node at (-1,1.5) {\small $n$};

\node at (-1.2,0) {$\bm{v}_0$};

\end{tikzpicture}
\qquad
\qquad
\begin{tikzpicture}[x=.3cm,y=.3cm]

\draw[step=.3cm,gray,very thin] (0,0) grid (6,13);

\draw[ligne,black] (0,0)--(3,3)--(6,3)--(6,13)--(3,10)--(0,13)--(0,0);
\draw[black] (3,3)--(3,10);

\node[bd] at (0,0) {};
\node[wd] at (1,1) {};
\node[wd] at (2,2) {};
\node[bd] at (3,3) {};
\node[wd] at (4,3) {};
\node[wd] at (5,3) {};
\node[bd] at (6,3) {};
\node[wd] at (6,4) {};
\node[wd] at (6,5) {};
\node[bd] at (6,6) {};
\node[wd] at (6,7) {};
\node[wd] at (6,8) {};
\node[bd] at (6,9) {};
\node[wd] at (6,10) {};
\node[wd] at (6,11) {};
\node[bd] at (6,12) {};
\node[bd] at (6,13) {};
\node[wd] at (5,12) {};
\node[wd] at (4,11) {};
\node[bd] at (3,10) {};
\node[wd] at (2,11) {};
\node[wd] at (1,12) {};
\node[bd] at (0,13) {};
\node[bd] at (0,12) {};
\node[wd] at (0,11) {};
\node[wd] at (0,10) {};
\node[bd] at (0,9) {};
\node[wd] at (0,8) {};
\node[wd] at (0,7) {};
\node[bd] at (0,6) {};
\node[wd] at (0,5) {};
\node[wd] at (0,4) {};
\node[bd] at (0,3) {};
\node[wd] at (0,2) {};
\node[wd] at (0,1) {};

\node at (2.5,0.5) {\small $n$};
\node at (4.5,2) {\small $n$};
\node at (7,4.5) {\small $n$};
\node at (7,7.5) {\small $n$};
\node at (7,10.5) {\small $n$};
\node at (4,12.5) {\small $n$};
\node at (2,12.5) {\small $n$};
\node at (-1,10.5) {\small $n$};
\node at (-1,7.5) {\small $n$};
\node at (-1,4.5) {\small $n$};
\node at (-1,1.5) {\small $n$};

\node at (-1.2,0) {$\bm{v}_0$};

\end{tikzpicture}
\ee
for $N=2n$ and $N=2n+1$, respectively. Again we take the bottom left corner to be the origin. The polygon vectors are
\be
\ba
N&=2n: \qquad &&\bm{u}=\left[(n,n),(1,0),(n-1,0),\underbrace{(0,n)}_{3},(-n,-n),(-1,1),(-n+1,n-1),\underbrace{(0,-n)}_4 \right]\\
N&=2n+1: \qquad &&\bm{u}=\left[(n,n),(n,0),\underbrace{(0,n)}_{3},(0,1),(-n,-n),(-n,n),(0,-1),\underbrace{(0,-n)}_4 \right]
\ea
\ee
For odd $N$ the edge-moves are as for \eqref{MonoSQCD} but for even $N$ they are different
\be
\ba
N&=2n: \qquad &&\mathfrak{T}= \left(\mathfrak{M}^-_{5} \mathfrak{M}^-_{6} \mathfrak{M}^-_{7} \mathfrak{M}^-_{8}\right) \left(\mathfrak{M}^-_{4} \mathfrak{M}^-_{5} \mathfrak{M}^-_{6} \mathfrak{M}^-_{7}\right)\\
N&=2n+1: \qquad &&\mathfrak{T}=(\mathfrak{M}_4^+ \cdots \mathfrak{M}_7^+) \mathfrak{M}^-_2 \mathfrak{M}^+_1 (\mathfrak{M}_3^+ \cdots \mathfrak{M}_6^+) \mathfrak{M}^-_7
\ea
\ee
This leads to the transformed vectors
\be
\ba
N&=2n: \qquad &&\mathfrak{T}\bm{u}=\left[\underbrace{(1,0)}_2,\underbrace{(n-1,0)}_2,\underbrace{(n,0)}_{4},\underbrace{(0,2n)}_3,\underbrace{(-3n,-3n)}_2 \right]\\
N&=2n+1: \qquad &&\mathfrak{T}\bm{u}=\left[(2n,0),(4n+1,8n+2),(0,5n+1),(-6n-1,-6n-1),\underbrace{(0,-n)}_7,\underbrace{(0,-1)}_2 \right]
\ea
\ee

\paragraph{Magnetic Quivers and Decoupling.}

The magnetic quiver computed from the GTP is:
\be \label{MQk1/22AS7F}
\begin{tikzpicture}[x=.8cm,y=.8cm]
\node (g1) at (0,0) [gauge,label=below:{1}] {};
\node (g2) at (1.5,0) [gaugeb,label=below:{2}] {};
\node (g3) at (3,0) [gauge,label=below:{$n$+1}] {};
\node (g4) at (4.5,0) [gaugeb,label=below:{$2n$}] {};
\node (g5) at (6,0) [gauge,label=below:{$3n$}] {};
\node (g6) at (7.5,0) [gauge,label=below:{$4n$}] {};
\node (g7) at (9,0) [gauge,label=below:{$5n$}] {};
\node (g8) at (10.5,0) [gauge,label=below:{$6n$}] {};
\node (g9) at (12,0) [gauge,label=below:{$4n$}] {};
\node (g10) at (13.5,0) [gauge,label=below:{$2n$}] {};
\node (g11) at (10.5,1) [gauge,label=right:{$3n$}] {};
\draw (g1)--(g2)--(g3)--(g4)--(g5)--(g6)--(g7)--(g8)--(g9)--(g10);
\draw (g8)--(g11);

\node (h1) at (0,-3) [gauge,label=below:{1}] {};
\node (h2) at (1.5,-3) [gaugeb,label=below:{2}] {};
\node (h3) at (3,-3) [gauge,label=below:{$n$+2}] {};
\node (h4) at (4.5,-3) [gauge,label=below:{$2n$+2}] {};
\node (h5) at (6,-3) [gauge,label=below:{$3n$+2}] {};
\node (h6) at (7.5,-3) [gauge,label=below:{$4n$+2}] {};
\node (h7) at (9,-3) [gauge,label=below:{$5n$+2}] {};
\node (h8) at (10.5,-3) [gauge,label=below:{$6n$+2}] {};
\node (h9) at (12,-3) [gauge,label=below:{$4n$+1}] {};
\node (h10) at (13.5,-3) [gaugeb,label=below:{$2n$}] {};
\node (h11) at (10.5,-2) [gauge,label=right:{$3n$+1}] {};
\draw (h1)--(h2)--(h3)--(h4)--(h5)--(h6)--(h7)--(h8)--(h9)--(h10);
\draw (h8)--(h11);

\end{tikzpicture}
\ee
for $N=2n$ and $N=2n+1$ respectively.
We see that the flavor symmetry coming from the balanced nodes depends on the parity of $N$. For $N$ even it is $\e_7 \oplus \su(2) \oplus \su(2) \oplus \uu(1)$ whereas for $N$ odd it is $\so(16) \oplus \su(2) \oplus \uu(1)$. In both cases, the $\uu(1)$ is expected to enhance to $\su(2)$.
For $N=4$ the flavor symmetry seems to be enhanced to $\e_7 \oplus \su(4)$, where the $\su(4)$ is known to enhance further to $\so(7)$ \cite{Jefferson:2017ahm,Mekareeya:2017jgc,Yonekura:2015ksa,Bhardwaj:2020avz}. We will discuss all these enhancements in section \ref{sec:NSLQuivers}.

\subsection[\texorpdfstring{$SU(N)_2 + 2 AS + 6F$}{SU(N)2 + 2 AS + 6 F}]{\boldmath{$SU(N)_2 + 2 AS + 6F$}}

\label{sec:2AS6F}

\paragraph{GTP for Tree Tops.}

Finally, there is the strongly coupled $SU(N)_{2} + 2 \bm{AS} + 6 \bm{F}$. Again, we have to distinguish between $N$ even and odd. The weakly coupled pGTP is given by (drawn for $n=3$)
\be
\begin{tikzpicture}[x=.3cm,y=.3cm]

\draw[step=.3cm,gray,very thin] (0,-3) grid (6,9);

\draw[ligne,black] (0,0)--(3,-3)--(6,-3)--(6,6)--(3,3)--(0,9)--(0,0);
\draw[black] (3,-3)--(3,3);

\node[bd] at (0,0) {};
\node[wd] at (1,-1) {};
\node[wd] at (2,-2) {};
\node[bd] at (3,-3) {};
\node[bd] at (4,-3) {};
\node[wd] at (5,-3) {};
\node[bd] at (6,-3) {};
\node[wd] at (6,-2) {};
\node[wd] at (6,-1) {};
\node[bd] at (6,0) {};
\node[wd] at (6,1) {};
\node[wd] at (6,2) {};
\node[bd] at (6,3) {};
\node[wd] at (6,4) {};
\node[wd] at (6,5) {};
\node[bd] at (6,6) {};
\node[wd] at (5,5) {};
\node[wd] at (4,4) {};
\node[bd] at (3,3) {};
\node[bd] at (2,5) {};
\node[wd] at (1,7) {};
\node[bd] at (0,9) {};
\node[wd] at (0,8) {};
\node[wd] at (0,7) {};
\node[bd] at (0,6) {};
\node[wd] at (0,5) {};
\node[wd] at (0,4) {};
\node[bd] at (0,3) {};
\node[wd] at (0,2) {};
\node[wd] at (0,1) {};

\node at (0.5,-2.5) {\small $n$};
\node at (5,-4) {\small $n$-1};
\node at (7,-1.5) {\small $n$};
\node at (7,1.5) {\small $n$};
\node at (7,4.5) {\small $n$};
\node at (4,5.5) {\small $n$};
\node at (2.5,7) {\small $n$-1};
\node at (-1,7.5) {\small $n$};
\node at (-1,4.5) {\small $n$};
\node at (-1,1.5) {\small $n$};

\node at (-1.2,0) {$\bm{v}_0$};

\end{tikzpicture}
\qquad
\qquad
\begin{tikzpicture}[x=.3cm,y=.3cm]

\draw[step=.3cm,gray,very thin] (0,-3) grid (6,10);

\draw[ligne,black] (0,0)--(3,-3)--(6,-3)--(6,7)--(3,4)--(0,10)--(0,0);
\draw[black] (3,-3)--(3,4);

\node[bd] at (0,0) {};
\node[wd] at (1,-1) {};
\node[wd] at (2,-2) {};
\node[bd] at (3,-3) {};
\node[wd] at (4,-3) {};
\node[wd] at (5,-3) {};
\node[bd] at (6,-3) {};
\node[wd] at (6,-2) {};
\node[wd] at (6,-1) {};
\node[bd] at (6,0) {};
\node[wd] at (6,1) {};
\node[wd] at (6,2) {};
\node[bd] at (6,3) {};
\node[wd] at (6,4) {};
\node[wd] at (6,5) {};
\node[bd] at (6,6) {};
\node[bd] at (6,7) {};
\node[wd] at (5,6) {};
\node[wd] at (4,5) {};
\node[bd] at (3,4) {};
\node[wd] at (2,6) {};
\node[wd] at (1,8) {};
\node[bd] at (0,10) {};
\node[bd] at (0,9) {};
\node[wd] at (0,8) {};
\node[wd] at (0,7) {};
\node[bd] at (0,6) {};
\node[wd] at (0,5) {};
\node[wd] at (0,4) {};
\node[bd] at (0,3) {};
\node[wd] at (0,2) {};
\node[wd] at (0,1) {};

\node at (0.5,-2.5) {\small $n$};
\node at (4.5,-4) {\small $n$};
\node at (7,-1.5) {\small $n$};
\node at (7,1.5) {\small $n$};
\node at (7,4.5) {\small $n$};
\node at (4.5,6.5) {\small $n$};
\node at (2.5,7) {\small $n$};
\node at (-1,7.5) {\small $n$};
\node at (-1,4.5) {\small $n$};
\node at (-1,1.5) {\small $n$};

\node at (-1.2,0) {$\bm{v}_0$};

\end{tikzpicture}
\ee
The vectors of the polygons are
\be
\ba
N&=2n: \qquad &&\bm{u}=\left[(n,-n),(1,0),(n-1,0),\underbrace{(0,n)}_{3},(-n,-n),(-1,2),(-n+1,2n-2),\underbrace{(0,-n)}_3 \right]\\
N&=2n+1: \qquad &&\bm{u}=\left[(n,-n),(n,0),\underbrace{(0,n)}_{3},(0,1),(-n,-n),(-n,2n),(0,-1),\underbrace{(0,-n)}_3 \right]
\ea
\ee
To make these polygons convex we apply the operations
\be
\ba
N&=2n: \qquad &&\mathfrak{T}=(\mathfrak{M}_{4}^- \mathfrak{M}_{3}^-) (\mathfrak{M}_{2}^+ \mathfrak{M}_{1}^+) (\mathfrak{M}_{4}^- \cdots \mathfrak{M}_{8}^-)(\mathfrak{M}_{3}^- \cdots \mathfrak{M}_{7}^-) \\
N&=2n+1: \qquad &&\mathfrak{T}=\mathfrak{M}_{11}^- (\mathfrak{M}_{10}^+ \cdots \mathfrak{M}_2^+)\mathfrak{M}_1^- \mathfrak{M}_2^+ (\mathfrak{M}_3^- \cdots \mathfrak{M}_7^-)
\ea
\ee
leading to the transformed vectors
\be
\ba
N&=2n: \qquad &&\mathfrak{T}\bm{u}=\left[(4n,0),\underbrace{(0,3n)}_{3},(-4n,0),\underbrace{(0,-n)}_3,\underbrace{(0,-n-1)}_2,\underbrace{(0,-2n+1)}_2 \right]\\
N&=2n+1: \qquad &&\mathfrak{T}\bm{u}=\left[(10n,-10n),\underbrace{(0,8n)}_{3},(0,7n+1),\underbrace{(-5n,-5n)}_2,(0,-4n),\underbrace{(0,-2n)}_3,(0,-n-1) \right]
\ea
\ee

\paragraph{Magnetic Quivers and Decoupling.}

We can then compute the magnetic quiver for $SU(N)_{2} + 2 \bm{AS} + 6 \bm{F}$ using \eqref{SimoneDuality}
\be \label{MQk22AS6F}
\begin{tikzpicture}[x=.8cm,y=.8cm]
\node (g1) at (0,0) [gaugeb,label=below:{2}] {};
\node (g2) at (1.5,0) [gauge,label=below:{$n$+2}] {};
\node (g3) at (3,0) [gauge,label=below:{$2n$+2}] {};
\node (g4) at (4.5,0) [gauge,label=below:{$3n$+2}] {};
\node (g5) at (6,0) [gauge,label=below:{$4n$+2}] {};
\node (g6) at (7.5,0) [gauge,label=below:{$5n$+2}] {};
\node (g7) at (9,0) [gauge,label=below:{$6n$+2}] {};
\node (g8) at (10.5,0) [gauge,label=below:{$4n$+1}] {};
\node (g9) at (12,0) [gaugeb,label=below:{$2n$}] {};
\node (g10) at (9,1) [gauge,label=right:{$3n$+1}] {};
\draw (g1)--(g2)--(g3)--(g4)--(g5)--(g6)--(g7)--(g8)--(g9);
\draw (g7)--(g10);

\node (h1) at (0,-3) [gaugeb,label=below:{2}] {};
\node (h2) at (1.5,-3) [gauge,label=below:{$n$+2}] {};
\node (h3) at (3,-3) [gaugeb,label=below:{$2n$+2}] {};
\node (h4) at (4.5,-3) [gauge,label=below:{$3n$+3}] {};
\node (h5) at (6,-3) [gauge,label=below:{$4n$+4}] {};
\node (h6) at (7.5,-3) [gauge,label=below:{$5n$+5}] {};
\node (h7) at (9,-3) [gauge,label=below:{$6n$+6}] {};
\node (h8) at (10.5,-3) [gauge,label=below:{$4n$+4}] {};
\node (h9) at (12,-3) [gauge,label=below:{$2n$+2}] {};
\node (h10) at (9,-2) [gauge,label=right:{$3n$+3}] {};
\draw (h1)--(h2)--(h3)--(h4)--(h5)--(h6)--(h7)--(h8)--(h9);
\draw (h7)--(h10);

\end{tikzpicture}
\,,
\ee
for $N=2n$ and $N=2n+1$, respectively.
For $N$ even and odd the visible flavor symmetry is $\so(16) \oplus \uu(1)$ and $\e_7 \oplus \su(2) \oplus \uu(1)$, respectively. Again, the $\uu(1)$ is known to enhance to $\su(2)$.
For low rank, the flavor symmetry is no longer simply laced. Concretely, we find $\so(18)\to \so(19)$ for $N=4$ and $\e_7 \oplus \su(3)\to \e_7 \oplus \mathfrak{g}_2$ for $N=5$. Again, this will be discussed in section \ref{sec:NSLQuivers}.

\subsection{5d Quiver Gauge Theories and UV MQs}
\label{sec:quivertheories}

We study 5d $[n_1]-SU(2)^m-[n_2]$ quiver gauge theories, with $m$ $SU(2)$ gauge nodes, $m-1$ bifundamental matter multiplets, and $n_1$, respectively $n_2$, matter multiplets in the fundamental of the first and last $SU(2)$ gauge groups. For generic $m$, the theories with $n_1 \leq 3, n_2 \leq 4$ are descendants of $(D_{m+3},D_{m+3})$-conformal matter theories. For rank $m=2$ there exists an additional tree of theories with $n_1\leq1, n_2\leq5$, which descend from the E-string, that merges with the $(D_{5},D_{5})$-conformal matter tree for $n_1 \leq 1,n_2\leq4$. The polygon for this class of theories is (drawn for $m=3, n_1=3, n_2=4$)

\be 
\begin{tikzpicture}[x=.5cm,y=.5cm]
\draw[step=.5cm,gray,very thin] (0,-1) grid (4,3);

\draw[ligne,black] (0,0)--(3,0)--(4,-1)--(4,3)--(3,2)--(1,2)--(0,3)--(0,0);
\foreach \x in {1,2,3}
\draw[black] (\x,0)--(\x,2);

 \foreach \x in {0,1,...,3}
    \node[bd] at (\x,0) {};
 \foreach \x in {0,1,...,4}
    \node[bd] at (4,\x-1) {};
    
 \foreach \x in {0,1,2}
    \node[bd] at (3-\x,2) {};
 \foreach \x in {0,1,2,3}
    \node[bd] at (0,\x) {};

    \node at (2,2.6) {$m-1$};
    
    \draw[|-] (-1,0)--(-1,1) node[left, midway] {$\lfloor \frac{n_1}{2} \rfloor$};
    \draw[|-|] (-1,1)--(-1,3) node[left,midway] {$\lceil \frac{n_1}{2} \rceil$};;

    \draw[|-] (5,-1)--(5,1) node[right, midway] {$\lfloor \frac{n_2}{2} \rfloor$};
    \draw[|-|] (5,1)--(5,3) node[right,midway] {$\lceil \frac{n_2}{2} \rceil$};
    
    \node at (0,-.6) {$\bm{v}_0$};
      
\end{tikzpicture}
\ee
The dual web of this polygon is given in appendix \ref{app:QuiverWebs}. We take the origin $\bm{v}_0$ to be the bottom left vertex. Then the corresponding vectors are given by
\be
\bm{u}=\left[ (1,\lfloor \frac{n_1}{2} \rfloor-1), \underbrace{(1,0)}_{m-1}, (1,1-\lfloor \frac{n_2}{2} \rfloor),\underbrace{(0,1)}_{n_2},(-1,1-\lceil \frac{n_2}{2} \rceil),\underbrace{(-1,0)}_{m-1},(-1,\lceil \frac{n_1}{2} \rceil-1),\underbrace{(0,-1)}_{n_1} \right]\,.
\ee
For $n_i \leq 2$ this is a convex toric diagram, in which we can go to the strong coupling limit by removing the internal lines. When either of the $n_i$ is greater than 2, the polygon is non-convex and we must perform edge-moves to obtain the UV theory. 

In the case where one (or both) $n_i=0$, i.e. there is no fundamental matter associated to one (or both) of the $SU(2)$s, there are two choices for the angles of the leftmost/rightmost edges, corresponding to $\theta=0,\pi$ for the associated $SU(2)$. In particular, a set of $(\pm 1,-1)$-edges gives rise to a $\theta=\pi$ gauge theory, whereas a $(-1,0)$- and $(1,\pm 2)$-edge corresponds to a $\theta=0$ theory (drawn for $m=3, n_2=4$):
\be 
\begin{tikzpicture}[x=.5cm,y=.5cm]
\draw[step=.5cm,gray,very thin] (0,-1) grid (4,3);

\draw[ligne,black] (1,0)--(3,0)--(4,-1)--(4,3)--(3,2)--(0,2)--(1,0); 
\foreach \x in {1,2,3}
\draw[black] (\x,0)--(\x,2);
 \foreach \x in {1,2,3}
    \node[bd] at (\x,0) {};
 \foreach \x in {0,1,...,4}
    \node[bd] at (4,\x-1) {};
    
 \foreach \x in {0,1,2,3}
    \node[bd] at (3-\x,2) {};

    \node at (1.5,2.6) {$m$};

    \draw[|-] (5,-1)--(5,1) node[right, midway] {$\lfloor \frac{n_2}{2} \rfloor$};
    \draw[|-|] (5,1)--(5,3) node[right,midway] {$\lceil \frac{n_2}{2} \rceil$};;
    
    \node at (-3,1) {$\theta=0:$};
    

\draw[step=.5cm,gray,very thin] (14,-1) grid (18,3);

\draw[ligne,black] (15,0)--(17,0)--(18,-1)--(18,3)--(17,2)--(15,2)--(14,1)--(15,0);
 \foreach \x in {0,1,2}
\draw[black] (15+\x,0)--(15+\x,2);

 \foreach \x in {1,2,3}
    \node[bd] at (14+\x,0) {};
 \foreach \x in {0,1,...,4}
    \node[bd] at (18,\x-1) {};
    
 \foreach \x in {1,2,3}
    \node[bd] at (18-\x,2) {};
    \node[bd] at (14,1) {};

    \node at (16,2.6) {$m-1$};

    \draw[|-] (19,-1)--(19,1) node[right, midway] {$\lfloor \frac{n_2}{2} \rfloor$};
    \draw[|-|] (19,1)--(19,3) node[right,midway] {$\lceil \frac{n_2}{2} \rceil$};;
    
    \node at (11,1) {$\theta=\pi:$};
      
\end{tikzpicture}
\ee
Notice that the $\theta=0$ polygon is inconsistent with an $SU(m+1)$ ruling, i.e. there exists no SQCD dual to these theories, which makes this subclass of quiver gauge theories particularly interesting to study in the strong coupling regime. 

\paragraph{GTP for Tree Tops.}

We will first study the submarginal theory, which is the first descendant of the circle reduction of $(D_{m+3},D_{m+3})$-conformal matter. The theory is $[3]-SU(2)^m-[4]$ with polygon (drawn for $m=3$)
\be 
\begin{tikzpicture}[x=.5cm,y=.5cm]
\draw[step=.5cm,gray,very thin] (0,-1) grid (4,3);

\draw[ligne,black] (0,0)--(3,0)--(4,-1)--(4,3)--(3,2)--(1,2)--(0,3)--(0,0);
 \foreach \x in {1,2,3}
    \draw[black] (\x,0)--(\x,2);

 \foreach \x in {0,1,2,3}
    \node[bd] at (\x,0) {};
 \foreach \x in {0,1,...,4}
    \node[bd] at (4,\x-1) {};
    
 \foreach \x in {1,2,3}
    \node[bd] at (4-\x,2) {};
 \foreach \x in {0,1,2,3}
    \node[bd] at (0,\x) {};

    \node at (2,2.6) {$m-1$};  
    \node at (0,-.6) {$\bm{v}_0$};
      
\end{tikzpicture}
\ee
The polygon consists of the set of vectors
\be
\bm{u}=\left[ \underbrace{(1,0)}_{m}, (1,-1),\underbrace{(0,1)}_{4},(-1,-1),\underbrace{(-1,0)}_{m-1},(-1,1),\underbrace{(0,-1)}_3 \right]\,.
\ee
The polygon has non-convexities at $\mathfrak{C}=\{m,m+6,2m+5\}$, which can be removed by applying the transformation given by
\be
\ba
\mathfrak{T}=(\mathfrak{M}_3^+ \cdots \mathfrak{M}_{2m+4}^+ \mathfrak{M}_{2m+5}^+)&(\mathfrak{M}_2^+ \cdots \mathfrak{M}_{2m+3}^+ \mathfrak{M}_{2m+4}^+) \times \\
&(\mathfrak{M}_{2m+5}^- \cdots \mathfrak{M}_{m+7}^- \mathfrak{M}_{m+6}^-)(\mathfrak{M}_1^+ \cdots \mathfrak{M}_{m-1}^+ \mathfrak{M}_{m}^+)\,,
\ea
\ee
leading to the transformed vectors
\be
\mathfrak{T}\bm{u}=\left[ (m+1,-m-1),(2m+4,6m+12),(-3m-5,-3m-5),\underbrace{(0,-1)}_{2m+6} \right]\,.
\ee
%
%
%
%
%
%

For rank $m=2$, there is an additional tree of quiver theories, that arises from a circle reduction of the E-string. The submarginal theory is $[1]-SU(2)-SU(2)-[5]$ with polygon
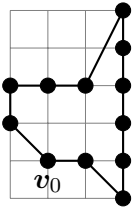
\be 
\begin{tikzpicture}[x=.5cm,y=.5cm]
\draw[step=.5cm,gray,very thin] (0,-1) grid (3,4);

\draw[ligne,black] (1,0)--(2,0)--(3,-1)--(3,4)--(2,2)--(0,2)--(0,1)--(1,0); 
\foreach \x in {0,1}
    \node[bd] at (1+\x,0) {};
 \foreach \x in {0,1,...,5}
    \node[bd] at (3,\x-1) {};    
 \foreach \x in {0,1,2}
    \node[bd] at (2-\x,2) {};
    \node[bd] at (0,1) {};
    
    \node at (1,-.6) {$\bm{v}_0$};
      
\end{tikzpicture}
\ee
build from the set of vectors given by
\be 
\bm{u}=\left[ (1,0),(1,-1),\underbrace{(0,1)}_5,(-1,-2),\underbrace{(-1,0)}_2,(0,-1),(1,-1) \right]\,.
\ee
The non-convexities lie at $\mathfrak{C}=\{1,8\}$. We apply the following edge-moves
\be 
\mathfrak{T} =\mathfrak{M}_1^+ \mathfrak{M}_2^- \mathfrak{M}_1^+ \mathfrak{M}_{12}^+(\mathfrak{M}_1^- \cdots \mathfrak{M}_8^- \mathfrak{M}_9^-)(\mathfrak{M}_8^+ \cdots \mathfrak{M}_3^+ \mathfrak{M}_2^+)(\mathfrak{M}_3^- \cdots \mathfrak{M}_7^- \mathfrak{M}_8^-)\mathfrak{M}_1^+\,,
\ee
which results in the transformed vectors
\be 
\mathfrak{T}\bm{u}=\left[(7,-14),\underbrace{(0,4)}_2,\underbrace{(0,5)}_5,(-7,-14),(0,-2),(0,-1),(0,-2) \right]\,.
\ee
%
%
%
%
%
%
%
%

\paragraph{Magnetic Quiver and Hasse Diagram for Tree Tops.}

The magnetic quiver for strongly coupled $[3]-SU(2)^m-[4]$ can be computed to be
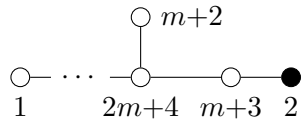
\be
\begin{tikzpicture}[x=.8cm,y=.8cm]
\node (g1) at (0,0) [gauge,label=below:{1}] {};
\node (g2) at (1,0) {$\cdots$};
\node (g3) at (2,0) [gauge,label=below:{$2m$+4}] {};
\node (g4) at (3.5,0) [gauge,label=below:{$m$+3}] {};
\node (g5) at (4.5,0) [gaugeb,label=below:{2}] {};
\node (g7) at (2,1) [gauge,label=right:{$m$+2}] {};
\draw (g1)--(g2)--(g3)--(g4)--(g5);
\draw (g3)--(g7);
\end{tikzpicture}
\ee
The balanced nodes of this magnetic quiver make up an $\so(4m+12)$ Dynkin diagram, corresponding to the flavor symmetry of the SCFT. The Hasse diagram for this theory is
\be
\begin{tikzpicture}
\node (1) [hasse] at (0,0) {};
\node (2) [hasse] at (0,-1) {};
\node (3) [hasse] at (0,-2) {};
\node (4) [hasse] at (0,-3) {};
\node (n1) at (0,-3.5) {};
\node (n2) at (0,-3.9) {$\vdots$};
\node (n3) at (0,-4.5) {};
\node (5) [hasse] at (0,-5) {};
\node (6) [hasse] at (0,-6) {};
\draw (1) edge [] node[label=left:$\mathfrak{e}_8$] {} (2);
\draw (2) edge [] node[label=left:$\mathfrak{d}_{10}$] {} (3);
\draw (3) edge [] node[label=left:$\mathfrak{d}_{12}$] {} (4);
\draw (4)--(n1);
\draw (n3)--(5);
\draw (5) edge [] node[label=left:$\mathfrak{d}_{2m+6}$] {} (6);
\end{tikzpicture}
\ee

The $[1]-SU(2)-SU(2)-[5]$ theory is UV dual to the rank two $E_8$ theory, so it magnetic quiver is given by
\be 
\begin{tikzpicture}[x=1cm,y=.8cm] 
\node (g0) at (-1,0) [gauge,label=below:{1}] {};
\node (g1) at (0,0) [gaugeb,label=below:{2}] {};
\node (g2) at (1,0) [gauge,label=below:{4}] {};
\node (g3) at (2,0) [gauge,label=below:{6}] {};
\node (g4) at (3,0) [gauge,label=below:{8}] {};
\node (g5) at (4,0) [gauge,label=below:{10}] {};
\node (g6) at (5,0) [gauge,label=below:{12}] {};
\node (g7) at (6,0) [gauge,label=below:{8}] {};
\node (g8) at (7,0) [gauge,label=below:{4}] {};
\node (g9) at (5,1) [gauge,label=right:{6}] {};
\draw (g0)--(g1)--(g2)--(g3)--(g4)--(g5)--(g6)--(g7)--(g8);
\draw (g6)--(g9);
\end{tikzpicture}
\ee

\paragraph{GTP for Quivers Without SQCD Duals.}

Most of the $[n_1]-SU(2)^m-[n_2]$ quiver gauge theories have a UV dual SQCD theory. For this reason, their UV brane-webs and associated MQs are captured by the analysis in \cite{Cabrera:2018jxt}. However, there is a subset of quiver theories that have a 5d UV fixed point, that are not UV dual to any other gauge theory. These are the $[0]_{\theta=0}-SU(2)^m-[n]$ theories, which must be treated separately.

For generic $m$ the top descendant is the $[0]_{0}-SU(2)^m-[4]$ theory with polygon (drawn for $m=3$)
\be 
\begin{tikzpicture}[x=.5cm,y=.5cm]
\draw[step=.5cm,gray,very thin] (0,-1) grid (4,3);

\draw[ligne,black] (1,0)--(3,0)--(4,-1)--(4,3)--(3,2)--(0,2)--(1,0); 
\foreach \x in {1,2,3}
\draw[black] (\x,0)--(\x,2);
 \foreach \x in {1,2,3}
    \node[bd] at (\x,0) {};
 \foreach \x in {0,1,...,4}
    \node[bd] at (4,\x-1) {};
    
 \foreach \x in {1,2,3,4}
    \node[bd] at (4-\x,2) {};

    \node at (1.5,2.6) {$m$};
    \node at (1,-.6) {$\bm{v}_0$};
      
\end{tikzpicture}
\ee
and corresponding vectors given by
\be
\bm{u}=\left[ \underbrace{(1,0)}_{m-1}, (1,-1),\underbrace{(0,1)}_{4},(-1,-1),\underbrace{(-1,0)}_{m},(1,-2) \right]\,.
\ee
The non-convexities are $\mathfrak{C}=\{m-1,m+5\}$. They can be removed by the following edge-moves,
\be
\mathfrak{T}=(\mathfrak{M}_{1}^+ \cdots \mathfrak{M}_{m-2}^+ \mathfrak{M}_{m-1}^+)(\mathfrak{M}_{2m+4}^- \cdots \mathfrak{M}_{m+6}^- \mathfrak{M}_{m+5}^-)\,,
\ee
producing a convex GTP given by (drawn for $m=3$)
\be \label{GTP40}
\begin{tikzpicture}[x=.5cm,y=.5cm]
\draw[step=.5cm,gray,very thin] (-1,-3) grid (3,6);

\draw[ligne,black] (0,0)--(3,-3)--(3,6)--(-1,2)--(0,0); 

    \node[bd] at (0,0) {};
    \node[bd] at (-1,2) {};
    
 \foreach \x in {1,2}
    \node[wd] at (\x,-\x) {};
    
 \foreach \x in {0,1,...,9}
    \node[bd] at (3,-3+\x) {};
    
 \foreach \x in {1,2,3}
    \node[wd] at (3-\x,6-\x) {};

    \node at (1,-2) {$m$};
    \node at (5,1.5) {$2m+3$};
    \node at (-.5,4.5) {$m+1$};
      
\end{tikzpicture}
\ee
\be
\mathfrak{T}\bm{u}=\left[ (m,-m),\underbrace{(0,1)}_{2m+3},(-m-1,-m-1),(1,-2) \right]\,.
\ee

In the case of rank $m=2$, there is an additional quiver theory with no SQCD dual, that descends from the E-string, namely $[0]_{0}-SU(2)^m-[5]$ with polygon
\be 
\begin{tikzpicture}[x=.5cm,y=.5cm]
\draw[step=.5cm,gray,very thin] (0,-1) grid (3,4);

\draw[ligne,black] (1,0)--(2,0)--(3,-1)--(3,4)--(2,2)--(0,2)--(1,0); 
\foreach \x in {0,1}
    \node[bd] at (1+\x,0) {};
 \foreach \x in {0,1,...,5}
    \node[bd] at (3,\x-1) {};    
 \foreach \x in {0,1,2}
    \node[bd] at (2-\x,2) {};
      
     \node at (1,-.6) {$\bm{v}_0$};

\end{tikzpicture}
\ee
and vectors 
\be 
\bm{u}=\left[ (1,0),(1,-1),\underbrace{(0,1)}_5,(-1,-2),\underbrace{(-1,0)}_2,(1,-2) \right]\,.
\ee
The non-convex edges are $\mathfrak{C}=\{1,8\}$. By applying the string of edge-moves given by
\be 
\mathfrak{T}=(\mathfrak{M}_4^+ \cdots \mathfrak{M}_8^+ \mathfrak{M}_9^+)(\mathfrak{M}_3^- \cdots \mathfrak{M}_6^- \mathfrak{M}_7^-)(\mathfrak{M}_6^+ \cdots \mathfrak{M}_3^+ \mathfrak{M}_2^+)(\mathfrak{M}_3^- \cdots \mathfrak{M}_7^- \mathfrak{M}_8^-)\mathfrak{M}_1^+\,,
\ee
we obtain the transformed vector
\be 
\mathfrak{T}\bm{u}=\left[\underbrace{(2,-2)}_2,(14,42),(-37,0),\underbrace{(3,-6)}_6,(1,-2)  \right]\,.
\ee

\paragraph{Magnetic Quiver and Hasse Diagram for Quivers Without SQCD Dual.}

The UV magnetic quiver computed from the GTP in \eqref{GTP40} of $[0]_{0}-SU(2)^m-[4]$ is
\be
\begin{tikzpicture}[x=.8cm,y=.8cm]
\node (g1) at (0,0) [gauge,label=below:{1}] {};
\node (g2) at (1,0) {$\cdots$};
\node (g3) at (2,0) [gauge,label=below:{$2m$}] {};
\node (g4) at (3.5,0) [gauge,label=below:{$2m$}] {};
\node (g5) at (4.5,0) [gauge,label=below:{$\phantom{1}m$}] {};
\node (g6) at (2,1) [gaugeb,label=left:{1}] {};
\node (g7) at (3.5,1) [gauge,label=right:{$m\phantom{1}$}] {};
\draw (g1)--(g2)--(g3)--(g4)--(g5);
\draw (g3)--(g6);
\draw (g4)--(g7);
\end{tikzpicture}
\ee
The corresponding Hasse diagram is
\be
\begin{tikzpicture}
\node (1) [hasse] at (0,0) {};
\node (2) [hasse] at (0,-1) {};
\node (3) [hasse] at (0,-2) {};
\node (n1) at (0,-2.5) {};
\node (n2) at (0,-2.9) {$\vdots$};
\node (n3) at (0,-3.5) {};
\node (4) [hasse] at (0,-4) {};
\node (5) [hasse] at (0,-5) {};
\draw (1) edge [] node[label=left:$\mathfrak{d}_5$] {} (2);
\draw (2) edge [] node[label=left:$\mathfrak{d}_{7}$] {} (3);
\draw (3)--(n1);
\draw (n3)--(4);
\draw (4) edge [] node[label=left:$\mathfrak{d}_{2m+3}$] {} (5);
\end{tikzpicture}
\ee
From this we see that the theory has $\so(4m+6)$ flavor symmetry.

We find the UV magnetic quiver for $[0]_{0}-SU(2)^m-[5]$ to be
\be 
\begin{tikzpicture}[x=1cm,y=.8cm] 
\node (g1) at (0,0) [gauge,label=below:{2}] {};
\node (g2) at (1,0) [gauge,label=below:{4}] {};
\node (g3) at (2,0) [gauge,label=below:{6}] {};
\node (g4) at (3,0) [gauge,label=below:{8}] {};
\node (g5) at (4,0) [gauge,label=below:{10}] {};
\node (g6) at (5,0) [gauge,label=below:{7}] {};
\node (g7) at (6,0) [gauge,label=below:{4}] {};
\node (g8) at (7,0) [gaugeb,label=below:{1}] {};
\node (g9) at (4,1) [gauge,label=right:{5}] {};
\draw (g1)--(g2)--(g3)--(g4)--(g5)--(g6)--(g7)--(g8);
\draw (g5)--(g9);
\end{tikzpicture}
\ee
The balanced nodes give rise to the Dynkin diagram of $\e_8$, and the Hasse diagram is given by
\be
\begin{tikzpicture}
\node (1) [hasse] at (0,0) {};
\node (2) [hasse] at (0,-1) {};
\node (3) [hasse] at (0,-2) {};
\draw (1) edge [] node[label=left:$\mathfrak{e}_7$] {} (2);
\draw (2) edge [] node[label=left:$\mathfrak{e}_{8}$] {} (3);
\end{tikzpicture}
\ee


\section{Enhanced Non-Simply-Laced Flavor Symmetry}

\label{sec:NSLQuivers}

In the previous section, we have computed a series of simply-laced unitary magnetic quivers. Furthermore, the tables in appendix \ref{sec:MQTables} include almost entirely simply-laced magnetic quivers. However, it is known that the flavor symmetry of some lower rank SCFTs, more precisely $SU(4)$ and $SU(5)$ with two antisymmetric and various fundamental matter multiplets, have a non-simply laced flavor symmetry. This section is devoted to the analysis of this apparent problem. 

We begin by recalling that certain simply laced quivers can have a Coulomb branch with a non-simply-laced isometry algebra. The quivers which we will use are summarized in table \ref{tab:NSLLeaves}. These quivers were studied in \cite{Gaiotto:2012uq,Cremonesi:2014vla,Mekareeya:2017jgc,Hanany:2018vph,Bourget:2020bxh} in which the computation of their Hilbert series and heighest weight generating functions (HWGs) are performed, thus identifying the corresponding nilpotent orbits and Hasse diagrams. The simplest occurrence of this phenomenon can be traced back to a the observation by Kostant and Brylinski \cite{brylinski1994nilpotent} that a $\mathbb{Z}_2$ quotient of the closure of the minimal nilpotent orbit of $\mathfrak{d}_{n+1}$ yields the closure of the next-to-minimal nilpotent orbit of $\mathfrak{b}_n$, combined with the fact that this quotient can be implemented on the quiver using the \emph{bouquet} construction \cite{Hanany:2018dvd} or \emph{wreathing} \cite{Bourget:2020bxh}. Note that magnetic quivers for the closures of the non-extremal leaves in the diagrams of table \ref{tab:NSLLeaves} are non-simply-laced, possibly with additional identifications (see section 5.2.4 in \cite{Bourget:2020bxh} for a detailed discussion of this subtle point). In the following, we do not represent these non-simply-laced quivers, and we apply the quiver subtraction algorithm using directly the non elementary slices of table \ref{tab:NSLLeaves}. 
 
\begin{table}[t]
\centering
\begin{tabular}{|c|c|c|c|}\hline 
Algebra & Singularity & Magnetic quiver & Hasse diagram \\
\hline\hline 
\begin{tabular}{l}
$\mathfrak{b}_n$
\end{tabular}
&
\begin{tabular}{l}
$\overline{\mathcal{O}}_{\mathrm{nmin}} (\mathfrak{so}(2n+1,\mathbb{C}))$
\end{tabular}
&
\begin{tabular}{l}
\begin{tikzpicture}[x=.8cm,y=.8cm]
\tikzset{every loop/.style={}}
\node (g1) at (0,0) [gauge,label=left:{2}] {};
\node (g2) at (1,0) [gauge,label=above:{2}] {};
\node (g3) at (2,0) [] {$\cdots$};
\node (g4) at (3,0) [gauge,label=above left:{2}] {};
\node (g5) at (4,0) [gauge,label=right:{1}] {};
\node (g7) at (3,1) [gauge,label=right:{1}] {};

\draw (g1)--(g2)--(g3)--(g4)--(g5);
\draw (g4)--(g7);
\path (g1) edge [loop] node {} (g1);
\draw[decoration={calligraphic brace,mirror,amplitude=5pt,}, decorate, line width=1.25pt] ($(g2)+(-.1,-.4)$) -- ($(g4)+(.1,-.4)$);
\node at (2,-1) [] {$n-2$};
\end{tikzpicture}
\end{tabular}
& 
\begin{tabular}{l}
\begin{tikzpicture}
\node (1) [hasse,label=right:$2n-1$] at (0,0) {};
\node (2) [hasse,label=right:$2n-2$] at (0,-1) {};
\node (3) [hasse] at (0,-2) {};
\draw (1) edge [] node[label=left:$\mathfrak{a}_1$] {} (2);
\draw (2) edge [] node[label=left:$\mathfrak{b}_n$] {} (3);
\end{tikzpicture}
\end{tabular}
\\
\hline
\begin{tabular}{l}
$\mathfrak{c}_2$
\end{tabular}
&
\begin{tabular}{l}
$\overline{\mathcal{O}}_{\mathrm{nmin}} (\mathfrak{sp}(2,\mathbb{C}))$
\end{tabular}
&
\begin{tabular}{l}
\begin{tikzpicture}[x=.8cm,y=.8cm]
	\tikzset{every loop/.style={}}
	\node (g1) at (0,0) [gauge,label=below:\large{1}] {};
	\node (g2) at (1,0) [gauge,label=below:\large{2}] {};
	\node (g3) at (2,0) [gauge,label=below:\large{1}] {};
	\draw (g1)--(g2)--(g3);
	\path (g2) edge [loop] node {} (g2);
\end{tikzpicture}
\end{tabular}
& 
\begin{tabular}{l}
\begin{tikzpicture}
\node (1) [hasse,label=right:3] at (0,0) {};
\node (2) [hasse,label=right:2] at (0,-1) {};
\node (3) [hasse] at (0,-2) {};
\draw (1) edge [] node[label=left:$\mathfrak{a}_1$] {} (2);
\draw (2) edge [] node[label=left:$\mathfrak{c}_2$] {} (3);
\end{tikzpicture}
\end{tabular}
\\
\hline
\begin{tabular}{l}
$\mathfrak{g}_2$
\end{tabular}
&
\begin{tabular}{l}
$\overline{\mathcal{O}}_{\mathrm{subreg}} (\mathfrak{g}_2)$
\end{tabular}
&
\begin{tabular}{l}
\begin{tikzpicture}[x=.8cm,y=.8cm]
\tikzset{every loop/.style={}}
\node (h0) at (-1,0) [gauge,label=below:\large{3}] {};
\node (h1) at (0,0) [gauge,label=below:\large{2}] {};
\node (h2) at (1,0) [gauge,label=below:\large{1}] {};
\draw (h0)--(h1)--(h2);
\path (h0) edge [loop] node {} (h0);
\end{tikzpicture}
\end{tabular}
& 
\begin{tabular}{l}
\begin{tikzpicture}
\node (1) [hasse,label=right:5] at (0,0) {};
\node (2) [hasse,label=right:4] at (0,-1) {};
\node (3) [hasse,label=right:3] at (0,-2) {};
\node (4) [hasse] at (0,-3) {};
\draw (1) edge [] node[label=left:$\mathfrak{a}_1$] {} (2);
\draw (2) edge [] node[label=left:$m$] {} (3);
\draw (3) edge [] node[label=left:$\mathfrak{g}_2$] {} (4);
\end{tikzpicture}
\end{tabular}
\\
\hline
\end{tabular}
\caption{Summary of non-simply laced symplectic leaves and their magnetic quivers. The quivers for the closure of the next-to-minimal orbit of $\mathfrak{b}_n$ is obtained from the quiver for the closure of the minimal nilpotent orbit of $\mathfrak{d}_{n+1}$ after $S_2$ wreathing acting on the two spinor nodes, which can be depicted equivalently by an adjoint hyper. The $\mathfrak{c}_2$ quiver is obtained by identification with the $\mathfrak{b}_2$ quiver. Finally the $\mathfrak{g}_2$ quiver is obtained from $S_3$ wreathing of the $\mathfrak{d}_{4}$ closure of the minimal nilpotent orbit quiver. Note that the closures of next-to-minimal nilpotent orbits of the other non-simply laced algebras $\mathfrak{c}_n$ and $\mathfrak{f}_4$ are also available in terms of wreathed quivers, but the wreathing acts on more than $\uu(1)$ nodes, so they can not be rephrased as quivers with adjoint matter.  \label{tab:NSLLeaves}}
\end{table}

We now illustrate on examples how these quivers intervene in the process of the quiver subtraction algorithm \cite{Bourget:2019aer,Bourget:2020mez} and allow to identify non-simply-laced symmetries in the Coulomb branch of simply laced quivers, even with no adjoint matter. The results agree with the conclusions of \cite{Mekareeya:2017jgc}. 

\subsection{\texorpdfstring{$\mathfrak{c}_2$ Quiver}{c2 quiver}}

Let us start with a particularly simple theory, $SU(4)_1 + 2\bm{AS}$. The flavor symmetry is known not to enhance at strong coupling and is given by $\spp(2) \oplus \uu(1)$, where the $\spp(2)$ appears as the antisymmetric representation of $\su(4)$ is real. The magnetic quiver at the SCFT point is conjectured to be 
\be \label{LeafC2}
\begin{tikzpicture}[x=.8cm,y=.8cm]
	\tikzset{every loop/.style={}}
	\node (g1) at (0,0) [gauge,label=below:\large{1}] {};
	\node (g2) at (1,0) [gauge,label=below:\large{2}] {};
	\node (g3) at (2,0) [gauge,label=below:\large{1}] {};
	\draw (g1)--(g2)--(g3);
	\path (g2) edge [loop] node {} (g2);
\end{tikzpicture}
\ee
see table \ref{tab:UVMQs2ASSU4} entry $(0,1)$. This quiver appears in table \ref{tab:NSLLeaves} and provides the $\spp(2)$ symmetry. Note that the $\uu(1)$ factor of the flavor symmetry is invisible, a feature that is common to most low $N_F$ theories.

Let us turn to a more involved theory. Consider the strongly coupled $SU(4)_0 +2\bm{AS} + 6 \bm{F}$ with magnetic quiver
\be
\begin{tikzpicture}[x=.8cm,y=.8cm]
\node (g1) at (0,0) [gauge,label=below:\large{1}] {};
\node (g2) at (1,0) [gaugeb,label=below:\large{2}] {};
\node (g3) at (2,0) [gauge,label=below:\large{4}] {};
\node (g4) at (3,0) [gauge,label=below:\large{6}] {};
\node (g5) at (4,0) [gauge,label=below:\large{8}] {};
\node (g6) at (5,0) [gauge,label=below:\large{6}] {};
\node (g7) at (6,0) [gauge,label=below:\large{4}] {};
\node (g8) at (7,0) [gaugeb,label=below:\large{2}] {};
\node (g9) at (8,0) [gauge,label=below:\large{1}] {};
\node (g10) at (4,1) [gauge,label=right:\large{4}] {};
\draw (g1)--(g2)--(g3)--(g4)--(g5)--(g6)--(g7)--(g8)--(g9);
\draw (g5)--(g10);
\end{tikzpicture}\,.
\ee
The only possible quiver subtraction is with an $\mathfrak{e}_7$ diagram to obtain
\be
\begin{tikzpicture}[x=.8cm,y=.8cm]
\node (g1) at (0,0) [gauge,label=above:\large{1}] {};
\node (g2) at (1,0) [gaugeb,label=above:\large{1}] {};
\node (g3) at (2,0) [gauge,label=above:\large{2}] {};
\node (g4) at (3,0) [gauge,label=above:\large{3}] {};
\node (g5) at (4,0) [gauge,label=below:\large{4}] {};
\node (g6) at (5,0) [gauge,label=above:\large{3}] {};
\node (g7) at (6,0) [gauge,label=above:\large{2}] {};
\node (g8) at (7,0) [gaugeb,label=above:\large{1}] {};
\node (g9) at (8,0) [gauge,label=above:\large{1}] {};
\node (g10) at (4,1) [gauge,label=above:\large{2}] {};
\node (g11) at (4,-1.5) [gauge,label=below:\large{1}] {};
\draw (g1)--(g2)--(g3)--(g4)--(g5)--(g6)--(g7)--(g8)--(g9);
\draw (g5)--(g10);
\draw (g1)--(g11)--(g9);
\end{tikzpicture}\,.
\ee
From here, there are two possibilities. Either we can subtract an $\mathfrak{a}_9$ diagram, leading to the affine $\mathfrak{e}_6$ diagram. Alternatively, we can subtract another $\mathfrak{e}_7$ diagram. It was discussed in \cite{Bourget:2020mez} that subtracting two equivalent quivers leads to the two rebalancing nodes to be identified, together with the coupling to an adjoint. Thus, this subtraction leads precisely to the quiver in \eqref{LeafC2} and we deduce that the Hasse diagram is conjectured to be 
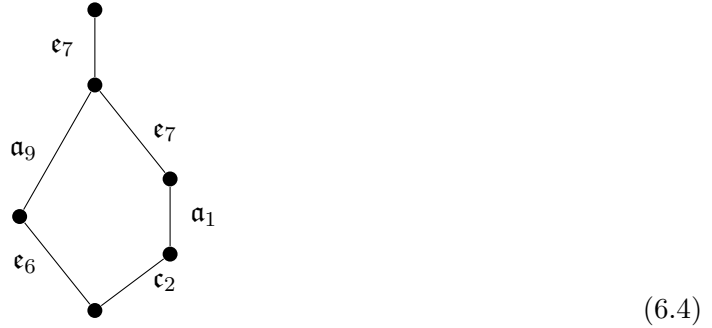
\be
\begin{tikzpicture}
\node (1) [hasse] at (0,0) {};
\node (2) [hasse] at (0,-1) {};
\node (3) [hasse] at (-1,-2.75) {};
\node (4) [hasse] at (1,-2.25) {};
\node (6) [hasse] at (1,-3.25) {};
\node (5) [hasse] at (0,-4) {};
\draw (1) edge [] node[label=left:$\mathfrak{e}_7$] {} (2);
\draw (2) edge [] node[label=left:$\mathfrak{a}_{9}$] {} (3);
\draw (2) edge [] node[label=right:$\mathfrak{e}_{7}$] {} (4);
\draw (3) edge [] node[label=left:$\mathfrak{e}_{6}$] {} (5);
\draw (4) edge [] node[label=right:$\mathfrak{a}_{1}$] {} (6);
\draw (6) edge [] node[label=right:$\mathfrak{c}_{2}$] {} (5);
\end{tikzpicture}
\ee
The flavor symmetry is $\e_6 \oplus \spp(2) \oplus \uu(1)$, as expected.

\subsection{\texorpdfstring{$\mathfrak{b}_9$ Quiver}{b9 quiver}}
We now check two other interesting theories, the strongly coupled $SU(N)_2 + 2\bm{AS} + 6 \bm{F}$, i.e. the submarginal theory discussed in section \ref{sec:2AS6F}, for $N=4$ and $N=5$. They have enhanced flavor symmetries $\so(19)$ and $\e_7\oplus \mathfrak{g}_2$ respectively. The magnetic quiver of the former is given by
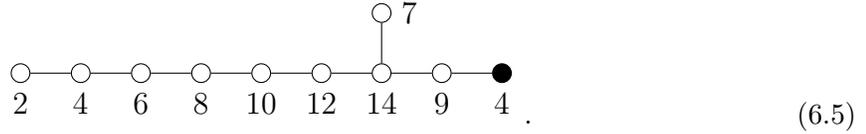
\be
\begin{tikzpicture}[x=.8cm,y=.8cm]
\node (g1) at (0,0) [gauge,label=below:\large{2}] {};
\node (g2) at (1,0) [gauge,label=below:\large{4}] {};
\node (g3) at (2,0) [gauge,label=below:\large{6}] {};
\node (g4) at (3,0) [gauge,label=below:\large{8}] {};
\node (g5) at (4,0) [gauge,label=below:\large{10}] {};
\node (g6) at (5,0) [gauge,label=below:\large{12}] {};
\node (g7) at (6,0) [gauge,label=below:\large{14}] {};
\node (g8) at (7,0) [gauge,label=below:\large{9}] {};
\node (g9) at (8,0) [gaugeb,label=below:\large{4}] {};
\node (g10) at (6,1) [gauge,label=right:\large{7}] {};
\draw (g1)--(g2)--(g3)--(g4)--(g5)--(g6)--(g7)--(g8)--(g9);
\draw (g7)--(g10);
\end{tikzpicture}\,.
\ee
We see that we can subtract two equivalent $\mathfrak{e}_8$ diagrams, leading to
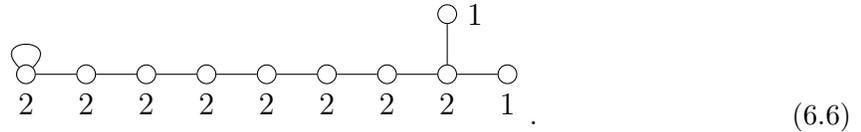
\be\label{LeafB9}
\begin{tikzpicture}[x=.8cm,y=.8cm]
\tikzset{every loop/.style={}}
\node (g0) at (-1,0) [gauge,label=below:\large{2}] {};
\node (g1) at (0,0) [gauge,label=below:\large{2}] {};
\node (g2) at (1,0) [gauge,label=below:\large{2}] {};
\node (g3) at (2,0) [gauge,label=below:\large{2}] {};
\node (g4) at (3,0) [gauge,label=below:\large{2}] {};
\node (g5) at (4,0) [gauge,label=below:\large{2}] {};
\node (g6) at (5,0) [gauge,label=below:\large{2}] {};
\node (g7) at (6,0) [gauge,label=below:\large{2}] {};
\node (g8) at (7,0) [gauge,label=below:\large{1}] {};
\node (g10) at (6,1) [gauge,label=right:\large{1}] {};
\draw (g0)--(g1)--(g2)--(g3)--(g4)--(g5)--(g6)--(g7)--(g8);
\draw (g7)--(g10);
\path (g0) edge [loop] node {} (g0);
\end{tikzpicture}\, . 
\ee
The Hasse diagram of the $SU(4)_2 + 2\bm{AS} + 6\bm{F}$ is then conjectured to be 
\be
\begin{tikzpicture}
\node (1) [hasse] at (0,0) {};
\node (2) [hasse] at (0,-1) {};
\node (3) [hasse] at (0,-2) {};
\node (4) [hasse] at (0,-3) {};
\node (5) [hasse] at (0,-4) {};
\draw (1) edge [] node[label=left:$\mathfrak{e}_8$] {} (2);
\draw (2) edge [] node[label=left:$\mathfrak{e}_8$] {} (3);
\draw (3) edge [] node[label=left:$\mathfrak{a}_1$] {} (4);
\draw (4) edge [] node[label=left:$\mathfrak{b}_9$] {} (5);
\end{tikzpicture}
\ee

\subsection{\texorpdfstring{$\mathfrak{g}_2$ Quiver}{g2 quiver}}
Finally, consider the magnetic quiver
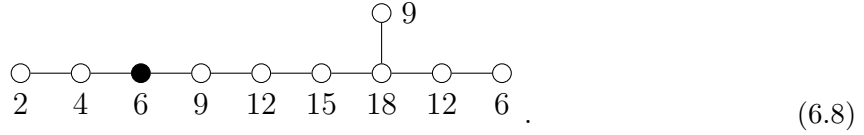
\be
\begin{tikzpicture}[x=.8cm,y=.8cm]
\node (h1) at (0,0) [gauge,label=below:\large{2}] {};
\node (h2) at (1,0) [gauge,label=below:\large{4}] {};
\node (h3) at (2,0) [gaugeb,label=below:\large{6}] {};
\node (h4) at (3,0) [gauge,label=below:\large{9}] {};
\node (h5) at (4,0) [gauge,label=below:\large{12}] {};
\node (h6) at (5,0) [gauge,label=below:\large{15}] {};
\node (h7) at (6,0) [gauge,label=below:\large{18}] {};
\node (h8) at (7,0) [gauge,label=below:\large{12}] {};
\node (h9) at (8,0) [gauge,label=below:\large{6}] {};
\node (h10) at (6,1) [gauge,label=right:\large{9}] {};
\draw (h1)--(h2)--(h3)--(h4)--(h5)--(h6)--(h7)--(h8)--(h9);
\draw (h7)--(h10);
\end{tikzpicture}\,.
\ee
After subtracting two $\mathfrak{e}_8$ diagrams this becomes
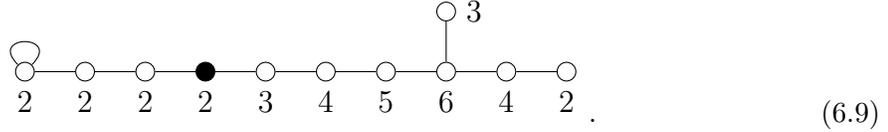
\be
\begin{tikzpicture}[x=.8cm,y=.8cm]
\tikzset{every loop/.style={}}
\node (h0) at (-1,0) [gauge,label=below:\large{2}] {};
\node (h1) at (0,0) [gauge,label=below:\large{2}] {};
\node (h2) at (1,0) [gauge,label=below:\large{2}] {};
\node (h3) at (2,0) [gaugeb,label=below:\large{2}] {};
\node (h4) at (3,0) [gauge,label=below:\large{3}] {};
\node (h5) at (4,0) [gauge,label=below:\large{4}] {};
\node (h6) at (5,0) [gauge,label=below:\large{5}] {};
\node (h7) at (6,0) [gauge,label=below:\large{6}] {};
\node (h8) at (7,0) [gauge,label=below:\large{4}] {};
\node (h9) at (8,0) [gauge,label=below:\large{2}] {};
\node (h10) at (6,1) [gauge,label=right:\large{3}] {};
\draw (h0)--(h1)--(h2)--(h3)--(h4)--(h5)--(h6)--(h7)--(h8)--(h9);
\draw (h7)--(h10);
\path (h0) edge [loop] node {} (h0);
\end{tikzpicture}\,.
\ee
Comparing with \eqref{LeafB9} this quiver allows for the subtraction of a $\mathfrak{b}_9$ leaf, leading to the affine $\mathfrak{e}_7$ Dynkin diagram. Alternatively, we can subtract a third equivalent copy of the $\mathfrak{e}_8$ leaf, which gives
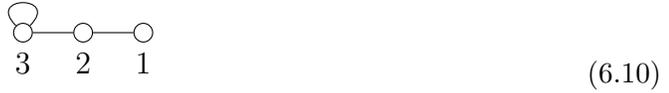
\be
\begin{tikzpicture}[x=.8cm,y=.8cm]
\tikzset{every loop/.style={}}
\node (h0) at (-1,0) [gauge,label=below:\large{3}] {};
\node (h1) at (0,0) [gauge,label=below:\large{2}] {};
\node (h2) at (1,0) [gauge,label=below:\large{1}] {};
\draw (h0)--(h1)--(h2);
\path (h0) edge [loop] node {} (h0);
\end{tikzpicture}
\ee
which we therefore identify with the $\mathfrak{g}_2$ leaf. To summarise, the Hasse diagram of the $SU(5)_2 +2\bm{AS} + 6 \bm{F}$ is conjectured to be 
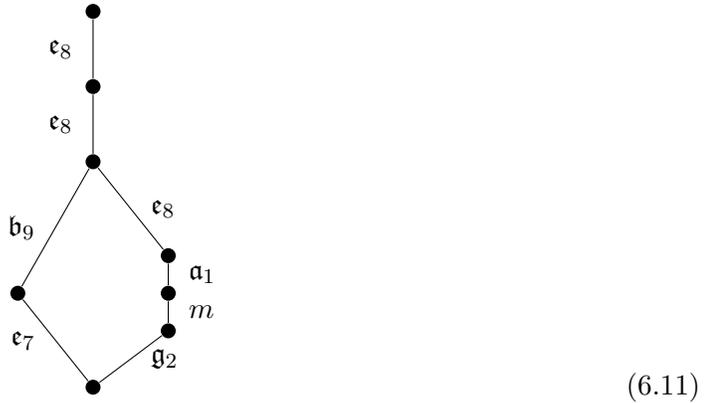
\be
\begin{tikzpicture}
\node (0) [hasse] at (0,1) {};
\node (1) [hasse] at (0,0) {};
\node (2) [hasse] at (0,-1) {};
\node (3) [hasse] at (-1,-2.75) {};
\node (4) [hasse] at (1,-2.25) {};
\node (5) [hasse] at (1,-2.75) {};
\node (6) [hasse] at (1,-3.25) {};
\node (7) [hasse] at (0,-4) {};
\draw (0) edge [] node[label=left:$\mathfrak{e}_8$] {} (1);
\draw (1) edge [] node[label=left:$\mathfrak{e}_8$] {} (2);
\draw (2) edge [] node[label=left:$\mathfrak{b}_{9}$] {} (3);
\draw (2) edge [] node[label=right:$\mathfrak{e}_{8}$] {} (4);
\draw (3) edge [] node[label=left:$\mathfrak{e}_{7}$] {} (7);
\draw (4) edge [] node[label=right:$\mathfrak{a}_1$] {} (5);
\draw (5) edge [] node[label=right:$m$] {} (6);
\draw (6) edge [] node[label=right:$\mathfrak{g}_{2}$] {} (7);
\end{tikzpicture}
\ee

\subsection{\texorpdfstring{Hidden $\su(2)$ Symmetries and E-Strings}{Hidden su(2) Symmetries and E-Strings}}
As a final example let us look at the submarginal $SU(5)_\half +2\bm{AS} + 7 \bm{F}$. It was discussed that his theory has a $\so(16)\oplus \su(2)^2$ flavor symmetry, although one of the $\su(2)$s can not immediately be seen from the magnetic quiver
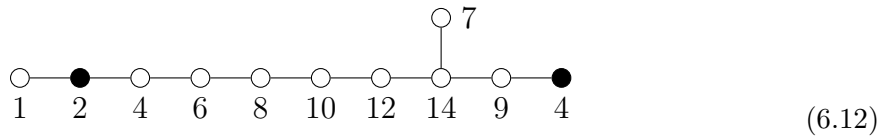
\be
\begin{tikzpicture}[x=.8cm,y=.8cm]
\node (g1) at (1,0) [gauge,label=below:\large{1}] {};
\node (g2) at (2,0) [gaugeb,label=below:\large{2}] {};
\node (g3) at (3,0) [gauge,label=below:\large{4}] {};
\node (g4) at (4,0) [gauge,label=below:\large{6}] {};
\node (g5) at (5,0) [gauge,label=below:\large{8}] {};
\node (g6) at (6,0) [gauge,label=below:\large{10}] {};
\node (g7) at (7,0) [gauge,label=below:\large{12}] {};
\node (g8) at (8,0) [gauge,label=below:\large{14}] {};
\node (g9) at (9,0) [gauge,label=below:\large{9}] {};
\node (g10) at (10,0) [gaugeb,label=below:\large{4}] {};
\node (g11) at (8,1) [gauge,label=right:\large{7}] {};
\draw (g1)--(g2)--(g3)--(g4)--(g5)--(g6)--(g7)--(g8)--(g9)--(g10);
\draw (g8)--(g11);
\end{tikzpicture}
\ee
Note that this enhancement occurs for all theories in tables \ref{tab:UVMQs2ASSU2n} and \ref{tab:UVMQs2ASSU2n+1}, but we consider the case at hand which is the simplest. First, we need to subtract an $\mathfrak{e}_8$ diagram, leading to
\be
\begin{tikzpicture}[x=.8cm,y=.8cm]
\node (g1) at (1,0) [gauge,label=below:\large{1}] {};
\node (g2) at (2,0) [gaugeb,label=below:\large{2}] {};
\node (g3) at (3,0) [gauge,label=below:\large{3}] {};
\node (g4) at (4,0) [gauge,label=below:\large{4}] {};
\node (g5) at (5,0) [gauge,label=below:\large{5}] {};
\node (g6) at (6,0) [gauge,label=below:\large{6}] {};
\node (g7) at (7,0) [gauge,label=below:\large{7}] {};
\node (g8) at (8,0) [gauge,label=below:\large{8}] {};
\node (g9) at (9,0) [gauge,label=below:\large{5}] {};
\node (g10) at (10,0) [gaugeb,label=below:\large{2}] {};
\node (g11) at (8,1) [gauge,label=right:\large{4}] {};
\node (g12) at (2,1) [gauge,label=right:\large{1}] {};
\draw (g1)--(g2)--(g3)--(g4)--(g5)--(g6)--(g7)--(g8)--(g9)--(g10);
\draw (g8)--(g11);
\draw (g2)--(g12);
\end{tikzpicture}
\ee
Subtracting the $\mathfrak{d}_{10}$ diagram leads us to a simply-laced diagram, from which an $\so(16) \oplus \su(2)$ flavor symmetry can be read off. The more interesting route is subtracting a second equivalent $\mathfrak{e}_8$ diagram
\be
\begin{tikzpicture}[x=.8cm,y=.8cm]
\tikzset{every loop/.style={}}
\node (g1) at (1,0) [gauge,label=below:\large{1}] {};
\node (g2) at (2,0) [gaugeb,label=below:\large{2}] {};
\node (g3) at (3,0) [gauge,label=below:\large{2}] {};
\node (g4) at (4,0) [gauge,label=below:\large{2}] {};
\node (g5) at (5,0) [gauge,label=below:\large{2}] {};
\node (g6) at (6,0) [gauge,label=below:\large{2}] {};
\node (g7) at (7,0) [gauge,label=below:\large{2}] {};
\node (g8) at (8,0) [gauge,label=below:\large{2}] {};
\node (g9) at (9,0) [gauge,label=below:\large{1}] {};
\node (g11) at (8,1) [gauge,label=right:\large{1}] {};
\node (g12) at (2,1) [gauge,label=right:\large{2}] {};
\draw (g1)--(g2)--(g3)--(g4)--(g5)--(g6)--(g7)--(g8)--(g9);
\draw (g8)--(g11);
\draw (g2)--(g12);
\path (g12) edge [loop] node {} (g12);
\end{tikzpicture}
\ee
We can now subtract either the $\mathfrak{b}_9$ diagram of table \ref{tab:NSLLeaves} or a $\mathfrak{d}_{10}$ diagram, each leading to an $\su(2)$. 
Thus, the full Hasse diagram is conjectured to be 
\be
\begin{tikzpicture}
\node (0) [hasse] at (0,1) {};
\node (1) [hasse] at (0,0) {};
\node (2) [hasse] at (1,-1) {};
\node (3) [hasse] at (2,-2) {};
\node (4) [hasse] at (-1.5,-1.5) {};
\node (5) [hasse] at (-0.5,-3) {};
\node (666) [hasse] at (-1.5,-2.5) {};
\node (6) [hasse] at (-1.5,-3.5) {};
\node (7) [hasse] at (-0.5,-4.5) {};

\draw (0) edge [] node[label=left:$\mathfrak{e}_8$] {} (1);
\draw (1) edge [] node[label=right:$\mathfrak{d}_{10}$] {} (2);
\draw (2) edge [] node[label=right:$\mathfrak{e}_7$] {} (3);
\draw (1) edge [] node[label=left:$\mathfrak{e}_8$] {} (4);
\draw (4) edge [] node[label=right:$\mathfrak{d}_{10}$] {} (5);
\draw (2) edge [] node[label=right:$\mathfrak{e}_{8}$] {} (5);
\draw (4) edge [] node[label=left:$\mathfrak{a}_1$] {} (666);
\draw (666) edge [] node[label=left:$\mathfrak{b}_9$] {} (6);
\draw (3) edge [] node[label=right:$\mathfrak{d}_8$] {} (7);
\draw (5) edge [] node[label=right:$\mathfrak{a}_1$] {} (7);
\draw (6) edge [] node[label=left:$\mathfrak{a}_1$] {} (7);

\end{tikzpicture}
\ee
unveiling the full $\so(16) \oplus \su(2) \oplus \su(2)$ flavor symmetry.

\paragraph{A remark about the higher rank E-string}
We conclude this section by noticing that for certain theories, as far as we are aware, the Higgs branch Hasse diagram is still not known, but a ``hidden'' $\su(2)$ flavor symmetry can be tracked down. This is the case for the higher rank E-string theories, which are described by the Lagrangian content $SU(N+1)_{(N+N_F - 7)/2} + 1 \mathbf{AS} + N_F \mathbf{F}$ for $ N_F \leq 7$. For simplicity, consider the case $3 \leq N_F \leq 7$ so that the Higgs branch is described by a single magnetic quiver, which is schematically 
\be
 \begin{tikzpicture}[x=.8cm,y=.8cm]
\node (g1) at (1,0) [gauge,label=below:\large{1}] {};
\node (g2) at (4,0) [draw] {$N$ affine $E_{N_F +1}$};
\draw (g1)--(g2);
\end{tikzpicture}
\ee 
The flavor symmetry is expected to be $\e_{N_F +1} \oplus \su(2)$, but the $\su(2)$ factor is not visible from the balanced nodes (for generic $N$). 
A naive implementation of the quiver subtraction algorithm would lead among other quivers to 
\be
\label{quivSymProd}
 \begin{tikzpicture}[x=.8cm,y=.8cm]
\tikzset{every loop/.style={}}
\node (g1) at (1,0) [gauge,label=below:\large{1}] {};
\node (g2) at (2,0) [gauge,label=below:\large{$N$}] {};
\draw (g1)--(g2);
\path (g2) edge [loop] node {} (g2);
\end{tikzpicture}
\ee 
The Coulomb branch of this quiver is known to be the moduli space of $N$ $SU(1)$ instantons on $\mathbb{C}^2$ \cite{Cremonesi:2014xha}. $SU(1)$ instantons are simply particles, and the $N$ particles are indistinguishable, so this space is simply a symmetric product $\mathbb{H}^N/S_N$; it contains a trivial factor $\mathbb{H}$ corresponding to the center of mass, and a singular space of quaternionic dimension $N-1$ whose Hasse diagram can not, to the best of our knowledge, be computed using quiver techniques available at the moment. Note however that the flavor symmetry of the Coulomb branch of \eqref{quivSymProd} is readily identified (for instance from a Hilbert series computation) to be $\su(2) \oplus \su(2)$, where the first factor comes from the center of mass $\mathbb{H}$ contribution, and the second factor comes from the singularity, and is conjecturally the $\su(2)$ factor in the flavor symmetry of the original quiver.


\subsubsection*{Acknowledgements}
We thank Lakshya Bhardwaj, Simone Giacomelli, Julius Grimminger, Amihay Hanany, Hirotaka Hayashi, Diego Rodriguez-Gomez, Marcus Sperling, Futoshi Yagi and Zhenghao Zhong for many illuminating discussions. The work of SSN and JE, and in part MvB is supported by the European Union's Horizon 2020 Framework with the ERC
Consolidator Grant number 682608 ``Higgs bundles: Supersymmetric Gauge Theories and
Geometry (HIGGSBNDL)". The work of AB is supported by the STFC Consolidated Grants ST/P000762/1 and ST/T000791/1. 
SSN acknowledges support also from the Simons Foundation.

\clearpage
\appendix

\section{Magnetic Quivers for all Rank 2 Theories}
\label{sec:rank2}

In this appendix we return to rank 2 5d SCFTs, and provide a comprehensive decoupling tree for these models. The gauge theory descriptions, flavor symmetries and web trees were previously determined in \cite{Jefferson:2018irk, Apruzzi:2019opn,Apruzzi:2019enx, Hayashi:2018lyv}. 
Here we provide the one missing piece, namely the Higgs branch data: the entries in our decoupling trees are the magnetic quivers, and associated flavor symmetry algebras. 
Recall that the rank 2 theories fall into descendants of one of the following marginal theories (we list the 6d SCFTs)
\begin{enumerate}
\item $(D_5, D_5)$ conformal matter: $Sp(2) + {10}\bm{F}  \  \longleftrightarrow\  [4] -SU(2) - SU(2)- [4] $
\item  Rank 2 E-string: $Sp(2) + 1 \bm{AS} + 8 \bm{F}$
\item Model 3: $Sp(2) + 2 \bm{AS}  + 4 \bm{F} \  \longleftrightarrow\  G_2 + 6 \bm{F} $
\item Model 4:  $Sp(2)_0 + 3 \bm{AS}$ 
\end{enumerate}
Note that all descendants of model 4 are covered by the other theories, so we do not discuss this separately. 
In the tables we list the model numbers in red numerals, in agreement with the enumeration of rank 2 models in 
\cite{Apruzzi:2019opn}, as well as the flavor symmetry algebra.

\newcommand{\mo}[1]{%
    \IfEqCase{#1}{%
{1}{
\begin{tabular}{c}
\textcolor{red}{1}  \\ MARGINAL \\$(D_{10},I_1)$ \\ $Sp(2)+10F$ \\ $[4]-SU(2)-SU(2)-[4]$ 
	\end{tabular}}%
{2}{
\begin{tabular}{c}
\textcolor{red}{2} \\ \raisebox{-.5\height}{ \begin{tikzpicture}[x=.8cm,y=.8cm]
	\node (g1) at (0,0) [gauge,label=below:{1}] {};
	\node (g2) at (1,0) {$\cdots$};
	\node (g3) at (2,0) [gauge,label=below:{8}] {};
	\node (g4) at (3,0) [gauge,label=below:{5}] {};
	\node (g5) at (4,0) [gaugeb,label=below:{2}] {};
	\node (g6) at (2,1) [gauge,label=left:{4}] {};
	\draw (g1)--(g2)--(g3)--(g4)--(g5);
	\draw (g3)--(g6);
	\end{tikzpicture}} \\ $\so(20)$
	\end{tabular}
}	
{3}{
\begin{tabular}{c}
\textcolor{red}{3} \\ 
MARGINAL \\ $(E_{8},SU(2))$ \\ $Sp(2)+AS+8F$ 
\end{tabular}
}
{4}{
\begin{tabular}{c}
\textcolor{red}{4} \\ \raisebox{-.5\height}{\begin{tikzpicture}[x=.8cm,y=.8cm]
	\node (g1) at (0,0) [gauge,label=below:{1}] {};
	\node (g2) at (1,0) {$\cdots$};
	\node (g3) at (2,0) [gauge,label=below:{5}] {};
	\node (g4) at (3,0) {$\cdots$};
	\node (g5) at (4,0) [gauge,label=below:{1}] {};
	\node (g6) at (2,1) [gaugeb,label=left:{2}] {};
	\draw (g1)--(g2)--(g3)--(g4)--(g5);
	\draw (g3)--(g6);
	\end{tikzpicture}} \\ $\su(10)$
	\end{tabular}
}
{5}{
\begin{tabular}{c}
\textcolor{red}{5} \\ \raisebox{-.5\height}{\begin{tikzpicture}[x=.8cm,y=.8cm]
	\node (g1) at (0,0) [gauge,label=below:{1}] {};
	\node (g2) at (1,0) {$\cdots$};
	\node (g3) at (2,0) [gauge,label=below:{6}] {};
	\node (g4) at (3,0) [gauge,label=below:{4}] {};
	\node (g5) at (4,0) [gaugeb,label=below:{2}] {};
	\node (g7) at (5,0) [gauge,label=below:{1}] {};
	\node (g6) at (2,1) [gauge,label=left:{3}] {};
	\draw (g1)--(g2)--(g3)--(g4)--(g5)--(g7);
	\draw (g3)--(g6);
	\end{tikzpicture}} \\ $\so(16) \oplus \su(2)$
		\end{tabular}
}
{6}{
\begin{tabular}{c}
\textcolor{red}{6} \\ \raisebox{-.5\height}{\begin{tikzpicture}[x=.8cm,y=.8cm]
	\node (g1) at (0,0) [gauge,label=below:{1}] {};
	\node (g2) at (1,0) [gaugeb,label=below:{2}] {};
	\node (g3) at (2,0) [gauge,label=below:{4}] {};
	\node (g4) at (3,0) [gauge,label=below:{6}] {};
	\node (g5) at (4,0) [gauge,label=below:{8}] {};
	\node (g6) at (5,0) [gauge,label=below:{10}] {};
	\node (g7) at (6,0) [gauge,label=below:{12}] {};
	\node (g8) at (7,0) [gauge,label=below:{8}] {};
	\node (g9) at (8,0) [gauge,label=below:{4}] {};
	\node (g10) at (6,1) [gauge,label=left:{6}] {};
	\draw (g1)--(g2)--(g3)--(g4)--(g5)--(g6)--(g7)--(g8)--(g9);
	\draw (g7)--(g10);
	\end{tikzpicture}}  \\ $\e_8 \oplus \su(2)$
		\end{tabular}
}
{7}{
\begin{tabular}{c}
\textcolor{red}{7} \\ \raisebox{-.5\height}{\begin{tikzpicture}[x=.8cm,y=.8cm]
	\node (g1) at (0,0) [gauge,label=below:{2}] {};
	\node (g2) at (1,0) [gauge,label=below:{4}] {};
	\node (g3) at (2,0) [gauge,label=below:{6}] {};
	\node (g4) at (3,0) [gauge,label=below:{8}] {};
	\node (g5) at (4,0) [gauge,label=below:{10}] {};
	\node (g6) at (5,0) [gauge,label=below:{7}] {};
	\node (g7) at (6,0) [gauge,label=below:{4}] {};
	\node (g8) at (7,0) [gaugeb,label=below:{1}] {};
	\node (g10) at (4,1) [gauge,label=left:{5}] {};
	\draw (g1)--(g2)--(g3)--(g4)--(g5)--(g6)--(g7)--(g8);
	\draw (g5)--(g10);
	\end{tikzpicture}}  \\ $\e_8$
		\end{tabular}
}
{8}{
\begin{tabular}{c}
\textcolor{red}{8} \\ \raisebox{-.5\height}{\begin{tikzpicture}[x=.8cm,y=.8cm]
	\node (g1) at (0,0) [gauge,label=below:{1}] {};
	\node (g2) at (1,0) {$\cdots$};
	\node (g3) at (2,0) [gauge,label=below:{4}] {};
	\node (g4) at (3,0) {$\cdots$};
	\node (g5) at (4,0) [gauge,label=below:{1}] {};
	\node (g6) at (2,1) [gaugeb,label=left:{2}] {};
	\node (g7) at (3,1) [gauge,label=right:{1}] {};
	\draw (g1)--(g2)--(g3)--(g4)--(g5);
	\draw (g3)--(g6)--(g7);
	\end{tikzpicture}} \\ $\su(8) \oplus \su(2)$
		\end{tabular}
}
{9}{
\begin{tabular}{c}
\textcolor{red}{9} \\ \raisebox{-.5\height}{\begin{tikzpicture}[x=.8cm,y=.8cm]
	\node (g1) at (0,0) [gauge,label=below:{1}] {};
	\node (g2) at (1,0) {$\cdots$};
	\node (g3) at (2,0) [gauge,label=below:{5}] {};
	\node (g4) at (3,0) [gauge,label=below:{3}] {};
	\node (g5) at (4,0) [gaugeb,label=below:{1}] {};
	\node (g6) at (2,1) [gauge,label=left:{3}] {};
	\node (g7) at (3,1) [gaugeb,label=right:{1}] {};
	\draw (g1)--(g2)--(g3)--(g4)--(g5);
	\draw (g3)--(g6)--(g7);
	\end{tikzpicture}} \\ $\so(14) \oplus \uu(1)$
		\end{tabular}
}
{10}{
\begin{tabular}{c}
\textcolor{red}{10} \\  \raisebox{-.5\height}{\begin{tikzpicture}[x=.8cm,y=.8cm]
	\node (g1) at (0,0) [gauge,label=below:{1}] {};
	\node (g2) at (1,0) [gaugeb,label=below:{2}] {};
	\node (g3) at (2,0) [gauge,label=below:{4}] {};
	\node (g4) at (3,0) [gauge,label=below:{6}] {};
	\node (g5) at (4,0) [gauge,label=below:{8}] {};
	\node (g6) at (5,0) [gauge,label=below:{6}] {};
	\node (g7) at (6,0) [gauge,label=below:{4}] {};
	\node (g8) at (7,0) [gauge,label=below:{2}] {};
	\node (g10) at (4,1) [gauge,label=left:{4}] {};
	\draw (g1)--(g2)--(g3)--(g4)--(g5)--(g6)--(g7)--(g8);
	\draw (g5)--(g10);
	\end{tikzpicture}} \\ $\e_7 \oplus \su(2)$
		\end{tabular}
}
{11}{\begin{tabular}{c}
\textcolor{red}{11} \\  \raisebox{-.5\height}{\begin{tikzpicture}[x=.8cm,y=.8cm]
	\node (g1) at (0,0) [gauge,label=below:{1}] {};
	\node (g2) at (1,0) [gauge,label=below:{2}] {};
	\node (g3) at (2,0) [gauge,label=below:{3}] {};
	\node (g4) at (3,0) [gauge,label=below:{4}] {};
	\node (g5) at (4,0) [gauge,label=below:{4}] {};
	\node (g6) at (5,0) [gauge,label=below:{2}] {};
	\node (g10) at (3,1) [gaugeb,label=left:{1}] {};
	\node (g11) at (4,1) [gauge,label=right:{2}] {};
	\draw (g1)--(g2)--(g3)--(g4)--(g5)--(g6);
	\draw (g4)--(g10);
	\draw (g5)--(g11);
	\end{tikzpicture}}  \\ $\so(14)$
		\end{tabular}
}
{12}{\begin{tabular}{c}
\textcolor{red}{12} \\  \raisebox{-.5\height}{\begin{tikzpicture}[x=.8cm,y=.8cm]
	\node (g1) at (0,0) [gauge,label=below:{1}] {};
	\node (g2) at (1,0) [gauge,label=below:{2}] {};
	\node (g3) at (2,0) [gauge,label=below:{3}] {};
	\node (g4) at (3,0) [gauge,label=below:{2}] {};
	\node (g5) at (4,0) [gauge,label=below:{1}] {};
	\node (g6) at (1,1) [gauge,label=above:{1}] {};
	\node (g7) at (2,1) [gaugeb,label=above:{2}] {};
	\node (g8) at (3,1) [gauge,label=above:{1}] {};
	\draw (g1)--(g2)--(g3)--(g4)--(g5);
	\draw (g6)--(g7)--(g8);
	\draw (g3)--(g7);
	\end{tikzpicture}} \\ $\su(6) \oplus \su(2)^2$
		\end{tabular}
}
{13}{\begin{tabular}{c}
\textcolor{red}{13} \\  \raisebox{-.5\height}{\begin{tikzpicture}[x=.8cm,y=.8cm]
	\node (g1) at (0,0) [gauge,label=below:{1}] {};
	\node (g2) at (1,0) [gauge,label=below:{2}] {};
	\node (g3) at (2,0) [gauge,label=below:{3}] {};
	\node (g4) at (3,0) [gauge,label=below:{3}] {};
	\node (g5) at (4,0) [gauge,label=below:{2}] {};
	\node (g6) at (5,0) [gauge,label=below:{1}] {};
	\node (g7) at (2,1) [gaugeb,label=left:{1}] {};
	\node (g8) at (3,1) [gaugeb,label=right:{1}] {};
	\draw (g1)--(g2)--(g3)--(g4)--(g5)--(g6);
	\draw (g3)--(g7);
	\draw (g4)--(g8);
	\end{tikzpicture}} \\ $\su(7) \oplus \uu(1)$
		\end{tabular}
}
{14}{\begin{tabular}{c}
\textcolor{red}{14} \\  \raisebox{-.5\height}{\begin{tikzpicture}[x=.8cm,y=.8cm]
	\node (g1) at (0,0) [gauge,label=below:{1}] {};
	\node (g2) at (1,0) {$\cdots$};
	\node (g3) at (2,0) [gauge,label=below:{4}] {};
	\node (g4) at (3,0) [gauge,label=below:{3}] {};
	\node (g5) at (4,0) [gaugeb,label=below:{1}] {};
	\node (g6) at (2,1) [gauge,label=left:{2}] {};
	\node (g7) at (3,1) [gaugeb,label=right:{1}] {};
	\draw (g1)--(g2)--(g3)--(g4)--(g5);
	\draw (g3)--(g6);
	\draw (g4)--(g7);
	\end{tikzpicture}}  \\ $ \so(12) \oplus \uu(1)$
		\end{tabular}
}
{15}{\begin{tabular}{c}
\textcolor{red}{15} \\  \raisebox{-.5\height}{\begin{tikzpicture}[x=.8cm,y=.8cm]
	\node (g1) at (0,0) [gauge,label=below:{1}] {};
	\node (g2) at (1,0) [gaugeb,label=below:{2}] {};
	\node (g3) at (2,0) [gauge,label=below:{4}] {};
	\node (g4) at (3,0) [gauge,label=below:{6}] {};
	\node (g5) at (4,0) [gauge,label=below:{4}] {};
	\node (g6) at (5,0) [gauge,label=below:{2}] {};
	\node (g10) at (3,1) [gauge,label=left:{4}] {};
	\node (g11) at (4,1) [gauge,label=right:{2}] {};
	\draw (g1)--(g2)--(g3)--(g4)--(g5)--(g6);
	\draw (g4)--(g10)--(g11);
	\end{tikzpicture}} \\ $ \e_6 \oplus \su(2)$
		\end{tabular}
}
{16}{\begin{tabular}{c} \textcolor{red}{16}  \\  MARGINAL \\  $Sp(2)+2AS+4F$ \\ $G_2 + 6F$
\end{tabular}
}
{17}{\begin{tabular}{c}
\textcolor{red}{17} \\  \raisebox{-.5\height}{\begin{tikzpicture}[x=.8cm,y=.8cm]
	\node (g1) at (0,0) [gauge,label=below:{1}] {};
	\node (g2) at (1,0) [gauge,label=below:{2}] {};
	\node (g3) at (2,0) [gauge,label=below:{2}] {};
	\node (g4) at (3,0) [gauge,label=below:{2}] {};
	\node (g5) at (4,0) [gauge,label=below:{2}] {};
	\node (g6) at (5,0) [gauge,label=below:{1}] {};
	\node (g10) at (2.5,1) [gaugeb,label=left:{1}] {};
	\draw (g1)--(g2)--(g3)--(g4)--(g5)--(g6);
	\draw (g2)--(g10)--(g5);
	\end{tikzpicture}}  \\ $\su(7)$
		\end{tabular}
}
{18}{\begin{tabular}{c}
\textcolor{red}{18} \\  \raisebox{-.5\height}{\begin{tikzpicture}[x=.8cm,y=.8cm]
	\node (g1) at (0,0) [gauge,label=below:{1}] {};
	\node (g2) at (1,0) [gauge,label=below:{2}] {};
	\node (g3) at (2,0) [gauge,label=below:{2}] {};
	\node (g4) at (3,0) [gauge,label=below:{2}] {};
	\node (g6) at (1,.8) [gaugeb,label=above:{1}] {};
	\node (g7) at (1.5,1.6) [gauge,label=above:{2}] {};
	\node (g8) at (2,.8) [gaugeb,label=above:{1}] {};
	\draw (g1)--(g2)--(g3)--(g4);
	\draw (g2)--(g6)--(g7)--(g8)--(g3);
	\end{tikzpicture}} \\ $\su(5) \oplus \su(2) \oplus \uu(1)$
		\end{tabular}
}
{19}{\begin{tabular}{c}
\textcolor{red}{19} \\  \raisebox{-.5\height}{\begin{tikzpicture}[x=.8cm,y=.8cm]
	\node (g1) at (0,0) [gauge,label=below:{1}] {};
	\node (g2) at (1,0) [gauge,label=below:{2}] {};
	\node (g3) at (2,0) [gauge,label=below:{2}] {};
	\node (g4) at (3,0) [gauge,label=below:{2}] {};
	\node (g5) at (4,0) [gauge,label=below:{1}] {};
	\node (g7) at (1,1) [gaugeb,label=left:{1}] {};
	\node (g8) at (2,1) [gaugeb,label=right:{1}] {};
	\draw (g1)--(g2)--(g3)--(g4)--(g5);
	\draw (g2)--(g7)--(g8)--(g4);
	\end{tikzpicture}} \\ $\su(6) \oplus \uu(1)$
		\end{tabular}
}
{20}{\begin{tabular}{c}
\textcolor{red}{20} \\  \raisebox{-.5\height}{\begin{tikzpicture}[x=.8cm,y=.8cm]
	\node (g1) at (0,0) [gauge,label=below:{1}] {};
	\node (g2) at (1,0) [gauge,label=below:{2}] {};
	\node (g3) at (2,0) [gauge,label=below:{3}] {};
	\node (g4) at (3,0) [gauge,label=below:{2}] {};
	\node (g5) at (4,0) [gaugeb,label=below:{1}] {};
	\node (g6) at (2,1) [gauge,label=left:{2}] {};
	\node (g7) at (3,1) [gaugeb,label=right:{1}] {};
	\draw (g1)--(g2)--(g3)--(g4)--(g5)--(g7)--(g6)--(g3);
	\end{tikzpicture}}   \\ $\so(10) \oplus \uu(1)$
		\end{tabular}
}
{21}{\begin{tabular}{c}
\textcolor{red}{21} \\  \raisebox{-.5\height}{\begin{tikzpicture}[x=.8cm,y=.8cm]
	\node (g1) at (0,0) [gauge,label=below:{1}] {};
	\node (g2) at (1,0) [gaugeb,label=below:{2}] {};
	\node (g3) at (2,0) [gauge,label=below:{4}] {};
	\node (g4) at (3,0) [gauge,label=below:{4}] {};
	\node (g5) at (4,0) [gauge,label=below:{2}] {};
	\node (g10) at (2,1) [gauge,label=left:{2}] {};
	\node (g11) at (3,1) [gauge,label=right:{2}] {};
	\draw (g1)--(g2)--(g3)--(g4)--(g5);
	\draw (g3)--(g10);
	\draw (g4)--(g11);
	\end{tikzpicture}} \\ $\so(10) \oplus \su(2)$
		\end{tabular}
}
{22}{\textcolor{red}{22} \begin{tabular}{c}   \raisebox{-.5\height}{ \begin{tikzpicture}[x=.8cm,y=.8cm,decoration={
    markings,
    mark=at position 0.5 with {\arrow{>},arrowhead=1cm}}]
	\node (g1) at (0,0) [gauge,label=below:{1}] {};
	\node (g2) at (1,0) {$\cdots$};
	\node (g3) at (2,0) [gauge,label=below:{5}] {};
	\node (g4) at (3,0) [gauge,label=below:{6}] {};
	\node (g5) at (4,0) [gaugeb,label=below:{2}] {};
	\draw (g1)--(g2)--(g3);
	\draw (g5)--(g4);
	\draw[doublearrow] (g3)-- node {\midarrowrev} (g4); 
	\end{tikzpicture}} \\ $\spp(6)$
	\end{tabular}
}
{23}{
\begin{tabular}{c}
\textcolor{red}{23} \\  \raisebox{-.5\height}{\begin{tikzpicture}[x=.8cm,y=.8cm]
	\node (g1) at (0,0) [gauge,label=below:{1}] {};
	\node (g2) at (1,0) [gauge,label=below:{2}] {};
	\node (g3) at (2,0) [gauge,label=below:{2}] {};
	\node (g4) at (3,0) [gauge,label=below:{1}] {};
	\node (g10) at (1.5,1) [gaugeb,label=left:{1}] {};
	\draw (g1)--(g2)--(g3)--(g4);
	\draw (g2)--(g10)--(g3);
	\end{tikzpicture}} \\ 
	\raisebox{-.5\height}{\begin{tikzpicture}[x=.8cm,y=.8cm]
	\node (g1) at (0,0) [gauge,label=below:{1}] {};
	\node (g2) at (1,0) [gauge,label=below:{1}] {};
	\node (g3) at (2,0) [gauge,label=below:{1}] {};
	\node (g4) at (3,0) [gauge,label=below:{1}] {};
	\node (g10) at (1.5,1) [gaugeb,label=left:{1}] {};
	\node (g11) at (1.5,2) [gauge,label=left:{1}] {};
	\draw (g1)--(g2)--(g3)--(g4);
	\draw (g1)--(g10)--(g4);
	\draw[double distance=2pt] (g11)--(g10);
	\end{tikzpicture}} \\ $\su(5) \oplus \su(2)$
		\end{tabular}
}
{24}{\begin{tabular}{c}
\textcolor{red}{24} \\  \raisebox{-.5\height}{\begin{tikzpicture}[x=.8cm,y=.8cm]
	\node (g1) at (0,0) [gauge,label=below:{1}] {};
	\node (g2) at (1,0) [gauge,label=below:{2}] {};
	\node (g3) at (2,0) [gauge,label=below:{1}] {};
	\node (g6) at (.5,.8) [gaugeb,label=left:{1}] {};
	\node (g7) at (1.5,.8) [gaugeb,label=right:{1}] {};
	\draw (g1)--(g2)--(g3);
	\draw (g2)--(g6)--(g7)--(g2);
	\end{tikzpicture}} \\ 
\raisebox{-.5\height}{\begin{tikzpicture}[x=.8cm,y=.8cm]
	\node (g1) at (0,0) [gauge,label=below:{1}] {};
	\node (g2) at (1,0) [gauge,label=below:{1}] {};
	\node (g3) at (2,0) [gauge,label=below:{1}] {};
	\node (g6) at (.5,.8) [gaugeb,label=left:{1}] {};
	\node (g7) at (1.5,.8) [gaugeb,label=right:{1}] {};
	\draw (g6)--(g1)--(g2)--(g3)--(g7);
	\draw[double distance=2pt] (g6)--(g7);
	\end{tikzpicture}} \\ $\su(4) \oplus \uu(1) \oplus \uu(1)$
		\end{tabular}
}
{25}{\begin{tabular}{c}
\textcolor{red}{25} \\  \raisebox{-.5\height}{\begin{tikzpicture}[x=.8cm,y=.8cm]
	\node (g1) at (0,0) [gauge,label=below:{1}] {};
	\node (g2) at (1,0) [gauge,label=below:{1}] {};
	\node (g3) at (2,0) [gauge,label=below:{1}] {};
	\node (g6) at (.5,.8) [gaugeb,label=above:{1}] {};
	\node (g7) at (1,1.6) [gauge,label=above:{1}] {};
	\node (g8) at (1.5,.8) [gaugeb,label=above:{1}] {};
	\draw (g1)--(g2)--(g3)--(g8)--(g7)--(g6)--(g1);
	\draw (g6)--(g8);
	\end{tikzpicture}} \\
\raisebox{-.5\height}{\begin{tikzpicture}[x=.8cm,y=.8cm]
	\node (g1) at (0,0) [gauge,label=below:{1}] {};
	\node (g2) at (1,0) [gauge,label=below:{2}] {};
	\node (g3) at (2,0) [gauge,label=below:{1}] {};
	\node (g7) at (1,1) [gaugeb,label=left:{1}] {};
	\draw (g1)--(g2)--(g3);
	\draw[double distance=2pt] (g2)--(g7);
	\end{tikzpicture}} \\ $\su(4) \oplus \su(2) \oplus \uu(1)$
		\end{tabular}
}
{26}{\begin{tabular}{c}
\textcolor{red}{26} \\  \raisebox{-.5\height}{\begin{tikzpicture}[x=.8cm,y=.8cm]
	\node (g1) at (0,0) [gauge,label=below:{1}] {};
	\node (g2) at (1,0) [gauge,label=below:{2}] {};
	\node (g3) at (2,0) [gauge,label=below:{2}] {};
	\node (g4) at (3,0) [gauge,label=below:{1}] {};
	\node (g7) at (1,1) [gaugeb,label=left:{1}] {};
	\draw (g1)--(g2)--(g3)--(g4);
	\draw (g2)--(g7)--(g3);
	\end{tikzpicture}} \\
	\raisebox{-.5\height}{\begin{tikzpicture}[x=.8cm,y=.8cm]
	\node (g1) at (0,0) [gauge,label=below:{1}] {};
	\node (g2) at (1,0) [gauge,label=below:{1}] {};
	\node (g3) at (2,0) [gauge,label=below:{1}] {};
	\node (g4) at (3,0) [gauge,label=below:{1}] {};
	\node (g7) at (1,1) [gaugeb,label=left:{1}] {};
	\node (g8) at (2,1) [gaugeb,label=right:{1}] {};
	\draw (g7)--(g1)--(g2)--(g3)--(g4)--(g8);
	\draw[double distance=2pt] (g8)--(g7);
	\end{tikzpicture}} \\ $\su(5)  \oplus \uu(1)$
		\end{tabular}
}
{27}{\begin{tabular}{c}
\textcolor{red}{27} \\  \raisebox{-.5\height}{\begin{tikzpicture}[x=.8cm,y=.8cm]
	\node (g1) at (0,0) [gauge,label=below:{1}] {};
	\node (g2) at (1,0) [gauge,label=below:{2}] {};
	\node (g3) at (2,0) [gauge,label=below:{2}] {};
	\node (g4) at (3,0) [gaugeb,label=below:{1}] {};
	\node (g5) at (1,1) [gauge,label=left:{1}] {};
	\node (g6) at (2,1) [gaugeb,label=right:{1}] {};
	\draw (g1)--(g2)--(g3)--(g4);
	\draw (g2)--(g5);
	\draw (g3)--(g6)--(g4);
	\end{tikzpicture}}    \\ 
	\raisebox{-.5\height}{\begin{tikzpicture}[x=.8cm,y=.8cm]
	\node (g1) at (0,0) [gauge,label=below:{1}] {};
	\node (g2) at (1,0) [gauge,label=below:{2}] {};
	\node (g3) at (2,0) [gauge,label=below:{2}] {};
	\node (g5) at (1,1) [gauge,label=left:{1}] {};
	\node (g6) at (2,1) [gaugeb,label=right:{1}] {};
	\draw (g1)--(g2)--(g3);
	\draw (g2)--(g5);
	\draw[double distance=2pt] (g6)--(g3);
	\end{tikzpicture}}   \\ $\so(8)  \oplus \uu(1)$
		\end{tabular}
}
{28}{\begin{tabular}{c}
\textcolor{red}{28} \\  \raisebox{-.5\height}{\begin{tikzpicture}[x=.8cm,y=.8cm]
	\node (g1) at (0,0) [gauge,label=below:{1}] {};
	\node (g2) at (1,0) [gaugeb,label=below:{2}] {};
	\node (g3) at (2,0) [gauge,label=below:{2}] {};
	\node (g4) at (3,0) [gauge,label=below:{2}] {};
	\node (g10) at (1,1) [gauge,label=left:{2}] {};
	\node (g11) at (2,1) [gauge,label=right:{2}] {};
	\draw (g1)--(g2)--(g3)--(g4)--(g11)--(g10)--(g2);
	\end{tikzpicture}} \\ $\su(5)  \oplus \su(2)$
		\end{tabular}
}
{29}{\begin{tabular}{c} \textcolor{red}{29}  \\   \raisebox{-.5\height}{ \begin{tikzpicture}[x=.8cm,y=.8cm,decoration={
    markings,
    mark=at position 0.5 with {\arrow{>},arrowhead=1cm}}]
	\node (g1) at (0,0) [gauge,label=below:{1}] {};
	\node (g2) at (1,0) [gauge,label=below:{2}] {};
	\node (g3) at (2,0) [gauge,label=below:{3}] {};
	\node (g4) at (3,0) [gauge,label=below:{4}] {};
	\node (g5) at (4,0) [gaugeb,label=below:{2}] {};
	\node (g6) at (5,0) [gauge,label=below:{1}] {};
	\draw (g1)--(g2)--(g3);
	\draw (g6)--(g5)--(g4);
	\draw[doublearrow] (g3)-- node {\midarrowrev} (g4); 
	\draw[doublearrow] (g6)-- node {\midarrow} (g5); 
	\end{tikzpicture}} \\ $\spp(4)  \oplus \su(2)$
\end{tabular}
}
{30}{\begin{tabular}{c}
\textcolor{red}{30} \\  
\raisebox{-.5\height}{\begin{tikzpicture}[x=.8cm,y=.8cm]
	\node (g1) at (0,0) [gauge,label=below:{1}] {};
	\node (g2) at (1,0) [gauge,label=below:{2}] {};
	\node (g3) at (2,0) [gauge,label=below:{1}] {};
	\node (g7) at (1,1) [gaugeb,label=left:{1}] {};
	\draw (g1)--(g2)--(g3);
	\draw[double distance=2pt] (g2)--(g7);
	\end{tikzpicture}} \\ $\su(4)  \oplus \uu(1)$
		\end{tabular}
}
{31}{\begin{tabular}{c}
\textcolor{red}{31} \\  \raisebox{-.5\height}{\begin{tikzpicture}[x=.8cm,y=.8cm]
	\node (g1) at (0,0) [gauge,label=below:{1}] {};
	\node (g2) at (1,0) [gauge,label=below:{1}] {};
	\node (g6) at (0,.8) [gaugeb,label=left:{1}] {};
	\node (g7) at (1,.8) [gaugeb,label=right:{1}] {};
	\draw (g6)--(g1)--(g2)--(g7);
	\draw[double distance=2pt] (g6)--(g7);
	\end{tikzpicture}} \\ 
	\raisebox{-.5\height}{\begin{tikzpicture}[x=.8cm,y=.8cm]
	\node (g1) at (0,0) [gaugeb,label=below:{1}] {};
	\node (g2) at (1,0) [gaugeb,label=below:{1}] {};
	\draw[double distance=4pt] (g2)--(g1);
	\draw (g1)--(g2);
	\end{tikzpicture}} \\ $\su(3)  \oplus \uu(1) \oplus \uu(1)$
		\end{tabular}
}
{32}{\begin{tabular}{c}
\textcolor{red}{32} \\  \raisebox{-.5\height}{\begin{tikzpicture}[x=.8cm,y=.8cm]
	\node (g1) at (0,0) [gauge,label=below:{1}] {};
	\node (g2) at (1,0) [gauge,label=below:{1}] {};
	\node (g6) at (.5,.8) [gaugeb,label=left:{1}] {};
	\node (g7) at (.5,1.6) [gauge,label=above:{1}] {};
	\draw (g1)--(g2)--(g6)--(g1);
	\draw[double distance=2pt] (g6)--(g7);
	\end{tikzpicture}} \\ 
	\raisebox{-.5\height}{\begin{tikzpicture}[x=.8cm,y=.8cm]
	\node (g1) at (0,0) [gauge,label=below:{1}] {};
	\node (g2) at (1,0) [gaugeb,label=below:{1}] {};
	\node (g3) at (0,1) [gaugeb,label=left:{1}] {};
	\draw (g3)--(g1)--(g2);
	\draw[double distance=2pt] (g2)--(g3);
	\end{tikzpicture}}   \\ $\su(3)  \oplus \su(2) \oplus \uu(1)$
		\end{tabular}
}
{33}{\begin{tabular}{c}
\textcolor{red}{33} \\  \raisebox{-.5\height}{\begin{tikzpicture}[x=.8cm,y=.8cm]
	\node (g1) at (0,0) [gauge,label=below:{1}] {};
	\node (g2) at (1,0) [gauge,label=below:{2}] {};
	\node (g3) at (2,0) [gauge,label=below:{1}] {};
	\node (g7) at (1,1) [gaugeb,label=left:{1}] {};
	\draw (g1)--(g2)--(g3);
	\draw[double distance=2pt] (g2)--(g7);
	\end{tikzpicture}} \\ 
	\raisebox{-.5\height}{\begin{tikzpicture}[x=.8cm,y=.8cm]
	\node (g1) at (0,0) [gaugeb,label=below:{1}] {};
	\node (g2) at (1,0) [gaugeb,label=below:{1}] {};
	\draw[double distance=4pt] (g2)--(g1);
	\draw (g1)--(g2);
	\end{tikzpicture}} \\ $\su(4) \oplus \uu(1)$
		\end{tabular}
}
{34}{\begin{tabular}{c}
\textcolor{red}{34} \\  \raisebox{-.5\height}{\begin{tikzpicture}[x=.8cm,y=.8cm]
	\node (g1) at (0,0) [gauge,label=below:{1}] {};
	\node (g2) at (1,0) [gauge,label=below:{1}] {};
	\node (g3) at (2,0) [gaugeb,label=below:{1}] {};
	\node (g5) at (0,1) [gauge,label=left:{1}] {};
	\node (g6) at (1,1) [gaugeb,label=right:{1}] {};
	\draw (g6)--(g5)--(g1)--(g2)--(g3);
	\draw[double distance=2pt] (g6)--(g3);
	\end{tikzpicture}}     \\ $\su(4) \oplus \uu(1)$
		\end{tabular}
}
{35}{\begin{tabular}{c}
\textcolor{red}{35} \\  \raisebox{-.5\height}{\begin{tikzpicture}[x=.8cm,y=.8cm]
	\node (g1) at (0,0) [gauge,label=below:{1}] {};
	\node (g2) at (1,0) [gaugeb,label=below:{2}] {};
	\node (g3) at (2,0) [gauge,label=below:{2}] {};
	\node (g10) at (1,1) [gauge,label=left:{2}] {};
	\draw (g1)--(g2)--(g3)--(g10)--(g2);
	\end{tikzpicture}}  \\ 
	\raisebox{-.5\height}{\begin{tikzpicture}[x=.8cm,y=.8cm]
	\node (g1) at (0,0) [gauge,label=below:{1}] {};
	\node (g2) at (1,0) [gaugeb,label=below:{2}] {};
	\node (g10) at (1,1) [gauge,label=left:{2}] {};
	\draw (g1)--(g2);
	\draw[double distance=2pt] (g2)--(g10);
	\end{tikzpicture}}  \\ $\su(3) \oplus \su(2)^2$
		\end{tabular}
}
{36}{\begin{tabular}{c}
\textcolor{red}{36} \\    \raisebox{-.5\height}{ \begin{tikzpicture}[x=.8cm,y=.8cm,decoration={
    markings,
    mark=at position 0.5 with {\arrow{>},arrowhead=1cm}}]
	\node (g2) at (1,0) [gauge,label=below:{1}] {};
	\node (g3) at (2,0) [gauge,label=below:{2}] {};
	\node (g4) at (3,0) [gauge,label=below:{3}] {};
	\node (g5) at (4,-.5) [gaugeb,label=right:{1}] {};
	\node (g6) at (4,.5) [gaugeb,label=right:{1}] {};
	\draw (g2)--(g3);
	\draw (g6)--(g4)--(g5)--(g6);
	\draw[doublearrow] (g3)-- node {\midarrowrev} (g4); 
	\end{tikzpicture}} \\ $\spp(3) \oplus \uu(1)$
\end{tabular}
}
{37}{\begin{tabular}{c} \textcolor{red}{37}  \\  MARGINAL \\  $Sp(2)_0+3AS$
\end{tabular}
}
{38}{\begin{tabular}{c}
\textcolor{red}{38} \\  
\raisebox{-.5\height}{\begin{tikzpicture}[x=.8cm,y=.8cm]
	\node (g1) at (0,0) [gauge,label=below:{1}] {};
	\node (g2) at (1,0) [gauge,label=below:{2}] {};
	\node (g3) at (2,0) [gauge,label=below:{1}] {};
	\node (g7) at (1,1) [gaugeb,label=left:{1}] {};
	\draw (g1)--(g2)--(g3);
	\draw[double distance=2pt] (g2)--(g7);
	\end{tikzpicture}} \\ $\su(4)$
		\end{tabular}
}
{39}{\begin{tabular}{c}
\textcolor{red}{39} \\  \raisebox{-.5\height}{\begin{tikzpicture}[x=.8cm,y=.8cm]
	\node (g1) at (0,0) [gauge,label=below:{1}] {};
	\node (g2) at (1,0) [gauge,label=below:{1}] {};
	\node (g3) at (.5,.8) [gauge,label=above:{1}] {};
	\draw (g1)--(g2)--(g3)--(g1);
	\end{tikzpicture}}  \\ $\su(3) \oplus \uu(1)$
		\end{tabular}
}
{40}{\begin{tabular}{c}
\textcolor{red}{40} \\  \raisebox{-.5\height}{\begin{tikzpicture}[x=.8cm,y=.8cm]
	\node (g1) at (0,0) [gauge,label=below:{1}] {};
	\node (g6) at (-.5,.8) [gaugeb,label=left:{1}] {};
	\node (g7) at (.5,.8) [gaugeb,label=right:{1}] {};
	\draw (g6)--(g1)--(g7);
	\draw[double distance=2pt] (g6)--(g7);
	\end{tikzpicture}} \\ $\su(2) \oplus \uu(1) \oplus \uu(1)$
		\end{tabular}
}
{41}{\begin{tabular}{c}
\textcolor{red}{41} \\  \raisebox{-.5\height}{\begin{tikzpicture}[x=.8cm,y=.8cm]
	\node (g1) at (0,0) [gaugeb,label=below:{1}] {};
	\node (g2) at (1,0) [gaugeb,label=below:{1}] {};
	\draw[double distance=4pt] (g2)--(g1);
	\draw (g1)--(g2);
	\end{tikzpicture}}  \\
	\raisebox{-.5\height}{\begin{tikzpicture}[x=.8cm,y=.8cm]
	\node (g1) at (0,0) [gauge,label=below:{1}] {};
	\node (g2) at (1,0) [gauge,label=below:{1}] {};
	\draw[double distance=2pt] (g2)--(g1);
	\end{tikzpicture}}  	  \\ $\su(2) \oplus \uu(1) \oplus \uu(1)$ 
		\end{tabular}
}
{42}{\begin{tabular}{c}
\textcolor{red}{42} \\  \raisebox{-.5\height}{\begin{tikzpicture}[x=.8cm,y=.8cm]
	\node (g1) at (0,0) [gauge,label=below:{1}] {};
	\node (g6) at (0,.8) [gaugeb,label=left:{1}] {};
	\node (g7) at (0,1.6) [gauge,label=above:{1}] {};
	\draw[double distance=2pt] (g6)--(g7);
	\draw[double distance=2pt] (g6)--(g1);
	\end{tikzpicture}} \\ $\su(2)^2 \oplus \uu(1)$
		\end{tabular}
}
{43}{\begin{tabular}{c}
\textcolor{red}{43} \\  \raisebox{-.5\height}{\begin{tikzpicture}[x=.8cm,y=.8cm]
	\node (g1) at (0,0) [gauge,label=below:{1}] {};
	\node (g2) at (1,0) [gauge,label=below:{1}] {};
	\node (g7) at (0,1) [gauge,label=left:{1}] {};
	\draw (g1)--(g2)--(g7)--(g1);
	\end{tikzpicture}}  \\ $\su(3) \oplus \uu(1)$
		\end{tabular}
}
{44}{\begin{tabular}{c}
\textcolor{red}{44} \\  \raisebox{-.5\height}{\begin{tikzpicture}[x=.8cm,y=.8cm]
	\node (g1) at (0,0) [gauge,label=below:{1}] {};
	\node (g2) at (1,0) [gaugeb,label=below:{1}] {};
	\node (g3) at (0,1) [gaugeb,label=left:{1}] {};
	\draw (g3)--(g1)--(g2);
	\draw[double distance=2pt] (g2)--(g3);
	\end{tikzpicture}}     \\
	\raisebox{-.5\height}{\begin{tikzpicture}[x=.8cm,y=.8cm]
	\node (g1) at (0,0) [gauge,label=below:{1}] {};
	\node (g2) at (1,0) [gauge,label=below:{1}] {};
	\draw[double distance=2pt] (g2)--(g1);
	\end{tikzpicture}}   \\ $\su(2)^2 \oplus \uu(1)$
		\end{tabular}
}
{45}{\begin{tabular}{c}
\textcolor{red}{45} \\  \raisebox{-.5\height}{\begin{tikzpicture}[x=.8cm,y=.8cm]
	\node (g1) at (0,0) [gauge,label=below:{1}] {};
	\node (g2) at (1,0) [gaugeb,label=below:{2}] {};
	\node (g10) at (1,1) [gauge,label=left:{2}] {};
	\draw (g1)--(g2);
	\draw[double distance=2pt] (g2)--(g10);
	\end{tikzpicture}}  \\ $\su(2)^2 \oplus \uu(1)$
		\end{tabular}
}
{46}{ \begin{tabular}{c} \textcolor{red}{46}   \\ 
\raisebox{-.5\height}{ 
\begin{tikzpicture}[x=.8cm,y=.8cm,decoration={
    markings,
    mark=at position 0.5 with {\arrow{>},arrowhead=1cm}}]
	\node (g3) at (2,0) [gauge,label=below:{1}] {};
	\node (g4) at (3,0) [gauge,label=below:{2}] {};
	\node (g5) at (4,-.5) [gaugeb,label=right:{1}] {};
	\node (g6) at (4,.5) [gaugeb,label=right:{1}] {};
	\draw (g6)--(g4)--(g5);
	\draw[doublearrow] (g3)-- node {\midarrowrev} (g4); 
	\draw[double distance=2pt] (g5)--(g6);
	\end{tikzpicture}} \\   $\spp(2) \oplus \uu(1)$
\end{tabular}
}
{47}{\begin{tabular}{c} \textcolor{red}{47}  \\   \raisebox{-.5\height}{ \begin{tikzpicture}[x=.8cm,y=.8cm,decoration={
    markings,
    mark=at position 0.5 with {\arrow{>},arrowhead=1cm}}]
	\node (g2) at (1,0) [gauge,label=below:{1}] {};
	\node (g3) at (2,0) [gauge,label=below:{2}] {};
	\node (g4) at (3,0) [gauge,label=below:{3}] {};
	\node (g5) at (4,0) [gaugeb,label=below:{1}] {};
	\draw (g2)--(g3);
	\draw[doublearrow] (g3)-- node {\midarrowrev} (g4); 
	\draw[doublearrow] (g5)-- node {\midarrow} (g4); 
	\end{tikzpicture}} \\ $\spp(3)$
\end{tabular}
}
{48}{\textcolor{red}{48}
}
{49}{\begin{tabular}{c}
\textcolor{red}{49} \\  
\raisebox{-.5\height}{\begin{tikzpicture}[x=.8cm,y=.8cm]
	\node (g1) at (0,0) [gaugeb,label=below:{1}] {};
	\node (g2) at (1,0) [gaugeb,label=below:{1}] {};
	\draw[double distance=4pt] (g2)--(g1);
	\draw (g1)--(g2);
	\end{tikzpicture}}    \\ $\uu(1) \oplus \uu(1)$
		\end{tabular}
}
{50}{\begin{tabular}{c}
\textcolor{red}{50} \\  \raisebox{-.5\height}{\begin{tikzpicture}[x=.8cm,y=.8cm]
	\node (g1) at (0,0) [gauge,label=below:{1}] {};
	\end{tikzpicture}}  \\ $\uu(1) \oplus \uu(1)$ 
		\end{tabular}
}
{51}{\begin{tabular}{c}
\textcolor{red}{51} \\  \raisebox{-.5\height}{\begin{tikzpicture}[x=.8cm,y=.8cm]
	\node (g1) at (0,0) [gauge,label=below:{1}] {};
	\node (g6) at (0,.8) [gauge,label=above:{1}] {};
	\draw[double distance=2pt] (g6)--(g1);
	\end{tikzpicture}}   \\ $\su(2) \oplus \uu(1)$
		\end{tabular}
}
{52}{\begin{tabular}{c}
\textcolor{red}{52} \\  \raisebox{-.5\height}{\begin{tikzpicture}[x=.8cm,y=.8cm]
	\node (g1) at (0,0) [gauge,label=below:{1}] {};
	\node (g7) at (0,1) [gauge,label=left:{1}] {};
	\draw[double distance=2pt] (g1)--(g7);
	\end{tikzpicture}}   \\ $\su(2) \oplus \uu(1)$
		\end{tabular}
}
{53}{\begin{tabular}{c}
\textcolor{red}{53} \\  \raisebox{-.5\height}{\begin{tikzpicture}[x=.8cm,y=.8cm]
	\node (g1) at (0,0) [gaugeb,label=below:{1}] {};
	\node (g2) at (1,0) [gaugeb,label=below:{1}] {};
	\draw[double distance=4pt] (g2)--(g1);
	\draw (g1)--(g2);
	\end{tikzpicture}}     \\ $\uu(1) \oplus \uu(1)$
		\end{tabular}
}
{54}{\begin{tabular}{c}
\textcolor{red}{54} \\  \raisebox{-.5\height}{\begin{tikzpicture}[x=.8cm,y=.8cm]
	\node (g1) at (0,0) [gauge,label=below:{1}] {};
	\node (g2) at (1,0) [gauge,label=below:{2}] {};
	\draw (g1)--(g2);
	\draw (g2) to [out=45,in=-45,looseness=10] (g2);
	\end{tikzpicture}}   \\ $\su(2) \oplus \uu(1)$
		\end{tabular}
}
{55}{ \begin{tabular}{c} \textcolor{red}{55}  \\ 
\raisebox{-.5\height}{ 
\begin{tikzpicture}[x=.8cm,y=.8cm,decoration={
    markings,
    mark=at position 0.5 with {\arrow{>},arrowhead=1cm}}]
	\node (g4) at (3,0) [gauge,label=below:{1}] {};
	\node (g5) at (4,-.5) [gaugeb,label=right:{1}] {};
	\node (g6) at (4,.5) [gaugeb,label=right:{1}] {};
	\draw (g6)--(g4)--(g5);
	\draw[double distance=4pt] (g5)--(g6);
	\draw (g6)--(g5);
	\end{tikzpicture}}    \\ $\uu(1) \oplus \uu(1)$
\end{tabular}
}
{56}{\begin{tabular}{c}
\textcolor{red}{56} \\  \raisebox{-.5\height}{\begin{tikzpicture}[x=.8cm,y=.8cm]
	\node (g1) at (0,0) [gauge,label=below:{1}] {};
	\node (g2) at (1,0) [gaugeb,label=below:{2}] {};
	\node (g10) at (1,1) [gauge,label=left:{2}] {};
	\draw (g1)--(g2);
	\draw[double distance=2pt] (g2)--(g10);
	\end{tikzpicture}}  \\ $\su(2)^2$
		\end{tabular}
}
{57}{\textcolor{red}{57}
}
{58}{\begin{tabular}{c}
\textcolor{red}{58} \\  \raisebox{-.5\height}{\begin{tikzpicture}[x=.8cm,y=.8cm]
	\node (g1) at (0,0) [gaugeb,label=below:{1}] {};
	\node (g2) at (1,0) [gaugeb,label=below:{1}] {};
	\draw[double distance=4pt] (g2)--(g1);
	\draw (g1)--(g2);
	\end{tikzpicture}}    \\ $\uu(1)$
		\end{tabular}
}
{59}{\begin{tabular}{c}
\textcolor{red}{59} \\  \raisebox{-.5\height}{\begin{tikzpicture}[x=.8cm,y=.8cm]
	\node (g1) at (0,0) [gauge,label=below:{1}] {};
	\end{tikzpicture}}  \\ $\uu(1)$ 
		\end{tabular}
}
{60}{\begin{tabular}{c}
\textcolor{red}{60} \\  \raisebox{-.5\height}{\begin{tikzpicture}[x=.8cm,y=.8cm]
	\node (g1) at (0,0) [gauge,label=below:{1}] {};
	\end{tikzpicture}}  \\ $\uu(1)$
		\end{tabular}
}
{61}{\begin{tabular}{c}
\textcolor{red}{61} \\  \raisebox{-.5\height}{\begin{tikzpicture}[x=.8cm,y=.8cm]
	\node (g1) at (0,0) [gauge,label=below:{1}] {};
	\node (g6) at (0,.8) [gauge,label=above:{1}] {};
	\draw[double distance=2pt] (g6)--(g1);
	\end{tikzpicture}}  \\ $\su(2)$
		\end{tabular}
}
{62}{\begin{tabular}{c}
\textcolor{red}{62} \\  \raisebox{-.5\height}{\begin{tikzpicture}[x=.8cm,y=.8cm]
	\node (g1) at (0,0) [gauge,label=below:{1}] {};
	\end{tikzpicture}}  \\ $\uu(1)$
		\end{tabular}
}
{63}{\begin{tabular}{c}
\textcolor{red}{63} \\   \raisebox{-.5\height}{\begin{tikzpicture}[x=.8cm,y=.8cm]
	\node (g1) at (0,0) [gauge,label=below:{1}] {};
	\end{tikzpicture}}  \\ $\uu(1)$
		\end{tabular}
}
{64}{\begin{tabular}{c}
\textcolor{red}{64} \\  \raisebox{-.5\height}{\begin{tikzpicture}[x=.8cm,y=.8cm]
	\node (g1) at (0,0) [gauge,label=below:{1}] {};
	\end{tikzpicture}}  \\ $\su(2)$
		\end{tabular}
}
{65}{\begin{tabular}{c} \textcolor{red}{65}  \\ 
\raisebox{-.5\height}{ 
\begin{tikzpicture}
	\draw (-.09,.5)--(-.09,-.5);
	\draw (.09,.5)--(.09,-.5);
	\draw (-.03,.5)--(-.03,-.5);
	\draw (.03,.5)--(.03,-.5);
	\node (g5) at (0,-.5) [gaugeb,label=right:{1}] {};
	\node (g6) at (0,.5) [gaugeb,label=right:{1}] {};
	\end{tikzpicture}}   \\ $\uu(1)$
\end{tabular}
}
{66}{\begin{tabular}{c}
\textcolor{red}{66} \\  \raisebox{-.5\height}{\begin{tikzpicture}[x=.8cm,y=.8cm]
	\node (g1) at (0,0) [gaugeb,label=below:{1}] {};
	\node (g2) at (1,0) [gaugeb,label=below:{1}] {};
	\draw (0,.03)--(1,.03);
	\draw (0,.09)--(1,.09);
	\draw (0,-.03)--(1,-.03);
	\draw (0,-.09)--(1,-.09);
	\end{tikzpicture}}     \\ $\uu(1)$
		\end{tabular}
}
{67}{\textcolor{red}{67}
}
{68}{\textcolor{red}{68}
}
{69}{\textcolor{red}{69} \begin{tabular}{c} MARGINAL 
\end{tabular}
}
{70}{\textcolor{red}{70} \begin{tabular}{c} MARGINAL 
\end{tabular}
}
}}


\begin{table}
\vspace*{-1cm}\begin{adjustbox}{max width=.95\paperwidth, max height=10cm,center} 
\begin{tabular}{|c||c|c|c|c|c|c|c|c|} \hline 
   &   0   &   1   &   2   &   3   &   4   &   5   &   6   &   7     \\  \hline \hline 
10 &       &       &       &       &       &\mo{1} &       &         \\  \hline 
9 &       &       &       &   -   &   -   &\mo{2} &\mo{3} &         \\  \hline 
8 &       &       &   -   &   -   &\mo{4} &\mo{5} &\mo{6} &         \\  \hline 
7 &       &   -   &   -   &   -   &\mo{8} &\mo{9} &\mo{10}&         \\  \hline 
6 &   -   &  -    &  -    &\mo{12}&\mo{13}&\mo{14}&\mo{15}&\mo{16}  \\  \hline 
5 &  -    &  -    &   -   &\mo{18}&\mo{19}&\mo{20}&\mo{21}&\mo{22}  \\  \hline 
4 &  -    &  -    &\mo{24}&\mo{25}&\mo{26}&\mo{27}&\mo{28}&\mo{29}  \\  \hline 
3 & -     & -     &\mo{31}&\mo{32}&\mo{33}&\mo{34}&\mo{35}&\mo{36}  \\  \hline 
2 & -     &\mo{40}&\mo{41}&\mo{42}&\mo{43}&\mo{44}&\mo{45}&\mo{46}  \\  \hline 
1 & -     &\mo{49}&\mo{50}&\mo{51}&\mo{52}&\mo{53}&\mo{54}&\mo{55}  \\  \hline 
0 &\mo{58}&\mo{59}&\mo{60}&\mo{61}&\mo{62}&\mo{63}&\mo{64}&\mo{65}  \\  \hline 
\end{tabular} 
\end{adjustbox}
\caption{Decoupling tree for rank 2 5d SCFTs with an $SU(3)_k + N_F \bm{F}$ description: the marginal theories are $(D_5, D_5)$ conformal matter, rank 2 E-strings, and the model 3. We show the IR gauge theory description of these, and in each model list also the non-abelian flavor symmetry algebra. 
The rows are labeled by $N_F$, the columns are labeled by $\frac{N_F}{2} + k$. The red numerals correspond to the enumeration in Appendix A.2 of \cite{Apruzzi:2019opn}. Decoupling a fundamental hyper while increasing $k$ (respectively decreasing $k$) by $\frac{1}{2}$ corresponds to moving one box downwards (resp. one box diagonally to the bottom left). The dashes correspond to quivers which can be obtained from other boxes by the symmetry $k \rightarrow -k$. \label{tab:Rank2SU}}
\end{table}

\begin{table}
\begin{adjustbox}{max width=.95\paperwidth, max height=10cm,center} 
\begin{tabular}{|c||c|c|c|c|} \hline 
   & 0AS & 1AS  & 2AS  & 3AS    \\  \hline \hline 
 10F  & \mo{1}&   &    &     \\  \hline 
 9F   & \mo{2}&   &    &     \\  \hline 
 8F   & \mo{5} &   &    &     \\  \hline 
 7F   & \mo{9}  & \mo{6}&    &     \\  \hline 
 6F   & \mo{14}& \mo{10} &    &     \\  \hline 
 5F   & \mo{20} & \mo{15}&    &     \\  \hline 
 4F   & \mo{27}  & \mo{21} & \mo{16}   &     \\  \hline 
 3F   &\mo{34}  & \mo{28}  & \mo{22}   &     \\  \hline 
 2F   & \mo{44}  & \mo{35}& \mo{29}   &     \\  \hline 
 1F   & \mo{53}  & \mo{45}& \mo{36}   &     \\  \hline 
 0F with $\theta=\pi$   & \mo{63}  & \mo{54} & \mo{46}   &     \\  \hline 
 0F with $\theta=0$ & \mo{66}  & \mo{56}& \mo{47}   &   \mo{37}   \\  \hline 
\end{tabular}
\end{adjustbox}
\caption{Decoupling tree of Rank 2 continued: Shown are the magnetic quivers for SCFTs, which have IR descriptions as $Sp(2)$ gauge theories. The lines correspond to the number of fundamentals and the columns to the number of antisymmetric hypermultiplets. When there are no fundamentals, we have to distinguish according to the value of the $\theta$-angle.
\label{tab:Rank2Sp2}}
\end{table}

\begin{table}
\begin{adjustbox}{max width=.95\paperwidth, max height=10cm,center} 
\begin{tabular}{|c||c|c|c|c|c|c|c|} \hline 
   & 5F & 4F  & 3F  & 2F & 1F & 0F with $\theta = \pi$  & 0F with $\theta = 0$    \\  \hline \hline 
 4F  &   & \mo{1} &  - & - & - & -  &  -\\  \hline 
 3F  &   & \mo{2}  & \mo{4}  & - & - & -  & - \\  \hline 
 2F  &   & \mo{5}  & \mo{8}  & \mo{12} & - &  -  & -\\  \hline 
 1F  &   & \mo{9}  & \mo{13}  & \mo{18} & \mo{24} &  - & - \\  \hline 
 \begin{tabular}{c}
 0F \\ with \\ $\theta = \pi$
\end{tabular}  & \mo{10} & \mo{14} & \mo{19}  & \mo{25} &\mo{31} & \mo{40}& \mo{39}   \\  \hline 
 \begin{tabular}{c}
 0F \\ with \\ $\theta = 0$
\end{tabular}  & \mo{7} & \mo{11} & \mo{17}  & \mo{23} & \mo{30} & \mo{39} & \mo{38}   \\  \hline 
\end{tabular}
\end{adjustbox}
\caption{Rank 2 decoupling tree continued: magnetic quivers for 5d SCFTs with $SU(2)-SU(2)$ quiver gauge theory IR description. The lines and columns correspond to the number of fundamentals on each gauge group.  The dashes correspond to quivers which can be obtained from other boxes by flipping the electric quiver.  \label{tab:Rank2SU2SU2}}
\end{table}

\begin{table}
\centering
\begin{tabular}{|c||c|} \hline 
  $N_f$ & Theory       \\  \hline \hline 
 6F  & \mo{16}    \\  \hline 
 5F  & \mo{22}    \\  \hline 
 4F  & \mo{29}    \\  \hline 
 3F  & \mo{36}    \\  \hline 
 2F  & \mo{46}    \\  \hline 
 1F  & \mo{55}    \\  \hline 
 0F  & \mo{65}    \\  \hline 
\end{tabular}
\caption{Rank 2 decoupling tree continued: $G_2$ theories with $N_f$ fundamentals.  \label{tab:Rank2G2}}
\end{table}



\clearpage

\section{Magnetic Quivers Decoupling Trees}
\label{sec:MQTables}

In this appendix we tabulate the magnetic quivers of the theories discussed in section \ref{sec:SUNasf}. In the magnetic quivers we assume large rank, when indicating the balance of a node. Additional nodes may be balanced for low rank. Furthermore, we give the enhanced flavour symmetry. Except for the higher rank $\e_{N_F+1}$-theories these can also be inferred from the magnetic quiver by computing the Hasse diagram, as discussed in section \ref{sec:NSLQuivers}.

\baselineskip=18pt  
\numberwithin{equation}{section}  
\allowdisplaybreaks  

\def \scale {.45}

We first look at the magnetic quivers for SCFTs with a weakly coupled description of the form $SU(N)_k + N_{AS} \bm{AS} + N_F \bm{F}$ for $N>3$ with sufficiently many flavours.\footnote{Note that for $N=2$ and $N=3$ the $\bm{AS}$ representation corresponds to the trivial and anti-fundamental representation respectively.} We will group them first depending on $N_{AS}=0,1,2$ and possibly their rank $r=N-1$. Each table is then structured in terms of $N_F$ and $\frac{N_F}{2}+ k$.
The SQCD theories are summarised in table \ref{tab:SQCD}. The descendants of the higher rank E-string are given in tables \ref{tab:UVMQsEStringEven} and \ref{tab:UVMQsEStringOdd}. In tables \ref{tab:UVMQs1ASEven} and \ref{tab:UVMQs1ASOdd} we give the other theories with $N_{AS}=1$. Finally, tables \ref{tab:UVMQs2ASSU2n} and \ref{tab:UVMQs2ASSU2n+1} show all magnetic quivers for theories with $N_{AS}=2$ for general rank.

As discussed in section \ref{sec:NSLQuivers} some of the $SU(N)_k + 2\bm{AS} + N_F \bm{F}$ theories with $N=4,5$ have enhanced non-simply laced flavour symmetry. The magnetic quivers and the enhanced flavour symmetries are summarised in tables \ref{tab:UVMQs2ASSU4} and \ref{tab:UVMQs2ASSU5}.

Another class of gauge theories with SCFTs have the form $[n_1]-SU(2)^m-[n_2]$, where we take $n_1\geq n_2$ without loss of generality. If either of the $n_i$ vanishes we also have to specify the discrete $\theta$-angle. However, they usually have a dual description as an SQCD theory. This means that we only have to consider the theories with $n_2=0$ and $\theta=0$. These are summarised in table \ref{tab:UVMQsSU2m}. 

\begin{sidewaystable}

\scalebox{\scale}{%
{\huge

}
}
\caption{UV-MQs for descendants of $SU(N)_0 + (2N+4) \bm{F}$. The lines and columns are labeled by $\frac{N_f}{2} \pm k$. The table is symmetric under $k \rightarrow -k$, so the dashes stand for the quivers in the corresponding symmetric box. 
\label{tab:SQCD}}

\end{sidewaystable}

\begin{sidewaystable}
\begin{center}
\scalebox{\scale}{%
{\huge
\begin{tabular}{ | c || c | c | c | c | c |}
\hline
$N_F \backslash \frac{N_F}{2}+k$ & $n+4$ & $n+3$ & $n+2$ & $n+1$ & $n$\\
\hline
\hline
8
&
Marginal
&

&

&

&

\\
&
&
&
&
&
\\
\hline
7
&
\begin{tikzpicture}[x=.8cm,y=.8cm]
\node (h1) at (0,0) [gaugeb,label=below:\large{1}] {};
\node (h2) at (1,0) [gaugeb,label=below:\large{$2n$-1}] {};
\node (h3) at (2.5,0) [gauge,label=below:\large{$4n$-2}] {};
\node (h4) at (4,0) [gauge,label=below:\large{$6n$-3}] {};
\node (h5) at (5.5,0) [gauge,label=below:\large{$8n$-4}] {};
\node (h6) at (7,0) [gauge,label=below:\large{$10n$-5}] {};
\node (h7) at (9,0) [gauge,label=below:\large{$12n$-6}] {};
\node (h8) at (10.5,0) [gauge,label=below:\large{$8n$-4}] {};
\node (h9) at (12,0) [gauge,label=below:\large{$4n$-2}] {};
\node (h10) at (9,1) [gauge,label=right:\large{$6n$-3}] {};
\draw (h1)--(h2)--(h3)--(h4)--(h5)--(h6)--(h7)--(h8)--(h9);
\draw (h7)--(h10);

\end{tikzpicture}
&
\begin{tikzpicture}[x=.8cm,y=.8cm]
\node (h1) at (0,0) [gauge,label=below:\large{1}] {};
\node (h2) at (1,0) [gauge,label=below:\large{2}] {};
\node (h3) at (2,0) [gauge,label=below:\large{3}] {};
\node (h4) at (3,0) [gauge,label=below:\large{4}] {};
\node (h5) at (4,0) [gauge,label=below:\large{5}] {};
\node (h6) at (5,0) [gauge,label=below:\large{6}] {};
\node (h7) at (6,0) [gauge,label=below:\large{4}] {};
\node (h8) at (7,0) [gaugeb,label=below:\large{2}] {};
\node (h9) at (5,1) [gaugeb,label=above:\large{3}] {};
\node (h10) at (6,1) [gaugeb,label=above:\large{1}] {};
\draw (h1)--(h2)--(h3)--(h4)--(h5)--(h6)--(h7)--(h8);
\draw (h6)--(h9)--(h10);
\draw (h10)--(h8) node[midway,above right=-.2] {\large $n$-2};
\end{tikzpicture}
&

&

&

\\
&
$\e_8\oplus \su(2)$
&
$\su(8)\oplus \uu(1)^2$
&
&
&
\\
\hline

\end{tabular}

}
}
\end{center}
\caption{UV-MQs for descendants of $SU(2n)_{n} +1 \bm{AS} + 8 \bm{F}$ with $N_F\geq 4$
 \label{tab:UVMQsEStringEven}}
\end{sidewaystable}


\begin{sidewaystable}
\begin{center}
\scalebox{\scale}{%
{\huge
\begin{tabular}{ | c || c | c | c | c | c |}
\hline
$N_F \backslash \frac{N_F}{2}+k$ & $n+9/2$ & $n+7/2$ & $n+5/2$ & $n+3/2$ & $n+1/2$\\
\hline
\hline
8
&
Marginal
&

&

&

&

\\
&
&
&
&
&
\\
\hline

\end{tabular}
}
}
\end{center}
\caption{UV-MQs for descendants of $SU(2n+1)_{n+\half} +1 \bm{AS} + 8 \bm{F}$ with $N_F\geq 4$
 \label{tab:UVMQsEStringOdd}}
\end{sidewaystable}


\begin{table}
\scalebox{\scale}{%
{\huge
\begin{tabular}{ | c || c | c | c | c | c |}
\hline
$N_F \backslash \frac{N_F}{2}+k$ & $n+3$ & $n+2$ & $n+1$ & $n$ & $n-1$\\
\hline
\hline
$2n+6$
&
Marginal
&

&

&

&

\\
&
&
&
&
&
\\
\hline
$2n+5$
&
\begin{tikzpicture}[x=.8cm,y=.8cm]
\node (g1) at (0,0) [gauge,label=below:\large{1}] {};
\node (g2) at (1,0) {\large$\cdots$};
\node (g3) at (2,0) [gauge,label=below:\large{$2n$+5}] {};
\node (g4) at (3.5,0) [gauge,label=below:\large{$n$+4}] {};
\node (g5) at (4.5,0) [gaugeb,label=below:\large{3}] {};
\node (g6) at (2,1) [gaugeb,label=right:\large{$n$+2}] {};
\draw (g1)--(g2)--(g3)--(g4)--(g5);
\draw (g3)--(g6);
\end{tikzpicture}
&
$-$
&

&

&

\\
&
$\su(2n+7)\oplus \uu(1)$
&
&
&
&
\\
\hline
$2n+4$
&
\begin{tikzpicture}[x=.8cm,y=.8cm]
\node (g1) at (0,0) [gauge,label=below:\large{1}] {};
\node (g2) at (1,0) {\large$\cdots$};
\node (g3) at (2,0) [gauge,label=below:\large{$2n$+3}] {};
\node (g4) at (3.5,0) [gauge,label=below:\large{$n$+3}] {};
\node (g5) at (4.5,0) [gaugeb,label=below:\large{3}] {};
\node (g6) at (5.5,0) [gaugeb,label=below:\large{1}] {};
\node (g7) at (2,1) [gaugeb,label=right:\large{$n$+1}] {};
\draw (g1)--(g2)--(g3)--(g4)--(g5)--(g6);
\draw (g3)--(g7);
\end{tikzpicture}
&
\begin{tikzpicture}[x=.8cm,y=.8cm]
\node (g1) at (0,0) [gauge,label=below:\large{1}] {};
\node (g2) at (1,0) {\large$\cdots$};
\node (g3) at (2,0) [gauge,label=below:\large{$2n$+2}] {};
\node (g4) at (3.5,0) [gaugeb,label=below:\large{$n$+2}] {};
\node (g5) at (4.5,0) [gaugeb,label=below:\large{3}] {};
\node (g6) at (5.5,0) [gauge,label=below:\large{2}] {};
\node (g7) at (6.5,0) [gauge,label=below:\large{1}] {};
\node (g8) at (2,1) [gauge,label=right:\large{$n$+1}] {};
\draw (g1)--(g2)--(g3)--(g4)--(g5)--(g6)--(g7);
\draw (g3)--(g8);
\end{tikzpicture}
&
$-$
&

&

\\
&
$\su(2n+5)\oplus \uu(1)^2$
&
$\su(2n+4)\oplus \su(3)\oplus \uu(1)$
&
&
&
\\
\hline
\end{tabular}
}
}
\caption{UV-MQs for descendants of $SU(2n)_0 +1 \bm{AS} + (2n+6) \bm{F}$ with $N_F\geq (2n-2)$ \label{tab:UVMQs1ASEven}}
\end{table}


\begin{table}

\scalebox{\scale}{%
{\huge
\begin{tabular}{ | c || c | c | c | c | c |}
\hline
$N_F \backslash \frac{N_F}{2}+k$ & $n+7/2$ & $n+5/2$ & $n+3/2$ & $n+1/2$ & $n-1/2$\\
\hline
\hline
$2n+7$
&
Marginal
&

&

&

&

\\
&
&
&
&
&
\\
\hline
$2n+6$
&
\begin{tikzpicture}[x=.8cm,y=.8cm]
\node (g1) at (0,0) [gauge,label=below:\large{1}] {};
\node (g2) at (1,0) {\large$\cdots$};
\node (g3) at (2,0) [gauge,label=below:\large{$2n$+6}] {};
\node (g4) at (3.5,0) [gaugeb,label=below:\large{$n$+4}] {};
\node (g5) at (4.5,0) [gaugeb,label=below:\large{3}] {};
\node (g6) at (2,1) [gauge,label=right:\large{$n$+3}] {};
\draw (g1)--(g2)--(g3)--(g4)--(g5);
\draw (g3)--(g6);
\end{tikzpicture}
&
$-$
&

&

&

\\
&
$\su(2n+8)\oplus \uu(1)$
&
&
&
&
\\
\hline
$2n+5$
&
\begin{tikzpicture}[x=.8cm,y=.8cm]
\node (g1) at (0,0) [gauge,label=below:\large{1}] {};
\node (g2) at (1,0) {\large$\cdots$};
\node (g3) at (2,0) [gauge,label=below:\large{$2n$+4}] {};
\node (g4) at (3.5,0) [gaugeb,label=below:\large{$n$+3}] {};
\node (g5) at (4.5,0) [gaugeb,label=below:\large{3}] {};
\node (g6) at (5.5,0) [gaugeb,label=below:\large{1}] {};
\node (g8) at (2,1) [gauge,label=right:\large{$n$+2}] {};
\draw (g1)--(g2)--(g3)--(g4)--(g5)--(g6);
\draw (g3)--(g8);
\end{tikzpicture}
&
\begin{tikzpicture}[x=.8cm,y=.8cm]
\node (g1) at (0,0) [gauge,label=below:\large{1}] {};
\node (g2) at (1,0) {\large$\cdots$};
\node (g3) at (2,0) [gauge,label=below:\large{$2n$+3}] {};
\node (g4) at (3.5,0) [gauge,label=below:\large{$n$+3}] {};
\node (g5) at (4.5,0) [gaugeb,label=below:\large{3}] {};
\node (g6) at (5.5,0) [gauge,label=below:\large{2}] {};
\node (g7) at (6.5,0) [gauge,label=below:\large{1}] {};
\node (g8) at (2,1) [gaugeb,label=right:\large{$n$+1}] {};
\draw (g1)--(g2)--(g3)--(g4)--(g5)--(g6)--(g7);
\draw (g3)--(g8);
\end{tikzpicture}
&
$-$
&

&

\\
&
$\su(2n+6)\oplus \uu(1)^2$
&
$\su(2n+5)\oplus \su(3)\oplus \uu(1)$
&
&
&
\\
\hline
\end{tabular}
}
}
\caption{UV-MQs for descendants of $SU(2n+1)_0 +1 \bm{AS} + (2n+7) \bm{F}$ with $N_F\geq (2n-2)$ \label{tab:UVMQs1ASOdd}}
\end{table}


\begin{sidewaystable}

\scalebox{\scale}{%
{\huge
\begin{tabular}{ | c || c | c | c | c | c | c |}
\hline
$N_F \backslash \frac{N_F}{2}+k$ & 5 & 4 & 3 & 2 & 1 & 0\\
\hline
\hline
8
&

&
Marginal
&

&

&

&

\\
&
&
&
&
&
&
\\
\hline
7
&
Marginal
&
\begin{tikzpicture}[x=.8cm,y=.8cm]
\node (g1) at (0,0) [gauge,label=below:\large{1}] {};
\node (g2) at (1,0) [gaugeb,label=below:\large{2}] {};
\node (g3) at (2,0) [gauge,label=below:\large{$n$+1}] {};
\node (g4) at (3,0) [gaugeb,label=below:\large{$2n$}] {};
\node (g5) at (4,0) [gauge,label=below:\large{$3n$}] {};
\node (g6) at (5,0) [gauge,label=below:\large{$4n$}] {};
\node (g7) at (6,0) [gauge,label=below:\large{$5n$}] {};
\node (g8) at (7,0) [gauge,label=below:\large{$6n$}] {};
\node (g9) at (8,0) [gauge,label=below:\large{$4n$}] {};
\node (g10) at (9,0) [gauge,label=below:{\large$2n$}] {};
\node (g11) at (7,1) [gauge,label=right:\large{$3n$}] {};
\draw (g1)--(g2)--(g3)--(g4)--(g5)--(g6)--(g7)--(g8)--(g9)--(g10);
\draw (g8)--(g11);
\end{tikzpicture}
&
$-$
&

&

&

\\
&
&
$\e_7\oplus \su(2)^3$
&
&
&
&
\\
\hline

\end{tabular}

}
}

\caption{UV-MQs for $SU(2n)_k +2 \bm{AS} + N_F \bm{F}$
 \label{tab:UVMQs2ASSU2n}}
\end{sidewaystable}


\begin{sidewaystable}

\scalebox{\scale}{%
{\huge
\begin{tabular}{ | c || c | c | c | c | c | c |}
\hline
$N_F \backslash \frac{N_F}{2}+k$ & 5 & 4 & 3 & 2 & 1 & 0\\
\hline
\hline
8
&

&
Marginal
&

&

&

&

\\
&
&
&
&
&
&
\\
\hline
7
&
Marginal
&
\begin{tikzpicture}[x=.8cm,y=.8cm]
\node (g1) at (1,0) [gauge,label=below:\large{1}] {};
\node (g2) at (2,0) [gaugeb,label=below:\large{2}] {};
\node (g3) at (3,0) [gauge,label=below:\large{$n$+2}] {};
\node (g4) at (4.5,0) [gauge,label=below:\large{$2n$+2}] {};
\node (g5) at (6,0) [gauge,label=below:\large{$3n$+2}] {};
\node (g6) at (7.5,0) [gauge,label=below:\large{$4n$+2}] {};
\node (g7) at (9,0) [gauge,label=below:\large{$5n$+2}] {};
\node (g8) at (10.5,0) [gauge,label=below:\large{$6n$+2}] {};
\node (g9) at (12,0) [gauge,label=below:\large{$4n$+1}] {};
\node (g10) at (13.5,0) [gaugeb,label=below:{\large$2n$}] {};
\node (g11) at (10.5,1) [gauge,label=right:\large{$3n$+1}] {};
\draw (g1)--(g2)--(g3)--(g4)--(g5)--(g6)--(g7)--(g8)--(g9)--(g10);
\draw (g8)--(g11);
\end{tikzpicture}
&
$-$
&

&

&

\\
&
&
$\so(16)\oplus \su(2)^2$
&
&
&
&
\\
\hline

\end{tabular}

}
}

\caption{UV-MQs for $SU(2n+1)_k +2 \bm{AS} + N_F \bm{F}$
 \label{tab:UVMQs2ASSU2n+1}}
\end{sidewaystable}


\begin{sidewaystable}

\scalebox{\scale}{%
{\huge
\begin{tabular}{ | c || c | c | c | c | c | c |}
\hline
$N_F \backslash \frac{N_F}{2}+k$ & 5 & 4 & 3 & 2 & 1 & 0\\
\hline
\hline
8
&

&
Marginal
&

&

&

&

\\
&
&
&
&
&
&
\\
\hline
7
&
Marginal
&
\begin{tikzpicture}[x=.8cm,y=.8cm]
\node (g1) at (0,0) [gauge,label=below:\large{1}] {};
\node (g2) at (1,0) [gauge,label=below:\large{2}] {};
\node (g3) at (2,0) [gauge,label=below:\large{3}] {};
\node (g4) at (3,0) [gaugeb,label=below:\large{4}] {};
\node (g5) at (4,0) [gauge,label=below:\large{6}] {};
\node (g6) at (5,0) [gauge,label=below:\large{8}] {};
\node (g7) at (6,0) [gauge,label=below:\large{10}] {};
\node (g8) at (7,0) [gauge,label=below:\large{12}] {};
\node (g9) at (8,0) [gauge,label=below:\large{8}] {};
\node (g10) at (9,0) [gauge,label=below:\large{4}] {};
\node (g11) at (7,1) [gauge,label=right:\large{6}] {};
\draw (g1)--(g2)--(g3)--(g4)--(g5)--(g6)--(g7)--(g8)--(g9)--(g10);
\draw (g8)--(g11);
\end{tikzpicture}
&
$-$
&

&

&

\\
&
&
$\e_7\oplus \so(7)$
&
&
&
&
\\
\hline

\end{tabular}

}
}

\caption{UV-MQs for $SU(4)_k +2 \bm{AS} + N_F \bm{F}$
 \label{tab:UVMQs2ASSU4}}
\end{sidewaystable}


\begin{sidewaystable}
\scalebox{\scale}{%
{\huge
\begin{tabular}{ | c ||  c | c | c | c | c | c | c | c |}
    \hline
    $N_F$ &  7 & 6 & 5 & 4 & 3 & 2 & 1 & 0 \\
\hline
\hline

\end{tabular}

}
}

\caption{UV-MQs for $SU(5)_{5-\frac{N_F}{2}} +2 \bm{AS} + N_F \bm{F}$ 
 \label{tab:UVMQs2ASSU5}}
 
 \bigskip
 \bigskip
 \bigskip
 \bigskip
  \bigskip
 \bigskip
 \bigskip
 \bigskip
 
 \scalebox{\scale}{%
{\huge
\begin{tabular}{ | c || c | c | c | c | c | c |}
    \hline
    $n_1$ & 4 & 3 & 2 & 1 & 0, \quad $\theta=\pi$ & 0,\quad $\theta=0$   \\
\hline
\hline
\end{tabular}

}
}

\caption{UV-MQs for $\left[n_1\right]-SU(2)^m-[0]_{\theta=0}$
 \label{tab:UVMQsSU2m}}

\end{sidewaystable}

\section{Brane-Webs} 
\label{app:webs}

\subsection{\texorpdfstring{${SU(N)}$ Gauge Theories}{SU(N) Gauge Theories}}
\label{app:SUNWebs}

In this appendix we collect various existing results in the literature, regarding the $(p,q)$ 5-brane-webs of $SU(N)$ gauge theories with anti-symmetric and fundamental matter. These were discussed in a variety of papers \cite{Bergman:2015dpa,Zafrir:2015rga,Jefferson:2017ahm,Hayashi:2018lyv}. We will use these results to construct the corresponding GTPs summarised in table \ref{tab:ToricPolygons}.
This discussion is ordered by the number of anti-symmetric hypermultiplets, $N_{AS}$.
First, for $N_{AS}=0$ the theories are SQCDs. The brane-web for the $SU(N)_k + N_F \bm{F}$ is given by\footnote{The drawing is for parameters $(N,k,N_L,N_R,a)=(N,\half,N-3,N-4,1)$}
\be
\begin{tikzpicture}

\draw[brane] (-.5,0)--(6.5,0); 
\draw[brane] (-3,0)--(-1.5,0);
\draw[brane] (7.5,0)--(9,0);
\draw[brane] (2,0)--(0,1);
\draw[brane] (2,0)--(1,-1);
\draw[brane] (4,0)--(7,1);
\draw[brane] (4,0)--(5,-1);

\node[D7] at (0,0) {};
\node at (-1,0) {$\hdots$};
\node[D7] at (-2,0) {};
\node[D7] at (-3,0) {};
\node[D7] at (6,0) {};
\node at (7,0) {$\hdots$};
\node[D7] at (8,0) {};
\node[D7] at (9,0) {};
\node[D7,label=above:{$(-k+\frac{N_L-N_R}{2}-a,1)$}] at (0,1) {};
\node[D7,label=below:{$(-N+k+\frac{N_L+N_R}{2}+a,-1)$}] at (1,-1) {};
\node[D7,label=above:{$(N-N_R-a,1)$}] at (7,1) {};
\node[D7,label=below:{$(a,-1)$}] at (5,-1) {};

\node at (-2.5,-.25) {1};
\node at (0.75,-.25) {$N_L$};
\node at (3,-.25) {$N$};
\node at (5.25,-.25) {$N_R$};
\node at (8.5,-.25) {1};

\end{tikzpicture}\ee
If not stated otherwise, the same number of D5-branes end on each of the D7-branes and unlabelled 5-branes have multiplicity one.
There are apparently two free parameters. First, we can choose the number of D7-branes on the left and on the right respectively, where
\be
N_L +N_R = N_F\,.
\ee
Secondly, there is an overall $SL(2,\Z)$ transformation with $\phi=T^n$, that leaves the D7-branes invariant.\footnote{In fact, we could also choose any other $SL(2,\Z)$-transformation.} This freedom is parametrised by $a$. However, the resulting theory is in fact invariant under both these operations. Thus, we can choose $N_L-N_R$ and $a$ to simplify the setup. In the following, we will always use an appropriate choice of $SL(2,\Z)$-frame.

Next, consider the case with a single anti-symmetric $N_{AS}=1$. It was found that the web can be written in terms of a (resolved) $O7^-$-plane and depends on whether $N$ is even or odd. 
Concretely, the web is\footnote{Both drawings are for parameters $(N,k,N_F)=(N,\half,N-3)$}
\be
\begin{tikzpicture}

\draw[brane] (-2,2)--(0,0)--(-1,-1);
\draw[brane] (0,0)--(4.5,0);
\draw[brane] (5.5,0)--(7,0);
\draw[brane] (4,1)--(2,0)--(3,-1);

\node[D7] at (-1,1) {};
\node[D7] at (-2,2) {};
\node[D7] at (-1,-1) {};
\node[D7] at (4,0) {};
\node at (5,0) {$\hdots$};
\node[D7] at (6,0) {};
\node[D7] at (7,0) {};
\node[D7,label=above:{$(n-k-\frac{N_F}{2}+1,1)$}] at (4,1) {};
\node[D7,label=below:{$(n+k-\frac{N_F}{2}-1,-1)$}] at (3,-1) {};

\node at (-1.75,1.5) {1};
\node at (-.75,.5) {$n$};
\node at (-.75,-.5) {$n$};
\node at (1,-.25) {$2n$};
\node at (3.25,-.25) {$N_F$};
\node at (6.5,-.25) {1};

\end{tikzpicture}
\ee
for even $N=2n$ and
\be
\begin{tikzpicture}

\draw[brane] (-1,1)--(0,0)--(-1,-1);
\draw[brane] (0,1)--(0,0);
\draw[brane] (0,0)--(4.5,0);
\draw[brane] (5.5,0)--(7,0);
\draw[brane] (5,1)--(2,0)--(2,-1);

\node[D7] at (-1,1) {};
\node[D7] at (0,1) {};
\node[D7] at (-1,-1) {};
\node[D7] at (4,0) {};
\node at (5,0) {$\hdots$};
\node[D7] at (6,0) {};
\node[D7] at (7,0) {};
\node[D7,label=above:{$(n-k-\frac{N_F}{2}+\frac{5}{2},1)$}] at (5,1) {};
\node[D7,label=below:{$(n+k-\frac{N_F}{2}-\frac{3}{2},-1)$}] at (2,-1) {};

\node at (.25,.5) {1};
\node at (-.75,.5) {$n$};
\node at (-1,-.5) {$n$+1};
\node at (1,-.25) {$2n$+1};
\node at (3.25,-.25) {$N_F$};
\node at (6.5,-.25) {1};

\end{tikzpicture}
\ee
for odd $N=2n+1$. In both cases we choose to have anti-symmetric and fundamental matter on different sides, i.e. $N_L=0$, but this is not strictly necessary.

Finally, there are theories with $N_{AS}=2$. Again, they depend on the parity of $N$ and are given by\footnote{Both drawings are for parameters $(N,k,N_L,N_R)=(N,1,1,3)$}
\be
\begin{tikzpicture}

\draw[brane] (-2.5,0)--(4.5,0);
\draw[brane] (-5,0)--(-3.5,0);
\draw[brane] (5.5,0)--(7,0);

\draw[brane] (0,0)--(0,1)--(0,2);
\draw[brane] (0,0)--(-1,-1);
\draw[brane] (2,0)--(1,1);
\draw[brane] (2,0)--(2,-1)--(2,-2);

\node[D7] at (-2,0) {};
\node at (-3,0) {$\hdots$};
\node[D7] at (-4,0) {};
\node[D7] at (-5,0) {};
\node[D7] at (4,0) {};
\node at (5,0) {$\hdots$};
\node[D7] at (6,0) {};
\node[D7] at (7,0) {};
\node[D7] at (0,1) {};
\node[D7,label=above:{$(k+\frac{N_L-N_R}{2},1)$}] at (0,2) {};
\node[D7,label=below:{$(-k+\frac{N_L+N_R}{2}-2,-1)$}] at (-1,-1) {};
\node[D7,label=above right:{$(2-N_R,1)$}] at (1,1) {};
\node[D7] at (2,-1) {};
\node[D7,label=below:{$(0,-1)$}] at (2,-2) {};

\node at (1,-.25) {$2n$};
\node at (3.25,-.25) {$N_R n$};
\node at (6.5,-.25) {$n$};
\node at (-1.25,-.25) {$N_L n$};
\node at (-4.5,-.25) {$n$};
\node at (-.25,.5) {$n$};
\node at (-.25,1.5) {1};
\node at (-.5,-.75) {$n$};
\node at (1.5,.75) {$n$};
\node at (2.25,-.5) {$n$};
\node at (2.25,-1.5) {1};

\end{tikzpicture}
\ee
for even $N=2n$ and
\be
\begin{tikzpicture}

\draw[brane] (-2.5,0)--(4.5,0);
\draw[brane] (-6,0)--(-3.5,0);
\draw[brane] (5.5,0)--(8,0);

\draw[brane] (0,0)--(0,1);
\draw[brane] (0,0)--(-1,-1);
\draw[brane] (2,0)--(1,1);
\draw[brane] (2,0)--(2,-1);

\node[D7] at (-2,0) {};
\node at (-3,0) {$\hdots$};
\node[D7] at (-4,0) {};
\node[D7] at (-5,0) {};
\node[D7] at (-6,0) {};
\node[D7] at (4,0) {};
\node at (5,0) {$\hdots$};
\node[D7] at (6,0) {};
\node[D7] at (7,0) {};
\node[D7] at (8,0) {};
\node[D7,label=above left:{$(k+\frac{N_L-N_R}{2},1)$}] at (0,1) {};
\node[D7,label=below:{$(-k+\frac{N_L+N_R}{2}-2,-1)$}] at (-1,-1) {};
\node[D7,label=above right:{$(2-N_R,1)$}] at (1,1) {};
\node[D7,label=below:{$(0,-1)$}] at (2,-1) {};

\node at (1,-.25) {$2n$+1};
\node at (3.25,-.25) {$N_R n$+1};
\node at (6.5,-.25) {$n$+1};
\node at (7.5,-.25) {1};
\node at (-1.25,-.25) {$N_L n$+1};
\node at (-4.5,-.25) {$n$+1};
\node at (-5.5,-.25) {1};
\node at (-.25,.5) {$n$};
\node at (-.5,-.75) {$n$};
\node at (1.5,.75) {$n$};
\node at (2.25,-.5) {$n$};

\end{tikzpicture}
\ee
for odd $N=2n+1$.

\subsection{Quiver Gauge Theories}
\label{app:QuiverWebs}

We consider the brane-web of a 5d $[n_1]-SU(2)^m-[n_2]$ quiver gauge theory, which was given for the rank $m=2$ case in \cite{Hayashi:2018lyv}. 
The web, seen in figure \ref{fig:SU2mweb}, consists of $m$ horizontally adjacent compact faces, a set of $2(m-1)$ external NS5-branes, $n_1+n_2$ external D5-branes, and 4 external branes at an angle in the $(x_5,x_6)$-plane. The height of each compact face corresponds to a Coulomb branch parameter $\phi_i$, i.e. there are $m$ local deformations of the web. 
The fundamental matter is represented by $n_1$ ($n_2$) external D5-branes attached to the leftmost (rightmost) compact face, where the vertical position of each D5-brane corresponds to a fundamental mass parameter $m_i$. Conventionally, the web is drawn in its most symmetrical phase, with $\lceil n_i/2 \rceil$ of the D5-branes attached above the compact faces and $\lfloor n_i/2 \rfloor$ below the compact faces. The bifundamental matter is represented by the external NS5-branes attached to the compact faces, and their mass parameters $m_b$ are related to the length of the $(1,\pm 1)$-5-branes in the junction.
The number of 7-branes in the web is 
\be 
\# \text{7-branes}=n_1+n_2+2(m-1)+4\,.
\ee

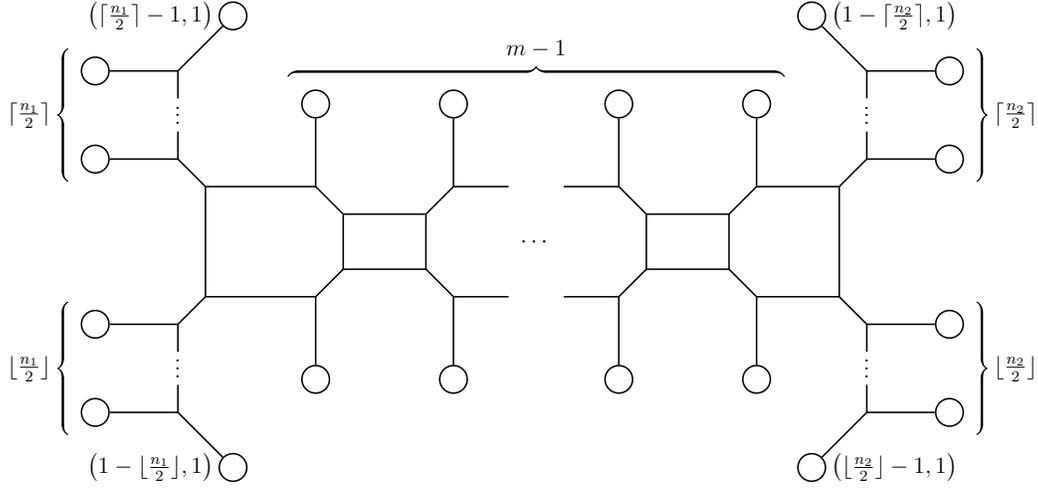
\begin{figure}[t]
\centering
\resizebox{0.9\textwidth}{!}{%
{\large
\begin{tikzpicture}
\node[roundnode] at (2,1.5) (1) {};
\node[roundnode] at (2,-3.5) (2) {};
\node[roundnode] at (13.5,0.5) (3) {};
\node[roundnode] at (13.5,2.1) (4) {};
\node[roundnode] at (11,3.1) (5) {};
\node[roundnode] at (13.5,-2.5) (6) {};
\node[roundnode] at  (13.5,-4.1) (7) {};
\node[roundnode] at  (11,-5.1) (8) {};
\node[roundnode] at  (-2,0.5) (9) {};
\node[roundnode] at  (-2,2.1) (10) {};
\node[roundnode] at  (0.5,3.1) (11) {};
\node[roundnode] at  (-2,-2.5) (12) {};
\node[roundnode] at  (-2,-4.1) (13) {};
\node[roundnode] at  (0.5,-5.1) (14) {};
\node[roundnode] at  (4.5,1.5) (15) {};
\node[roundnode] at  (4.5,-3.5) (16) {};
\node[roundnode] at  (7.5,1.5) (17) {};
\node[roundnode] at  (7.5,-3.5) (18) {};
\node[roundnode] at  (10,1.5) (19) {};
\node[roundnode] at  (10,-3.5) (20) {};

\node at (12,1.4) {$\vdots$};
\node at (12,-3.2) {$\vdots$};
\node at (-0.5,1.4) {$\vdots$};
\node at (-0.5,-3.2) {$\vdots$};
\node at (6,-1) {$\cdots$};
\node at (6,2.5) {$m-1$};

\draw[thick] (0,0)--(2,0)--(1);
\draw[thick] (0,0)--(0,-2)--(2,-2)--(2);
\draw[thick] (2,0)--(2.5,-0.5)--(2.5,-1.5);
\draw[thick] (2,-2)--(2.5,-1.5)--(4,-1.5)--(4,-0.5);
\draw[thick] (4.5,0)--(15);
\draw[thick] (2.5,-0.5)--(4,-0.5)--(4.5,0)--(5.5,0);
\draw[thick] (4,-1.5)--(4.5,-2)--(5.5,-2);
\draw[thick] (4.5,-2)--(16);
\draw[thick] (6.5,0)--(7.5,0)--(8,-0.5)--(9.5,-0.5)--(10,0)--(11.5,0);
\draw[thick] (7.5,0)--(17);
\draw[thick] (7.5,-2)--(18);
\draw[thick] (10,0)--(19);
\draw[thick] (10,-2)--(20);
\draw[thick] (6.5,-2)--(7.5,-2)--(8,-1.5)--(9.5,-1.5)--(10,-2)--(11.5,-2);
\draw[thick] (8,-0.5)--(8,-1.5);
\draw[thick] (9.5,-0.5)--(9.5,-1.5);
\draw[thick] (11.5,0)--(11.5,-2);
\draw[thick] (11.5,0)--(12,0.5)--(3);
\draw[thick] (12,0.5)--(12,1);
\draw[thick] (12,1.6)--(12,2.1)--(4);
\draw[thick] (12,2.1)--(5) node[xshift=1.5cm] {$\left(1- \lceil \frac{n_2}{2} \rceil, 1\right)$};
\draw[thick] (11.5,-2)--(12,-2.5)--(6);
\draw[thick] (12,-2.5)--(12,-3);
\draw[thick] (12,-3.6)--(12,-4.1)--(7);
\draw[thick] (12,-4.1)--(8) node[xshift=1.5cm] {$\left( \lfloor \frac{n_2}{2} \rfloor-1, 1\right)$};
\draw[thick] (0,0)--(-0.5,0.5)--(9);
\draw[thick] (-0.5,0.5)--(-0.5,1);
\draw[thick] (-0.5,1.6)--(-0.5,2.1)--(10);
\draw[thick] (-0.5,2.1)--(11) node[xshift=-1.5cm] {$\left( \lceil \frac{n_1}{2} \rceil-1, 1\right)$};
\draw[thick] (0,-2)--(-0.5,-2.5)--(12);
\draw[thick] (-0.5,-2.5)--(-0.5,-3);
\draw[thick] (-0.5,-3.6)--(-0.5,-4.1)--(13);
\draw[thick] (-0.5,-4.1)--(14) node[xshift=-1.5cm] {$\left( 1- \lfloor \frac{n_1}{2} \rfloor, 1\right)$};

\draw[decoration={calligraphic brace,amplitude=5pt}, decorate, line width=1.25pt] (1.5,2)--(10.5,2);
\draw[decoration={calligraphic brace,amplitude=5pt,mirror}, decorate, line width=1.25pt] ($(3)+(0.5,-0.4)$) -- ($(4)+(0.5,0.4)$);
\draw[decoration={calligraphic brace,amplitude=5pt}, decorate, line width=1.25pt] ($(6)+(0.5,0.4)$) --($(7)+(0.5,-0.4)$);
\draw[decoration={calligraphic brace,amplitude=5pt}, decorate, line width=1.25pt] ($(9)+(-0.5,-0.4)$) -- ($(10)+(-0.5,0.4)$);
\draw[decoration={calligraphic brace,amplitude=5pt,mirror}, decorate, line width=1.25pt] ($(12)+(-0.5,0.4)$) --($(13)+(-0.5,-0.4)$);
\node at ($(3)!0.5!(4)+(1.2,0)$) {$\lceil \frac{n_2}{2} \rceil$};
\node at ($(6)!0.5!(7)+(1.2,0)$) {$\lfloor \frac{n_2}{2} \rfloor$};
\node at ($(9)!0.5!(10)-(1.2,0)$) {$\lceil \frac{n_1}{2} \rceil$};
\node at ($(12)!0.5!(13)-(1.2,0)$) {$\lfloor \frac{n_1}{2} \rfloor$};

\end{tikzpicture}
}
}
\caption{Brane web for $[n_1]-SU(2)^m-[n_2]$ quiver gauge theory.
 \label{fig:SU2mweb}}
\end{figure}

When one (or both) $n_i=0$, i.e. there is no fundamental matter associated to one (or both) of the $SU(2)$s, there are two choices for the angles of the left/right external legs, corresponding to $\theta=0,\pi$ for the associated $SU(2)$. A set of $(1,\pm 1)$-5-branes gives rise to a $\theta=\pi$ gauge theory, whereas an NS5-brane and a $(2,\pm1)$-5-brane corresponds to a $\theta=0$ theory.

To reach the SCFT phase, we take all the matter to be massless and go to the origin of the Coulomb branch by collapsing the compact faces and taking all the D5-branes to lie along this line. This is shown in figure \ref{fig:SU2mmassless}.
Taking the strong coupling limit is then achieved by taking the length of the $m$ internal D5-branes to zero. This can be done immediately for $n_i \leq 2$, otherwise  it is necessary to perform Hanany-Witten moves before it is possible to take this limit.

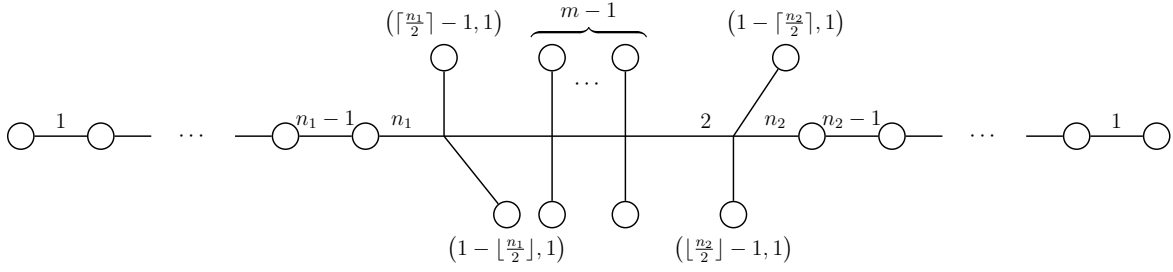
\begin{figure}[t]
\centering
\resizebox{\textwidth}{!}{%
{\large
\begin{tikzpicture}
\node[roundnode] (2) {};
\node[roundnode] (3) [right=of 2] {};
\node (4) [right=0.7 of 3] {};
\node[roundnode] (5) [right=3cm of 3] {};
\node (6) [left=0.7 of 5] {};
\node[roundnode] (7) [right=of 5] {};
\node[roundnode] (8) [right=8cm of 7] {};
\node[roundnode] (9) [right=of 8] {};
\node (10) [right=0.7 of 9] {};
\node[roundnode] (11) [right=3cm of 9] {};
\node (12) [left=0.7cm of 11] {};
\node[roundnode] (13) [right=of 11] {};

\node at ($(3)!0.5!(5)$) {$\cdots$};
\node at ($(9)!0.5!(11)$) {$\cdots$};
\node at ($(7)!0.5!(8)+(0,1)$) {$\cdots$};
\node at ($(7)!0.5!(8)+(0,2.4)$) {$m-1$};

\node[roundnode] at ($(7)!0.5!(8)+(-0.7,1.5)$) (15) {};
\node[roundnode] at ($(7)!0.5!(8)-(0.7,1.5)$) (16) {};
\node[roundnode] at ($(7)!0.5!(8)+(0.7,1.5)$) (21) {};
\node[roundnode] at ($(7)!0.5!(8)-(-0.7,1.5)$) (22) {};

\node[roundnode,label=above:{$\left( \lceil \frac{n_1}{2} \rceil-1, 1\right)$}] at ($(7)+(1.5,1.5)$) (17) {};
\node[roundnode,label=below:{$\left( 1- \lfloor \frac{n_1}{2} \rfloor, 1\right)$}] at ($(7)+(2.7,-1.5)$) (18) {};
\node[roundnode,label=above:{$\left(1- \lceil \frac{n_2}{2} \rceil, 1\right)$}] at ($(8)+(-0.5,1.5)$) (19) {};
\node[roundnode,label=below:{$\left( \lfloor \frac{n_2}{2} \rfloor-1, 1\right)$}] at ($(8)+(-1.5,-1.5)$) (20) {};

\draw[decoration={calligraphic brace,amplitude=5pt}, decorate, line width=1.25pt] ($(7)!0.5!(8)+(-1.1,1.9)$)--($(7)!0.5!(8)+(1.1,1.9)$);

\draw[thick] (2) -- (3) node[midway,above] {1};
\draw[thick] (3) -- (4);
\draw[thick] (5) -- (6);
\draw[thick] (5) -- (7) node[midway,above] {$n_1-1$};
\draw[thick] (7) node[above, xshift=0.7cm] {$n_1$} node[above,xshift=6.5cm] {2} -- (8) node[above, xshift=-0.7cm] {$n_2$};
\draw[thick] (8) -- (9) node[midway,above] {$n_2-1$};
\draw[thick] (9) -- (10);
\draw[thick] (11) -- (12);
\draw[thick] (11) -- (13)node[midway,above] {1};
\draw[thick] (15) -- (16);
\draw[thick] ($(7)+(1.5,0)$) -- (17);
\draw[thick] ($(7)+(1.5,0)$) -- (18);
\draw[thick] ($(8)-(1.5,0)$) -- (19);
\draw[thick] ($(8)-(1.5,0)$) -- (20);
\draw[thick] (21) -- (22);
\end{tikzpicture}
}
}
\caption{Brane web for $[n_1]-SU(2)^m-[n_2]$ quiver gauge theory with massless quarks at the origin of the Coulomb branch.
 \label{fig:SU2mmassless}}
\end{figure}

\section{\texorpdfstring{Rank $(N-1)$ E-Strings: Monodromies and Magnetic Quivers}{Rank (N-1) E-Strings: Monodromies and Magnetic Quivers}}
\label{app:HigherRankEString}

\subsection{General Analysis}

In this appendix we discuss the 5d SCFTs with IR description $SU(N)_{\frac{N+8-N_F}{2}} + 1 \bm{AS} + N_F \bm{F}$, i.e. the higher rank $E_{N_F+1}$-theories. We will discuss how to obtain the magnetic quiver, given by $N-1$ times the affine Dynkin diagram of $\e_{N_F+1}$ with a multiplicity one node attached to the affine node, from the pGTP in \eqref{pGTPEString}. 
Let us consider $N=2n$. In \cite{Jefferson:2018irk} this was discussed in terms of the weakly-coupled brane-web. Translating this into our language the polygon can be brought into the form
\be
\ba
\mathfrak{T}_0&=\left(\mathfrak{M}_{N_F+5}^+\dots\mathfrak{M}_{2}^+\right)\left(\mathfrak{M}_{3}^-\dots\mathfrak{M}_{N_F+2}^-\right)\mathfrak{M}_{-1}^+\mathfrak{M}_{-2}^-\\
\mathfrak{T}_0 \bm{u}&=\Big[(3n-3,-3n+3),(1,2n-5),(2n-1,(2n-1)(5-N_F)),\underbrace{(0,2n-1)}_{N_F},\\
&\qquad\qquad\qquad(-7n+5,-7n+5),(2n-1,-2n+1)\Big]
\ea
\ee
This form is universal to all choices of $n$ and $N_F$, but the polygon can still have a non-convexity at $i=2$ with $D_2=-2n-N_F+10$. It was argued that applying the monodromy
\be
\mathfrak{T}_{T^{8-N_F}}=\left(\mathfrak{M}_{1}^+\dots\mathfrak{M}_{2}^+\right)
\ee
acts on the non-convexity as
\be
D_2 \to D_2 + (8-N_F)\,,
\ee
while leaving all other $D_i$ unchanged.
For $N_F < 8$ this process terminates after a finite number of applications of $\mathfrak{T}_{T^{8-N_F}}$ for any choice of $n$, i.e. there is an SCFT.

Let us consider the case $N_F=6$, which turns out to be the simplest. One can check that $\mathfrak{T}_{T^2}$ needs to be applied exactly $n-2$ times, leading to a GTP with vertices
\be
\left(\mathfrak{T}_{T^2}\right)^{n-2} \mathfrak{T}_0 \bm{u}=\Big[(a-1,a-1),(1,1),(a+n-2,a+n-2),\underbrace{(0,a)}_{6},(-3a,-3a),(a-n+2,a-n+2)\Big]\,,
\ee
where $a=n^2-2n+3$.
This GTP allows us to compute the magnetic quiver  to be 
\be
\label{UVMQSUNeven+1AS+6FFromWeb}
\begin{tikzpicture}[x=.8cm,y=.8cm]
\node (g1) at (0,0) [gaugeb,label=below:{1}]{};
\node (g2) at (1.5,0) [gaugeb,label=below:{$a$-$n$+4}] {};
\node (g3) at (3.5,0) [gaugeb,label=below:{$2a$-$n$+3}] {};
\node (g4) at (5,0) [gauger,label=below:{$3a$}] {};
\node (g5) at (6.5,0) [gauge,label=below:{$4a$}] {};
\node (g6) at (8,0) [gauge,label=below:{$3a$}] {};
\node (g7) at (9.5,0) [gauge,label=below:{$2a$}] {};
\node (g8) at (11,0) [gauge,label=below:{$\phantom{1}a$}] {};
\node (g9) at (6.5,1) [gauge,label=right:{$2a$}] {};
\draw (g1)--(g2)--(g3)--(g4)--(g5)--(g6)--(g7)--(g8);
\draw (g5)--(g9);
\end{tikzpicture}\,.
\ee
It looks like we only see an $\su(6)$ flavor symmetry. However, we see that the balance of the fourth node from the left is $-(n-3)<0$, i.e. it is underbalanced, which we denote by coloring the node in red. After applying the local Seiberg duality in \eqref{SimoneDuality}, \eqref{UVMQSUNeven+1AS+6FFromWeb} becomes
\be
\begin{tikzpicture}[x=.8cm,y=.8cm]
\node (g1) at (0,0) [gaugeb,label=below:{1}] {};
\node (g2) at (1.5,0) [gaugeb,label=below:{$a$-$n$+4}] {};
\node (g3) at (3.5,0) [gaugeb,label=below:{$2a$-$n$+3}] {};
\node (g4) at (5.5,0) [gaugeb,label=below:{$3a$-$n$+3}] {};
\node (g5) at (7,0) [gauger,label=below:{$4a$}] {};
\node (g6) at (8.5,0) [gauge,label=below:{$3a$}] {};
\node (g7) at (10,0) [gauge,label=below:{$2a$}] {};
\node (g8) at (11.5,0) [gauge,label=below:{$\phantom{1}a$}] {};
\node (g9) at (7,1) [gauge,label=right:{$2a$}] {};
\draw (g1)--(g2)--(g3)--(g4)--(g5)--(g6)--(g7)--(g8);
\draw (g5)--(g9);
\end{tikzpicture}
\ee
coupled to $(n-3)$ free hypermultiplets.
However, now the central node is underbalanced. Applying \eqref{SimoneDuality} repeatedly we finally arrive at
\be  \label{MQE7HigherRank}
\begin{tikzpicture}[x=.8cm,y=.8cm]
\node (g1) at (0,0) [gaugeb,label=below:{1}] {};
\node (g2) at (1.5,0) [gaugeb,label=below:{$2n$-1}] {};
\node (g3) at (3,0) [gauge,label=below:{$4n$-2}] {};
\node (g4) at (4.5,0) [gauge,label=below:{$6n$-3}] {};
\node (g5) at (6,0) [gauge,label=below:{$8n$-4}] {};
\node (g6) at (7.5,0) [gauge,label=below:{$6n$-3}] {};
\node (g7) at (9,0) [gauge,label=below:{$4n$-2}] {};
\node (g8) at (10.5,0) [gauge,label=below:{$2n$-1}] {};
\node (g9) at (6,1) [gauge,label=right:{$4n$-2}] {};
\draw (g1)--(g2)--(g3)--(g4)--(g5)--(g6)--(g7)--(g8);
\draw (g5)--(g9);
\end{tikzpicture}
\ee
coupled to $(18n^2-74n+77)$ free hypermultiplets. For $n\leq3$ instead the second node from the left is underbalanced in \eqref{UVMQSUNeven+1AS+6FFromWeb} and after applying \eqref{SimoneDuality} we also end up with \eqref{MQE7HigherRank}. For $n=1$ the leftmost node is still underbalanced and a final application leads to the affine $\e_7$ Dynkin diagram.

For $N=2n+1$ we can apply very similar monodromies to arrive at a bare magnetic quiver
\be
\label{UVMQSUNodd+1AS+6FFromWeb}
\begin{tikzpicture}[x=.8cm,y=.8cm]
\node (g1) at (0,0) [gaugeb,label=below:{1}] {};
\node (g2) at (1.5,0) [gaugeb,label=below:{$b$-$n$+2}] {};
\node (g3) at (3.5,0) [gauge,label=below:{$2b$-$n$+2}] {};
\node (g4) at (5.5,0) [gauge,label=below:{$3b$-$n$+2}] {};
\node (g5) at (7.5,0) [gauge,label=below:{$4b$-$n$+2}] {};
\node (g6) at (9.5,0) [gauge,label=below:{$3b$-$n$+2}] {};
\node (g7) at (11.5,0) [gauge,label=below:{$2b$-$n$+2}] {};
\node (g8) at (13.5,0) [gaugeb,label=below:{$b$-$n$+2}] {};
\node (g9) at (7.5,1) [gauger,label=right:{$2b$}] {};
\draw (g1)--(g2)--(g3)--(g4)--(g5)--(g6)--(g7)--(g8);
\draw (g5)--(g9);
\end{tikzpicture}
\,,
\ee
where $b=n^2-n+2$.
After applying \eqref{SimoneDuality} this becomes
\be  
\begin{tikzpicture}[x=.8cm,y=.8cm]
\node (g1) at (0,0) [gaugeb,label=below:{1}] {};
\node (g2) at (1.5,0) [gaugeb,label=below:{$2n$}] {};
\node (g3) at (3,0) [gauge,label=below:{$4n$}] {};
\node (g4) at (4.5,0) [gauge,label=below:{$6n$}] {};
\node (g5) at (6,0) [gauge,label=below:{$8n$}] {};
\node (g6) at (7.5,0) [gauge,label=below:{$6n$}] {};
\node (g7) at (9,0) [gauge,label=below:{$4n$}] {};
\node (g8) at (10.5,0) [gauge,label=below:{$2n$}] {};
\node (g9) at (6,1) [gauge,label=right:{$4n$}] {};
\draw (g1)--(g2)--(g3)--(g4)--(g5)--(g6)--(g7)--(g8);
\draw (g5)--(g9);
\end{tikzpicture}
\,,
\ee
coupled to $\left(18n^2-61n+100\right)$ free hypermultiplets.

\subsection{Decoupling Trees for E-string GTPs}

In this subsection we illustrate the discussion above on a concrete example and we draw the GTPs explicitly. We focus on the analysis of the descendants of the $SU(N)_{\frac{N}{2}} + 1\bm{AS} + 8 \bm{F}$ theories for $N=6$.
A direct application of the formulas given in the main text give the pGTPs in table \ref{tabToricSU+ASmaxlevelEven}.

Once this is done, one has to apply appropriate monodromies to make the non-convex pGTPs convex. This is in general a daunting task, as the sequence of monodromies can be very involved. The hardest pGTPs to be made convex are the submarginal ones. In the present case, we focus on the pGTP in the box with labels $2$ and $-2n+1$ in table \ref{tabToricSU+ASmaxlevelEven}. It turns out a possible sequence of monodromies for that diagram is the following, which contains 50 terms
\begin{eqnarray}
\mathfrak{T} &=& 
\mathfrak{M}^{+}_{10}
\mathfrak{M}^{+}_{9}
\mathfrak{M}^{+}_{8}
\mathfrak{M}^{+}_{7}
\mathfrak{M}^{+}_{6}
\mathfrak{M}^{+}_{5}
\mathfrak{M}^{-}_{4}
\mathfrak{M}^{-}_{3}
\mathfrak{M}^{-}_{2}
\mathfrak{M}^{-}_{5}
\mathfrak{M}^{-}_{4}
\mathfrak{M}^{-}_{3}
\mathfrak{M}^{-}_{6}
\mathfrak{M}^{-}_{5}
\mathfrak{M}^{-}_{4}
\mathfrak{M}^{-}_{7}
\mathfrak{M}^{-}_{6}
\mathfrak{M}^{-}_{5}
\nonumber \\ && 
\mathfrak{M}^{-}_{8}
\mathfrak{M}^{-}_{7}
\mathfrak{M}^{-}_{6}
\mathfrak{M}^{-}_{9}
\mathfrak{M}^{-}_{8}
\mathfrak{M}^{-}_{7}
\mathfrak{M}^{-}_{10}
\mathfrak{M}^{-}_{9}
\mathfrak{M}^{+}_{11}
\mathfrak{M}^{-}_{10}
\mathfrak{M}^{-}_{9}
\mathfrak{M}^{-}_{8}
\mathfrak{M}^{-}_{11}
\mathfrak{M}^{-}_{10}
\mathfrak{M}^{-}_{9}
\mathfrak{M}^{-}_{12}
\mathfrak{M}^{-}_{1}
\mathfrak{M}^{-}_{2}
\nonumber \\ && 
\mathfrak{M}^{-}_{3}
\mathfrak{M}^{-}_{4}
\mathfrak{M}^{-}_{5}
\mathfrak{M}^{-}_{6}
\mathfrak{M}^{-}_{7}
\mathfrak{M}^{-}_{8}
\mathfrak{M}^{+}_{9}
\mathfrak{M}^{+}_{8}
\mathfrak{M}^{+}_{7}
\mathfrak{M}^{-}_{10}
\mathfrak{M}^{+}_{11}
\mathfrak{M}^{-}_{10}
\mathfrak{M}^{-}_{11}
\mathfrak{M}^{-}_{12}
\end{eqnarray}
This sequence is obtained by trial an error using an algorithm optimizing the area of the non-convex polygons obtained along the way. It is by no means claimed to be a shortest sequence. This monodromy leads to a GTP with vectors
\be
\mathfrak{T}\bm{u}=\left[\underbrace{(0,6)}_{5},(0,1),(-26,0),(0,-20),(0,-18),\underbrace{(13,0)}_{2},(0,7)\right]
\ee
One can easily compute the corresponding magnetic quiver, for instance by plugging the above set of coordinates into the code given in \cite{vanBeest:2020kou}. One then obtains the quiver which is reported in table \ref{tab:UVMQsEStringEven}. 
Once the sequence of monodromies is known for the submarginal theories, it can be used with little modification to all the other theories. For instance for the pGTP in box $(2,-2n)$ of Table \ref{tabToricSU+ASmaxlevelEven} one finds the GTP
\be
\mathfrak{T}\bm{u}=\left[(11,11),\underbrace{(0,17)}_{3},(-40,-40),\underbrace{(6,-6)}_{2},\underbrace{(5,-5)}_{2},(1,0),(6,0)\right]
\ee
for some modified monodromy $\mathfrak{T}$.
These GTPs, along with all other convex GTPs for the example at hand are reported in table \ref{tabToricSU+ASmaxlevelEvenMono}. We reproduce only the first two lines, as the other lines are already convex in table \ref{tabToricSU+ASmaxlevelEven} and stay the same. 
\begin{table}
\hspace*{-1cm}
\begin{tabular}{|c|c|c|c|c|c|c|c|c|c|}
 \hline
& $-2n+2$ & $-2n+1$ & $-2n$ & $-2n-1$ & $-2n-2$ & $-2n-3$ & $-2n-4$ & $-2n-5$ & $-2n-6$ \\
\hline
\end{tabular}
\caption{Theories of the form $SU(2n)_k+1\mathbf{AS}+ N_F \mathbf{F}$ which are descendants of $SU(2n)_n+1\mathbf{AS}+ 8 \mathbf{F}$. We label them by $\Delta_{\pm} = \frac{N_F}{2} -n-2 \pm k$.  To illustrate, we have chosen $n=3$, but the structure of the table does not depend on $n$.  }
\label{tabToricSU+ASmaxlevelEven}
\end{table}

\begin{sidewaystable}

\scalebox{.9}{%
\begin{tabular}{|c|c|c|c|c|c|c|c|c|c|}
 \hline
& $-2n+2$ & $-2n+1$ & $-2n$ & $-2n-1$ & $-2n-2$ & $-2n-3$ & $-2n-4$ & $-2n-5$ & $-2n-6$ \\
\hline
\end{tabular}
}
\caption{First two lines of table \ref{tabToricSU+ASmaxlevelEven} for $n=3$ after monodromies have been applied. For ease of read, we give the coordinates of the two largest polygons in the main text. }
\label{tabToricSU+ASmaxlevelEvenMono}
\end{sidewaystable}


\clearpage

\bibliography{FM}
\bibliographystyle{JHEP}

\end{document}